\documentclass[twocolumn]{aastex631}
\usepackage{mathtools}

\newcommand{\Gyr}{\,{\rm Gyr}}
\newcommand{\Myr}{\,{\rm Myr}}
\newcommand{\kms}{\,{\rm km}\,{\rm s}^{-1}}
\newcommand{\mgfe}[0]{[{\rm Mg/Fe}]}

\newcommand{\afe}[0]{[\alpha/{\rm Fe}]}

\newcommand{\mgh}{[{\rm Mg}/{\rm H}]}
\newcommand{\feh}[0]{[{\rm Fe/H}]}

\newcommand{\mgfecc}{[{\rm Mg}/{\rm Fe}]_{\rm cc}}

\newcommand{\Msun}{M_{\odot}}

\newcommand{\xh}{[{\rm X}/{\rm H}]} 

\newcommand{\yxcc}{y_{\rm X}^{\rm cc}}

\newcommand{\yfecc}{y_{\rm Fe}^{\rm cc}}

\newcommand{\ymgcc}{y_{\rm Mg}^{\rm cc}}

\newcommand{\yfeIa}{y_{\rm Fe}^{\rm Ia}}

\newcommand{\ironratio}{y_{\mathrm{Fe}}^{\mathrm{Ia}}/y_{\mathrm{Fe}}^{\mathrm{cc}}}

\newcommand{\Zfesun}{Z_{{\rm Fe},\odot}}

\newcommand{\Zmgsun}{Z_{{\rm Mg},\odot}}
\newcommand{\Zmg}{Z_{{\rm Mg}}}
\newcommand{\Zmgeq}{Z_{\rm Mg,eq}}

\newcommand{\taustar}{\tau_*}

\newcommand{\taubar}{\bar{\tau}}
\newcommand{\tauIa}{\tau_{\rm Ia}}

\newcommand{\Mdotstar}{\dot{M}_*}
\newcommand{\Mdotout}{\dot{M}_{\rm out}}

\newcommand{\meanage}{{\langle {\rm age} \rangle}}
\newcommand{\meanmgh}{{\langle [{\rm Mg}/{\rm H}] \rangle}}
\newcommand{\meanmgfe}{{\langle [{\rm Mg}/{\rm Fe}] \rangle}}
\newcommand{\meanfeh}{{\langle [{\rm Fe}/{\rm H}] \rangle}}
\newcommand{\Mgas}{M_{\rm gas}}
\newcommand{\tstart}{t_{\rm start}}
\newcommand{\tcut}{t_{\rm cut}}
\newcommand{\tform}{t_{\rm form}}
\newcommand{\tobs}{t_{\rm obs}}
\newcommand{\tdmin}{t_{{\rm d,min}}}
\newcommand{\Mstarform}{M_{*,{\rm form}}}

\newcommand{\Eqref}[1]{Equation~(\ref{eqn:#1})}

\defcitealias{Weinberg2024}{WGJT24}

\shortauthors{Gountanis et al.}

\graphicspath{{./}{figures/}}

\begin{document}

\title{Modeling the Ages and Chemical Abundances of Elliptical Galaxies}

%\correspondingauthor{Nicole Gountanis}
%\email{gountanis.1@osu.edu}

\author[0009-0008-8797-0865]{Nicole Marcelina Gountanis}
\author[0000-0001-7775-7261]{David H. Weinberg}
\affiliation{Department of Astronomy and Center for Cosmology and AstroParticle Physics, The Ohio State University, Columbus, OH 43210, USA}

\author[0000-0002-9861-4515]{Aliza G. Beverage}
\affiliation{Astronomy Department, University of California, Berkeley, CA 94720, USA}

\author[0000-0002-7393-3595]{Nathan R. Sandford}
\affiliation{Department of Astronomy and Astrophysics, University of Toronto, 50 St. George Street, Toronto, ON M5P 0A2, Canada}

\author[0000-0002-1590-8551]{Charlie Conroy}
\affiliation{Center for Astrophysics|Harvard \& Smithsonian, Cambridge, MA 02138, USA}

\author[0000-0002-7613-9872]{Mariska Kriek}
\affiliation{Leiden Observatory, Leiden University, P.O. Box 9513, 2300 RA Leiden, The Netherlands}

\begin{abstract}
\noindent Spectroscopic studies of elliptical galaxies show that their stellar population ages, mean metallicity, and $\alpha$-enhancement traced by $\mgfe$ all increase with galaxy stellar mass or velocity dispersion. We use one-zone galactic chemical evolution (GCE) models with a flexible star formation history (SFH) to model the age, $\mgh$, and $\mgfe$ inferred from SSP fits to observed ellipticals at $z \sim 0$ and $z \sim 0.7$. We show that an SSP fit to the spectrum computed from a full GCE model gives ages and abundances close to the light-weighted, logarithmically averaged values of the composite stellar population, $\meanage$, $\meanmgh$, and $\meanmgfe$. With supernova Mg and Fe yields fixed to values motivated by Milky Way stellar populations, we find that predicted $\meanmgh$-$\meanage$ and $\meanmgfe$-$\meanage$ relations are surprisingly insensitive to SFH parameters: older galaxies have higher $\meanmgfe$, but the detailed form of the SFH has limited impact. The star formation efficiency and outflow efficiency affect the early and late evolution of $\meanmgh$, respectively; explaining observed trends requires higher star formation efficiency and lower outflows in more massive galaxies. With core collapse supernova yields calibrated to the plateau $\mgfecc\approx0.45$ observed in many Milky Way studies, our models underpredict the observed $\meanmgfe$ ratios of ellipticals by 0.05-0.1 dex. Increasing the core collapse yield ratio to $\mgfecc=0.55$ improves the agreement, though the models remain below the data. We discuss potential resolutions of this discrepancy, including the possibility that many ellipticals terminate their star formation with a self-enriching, terminating burst that reduces the light-weighted age and boosts $\meanmgfe$. 
\end{abstract}

\keywords{}

\section{Introduction} \label{sec:intro}

Population synthesis modeling of the spectra of elliptical galaxies
%\footnote{\NMG{The galaxies examined in this study were chosen based on their morphology and star formation history. For simplicity, we refer to all of these galaxies as ellipticals throughout the paper.}} 
provides insights into the ages and chemical abundances of their stellar populations. At $z\sim 0$, ellipticals of higher stellar mass or higher velocity dispersion exhibit systematically older ages, higher metallicities, and larger enhancements of $\alpha$-elements relative to iron \citep{Thomas2005, Schiavon2007, Graves2009, Smith2009, Johansson2012, Conroy2014, Greene2019}. Recent studies of early-type galaxies at $z \sim 0.7$ show near-solar metallicity ($\feh \sim0$) and strong $\alpha$-enhancement ($\mgfe \sim 0.2-0.3$), but the trends with velocity dispersion are weaker than those at $z\sim 0$ \citep{Beverage2021,Beverage2023}. The elevated $\mgfe$ ratios of high-redshift ellipticals and the most massive low-redshift ellipticals are interpreted as a sign of rapid star formation, with many stars forming before Type Ia supernovae (SNIa) have had time to make large contributions to the abundance of iron-peak elements.

Traditional observational analyses fit the integrated spectrum of an early-type galaxy as a simple stellar population (SSP) of specified age and chemical composition, but of course, the stars form over time from an interstellar medium (ISM) with steadily evolving elemental abundances. In this paper, we use parameterized galactic chemical evolution (GCE) models to predict the light-weighted age, $\mgh$, and $\mgfe$ of observed elliptical populations. We also use these models to test how well the SSP fit recovers the true light-weighted quantities of the full population. We focus on Mg as a well-measured element that is expected to come entirely from core-collapse supernovae (CCSN), and Fe as a well-measured element with contributions from both CCSN and SNIa. We use $\mgh$ rather than $\feh$ as our primary measure of overall metallicity because our predictions for it do not depend on our assumptions about SNIa, and the two quantities are measured comparably well from observed spectra. We adopt one-zone GCE models that treat the star-forming gas reservoir as chemically homogeneous at a given time, an approximation that we consider adequate for predicting integrated properties, though it does not allow us to model radial gradients in age or abundance.

The population-averaged Mg and Fe yields of CCSN and the population-averaged Fe yield and delay time distribution (DTD) of SNIa are crucial inputs to our models. We adopt values that are empirically motivated by observations of Milky Way stellar abundances and cosmic SNIa rates, implicitly assuming that the stellar scale astrophysics that determines these quantities is similar in elliptical galaxies despite their distinctive star formation histories. We aim to address three broad questions:  
\begin{enumerate}
\item Given these yields motivated by independent observations, can GCE models with reasonable parameter values reproduce the ages, metallicities, and $\alpha$-enhancements of observed ellipticals at $z\sim 0$ and $z\sim 0.7$?
\item If so, what do the observed properties imply about the star formation histories, star formation efficiencies, and outflows of these galaxies?
\item If not, what modifications might be required to the stellar yields or other model assumptions, and what do they imply about the distinctive features of elliptical galaxies?
\end{enumerate}

We rely principally on analytic GCE models (\citealt{Weinberg2017equ}, hereafter WAF) because they are sufficient for our purposes here. We provide a {\tt Python} implementation of these analytic models, \texttt{fanCE}\footnote{\texttt{fanCE} (pronounced fan-C-E) for \textbf{F}lexible \textbf{An}alytic \textbf{C}hemical \textbf{E}volution} \citep{Gountanis2024}, so that others can easily adapt them to interpret other galaxies or stellar populations or explore alternative assumptions (see Appendix \ref{appx:code}). It would be straightforward to extend our approach using numerical models that allow more flexible assumptions such as bursty star formation histories and metallicity-dependent yields.

While the high metallicities and $\alpha$-enhancements of ellipticals are usually attributed to ``rapid star formation,'' this phrase is imprecise. In our models, there are four distinct timescales in the star formation history: the time at which star formation begins, the duration of an initial linear rise, the $e$-folding timescale of a subsequent decline, and the time at which star formation ceases entirely. The star formation efficiency, ${\rm SFE}\equiv \Mdotstar/\Mgas \equiv \taustar^{-1}$, defines another important timescale. One goal of our modeling is to tease apart the role of these distinct timescales in governing ages and chemical enrichment. While our flexible parameterization of star formation history (SFH) allows a variety of behaviors, all our models assume a sharp final transition from star formation to quiescence. 

Stellar population trends in the massive quiescent galaxy population were first identified in the low-redshift universe using Lick indices \citep{Worthey1994, Thomas1999, Trager2000b, Thomas2005}. These trends have since been confirmed using a variety of methods, including full-spectrum fitting with variable abundance patterns \citep{Walcher2009, Conroy2014}. By stacking thousands of SDSS spectra, \citet{Conroy2014} measure the average elemental abundances of 11 different elements as a function of velocity dispersion. Their analysis reproduces the well-established trends at $z\sim0$ with high precision and adds many individual elements.

One way to separate the star-formation histories from the assembly histories of massive quiescent galaxies is by linking the $z\sim0$ galaxies to their progenitors at higher redshift. However, measuring stellar population properties and chemical compositions becomes increasingly challenging at higher redshift, relying on already faint absorption lines that are shifted to near-infrared wavelengths. Studies of large samples of quiescent galaxies at $z\sim0.7$ find that their abundance patterns look remarkably similar to today's massive elliptical galaxies \citep{Beverage2023}. However, the lower velocity dispersion quiescent galaxy population at $z\sim0.7$ appears to have both higher [$\alpha$/Fe] and earlier formation redshifts compared to the $z\sim0$ population \citep{Leethochawalit2018, Beverage2023}. This finding shows that the $z\sim0$ ellipticals are not simply the passively evolved descendants of the $z\sim 0.7$ quiescent population.

Galaxies in observational samples such as \cite{Conroy2014} and \cite{Beverage2023} are selected by a combination of early-type (bulge-dominated) morphology and photometric or spectroscopic diagnostics that imply little or no ongoing star formation.  Throughout this paper we use the term ``ellipticals'' as shorthand for this combination of characteristics, even though many galaxies in these samples would not have a strictly elliptical morphological classification.  From the point of view of chemical evolution, the critical feature of these galaxies is their quiescent star formation history, with minimal or zero star formation for at least the last Gyr prior to the epoch at which they are observed.  Because the galaxies in our observational samples have a minimum velocity dispersion $\sigma \geq 80\kms$, we sometimes refer to them as ``massive ellipticals'' or ``massive quiescent galaxies'' to distinguish them from dwarf elliptical or dwarf spheroidal populations, which may have different evolutionary histories.

Full GCE modeling allows us to go beyond the empirical trends of SSP fits to the physical processes that drive them --- nucleosynthesis, SFH, SFE, feedback, and quenching. Section \ref{sec:data} describes our observational samples and SSP fitting tools and summarizes tests of our basic methodology, predicting SSP quantities from the light-weighted average quantities of composite stellar populations. Section \ref{sec:models} describes our GCE models, with particular attention to our flexible SFH parameterization, our choice of empirically motivated Mg and Fe yields, and the impact of model parameters on the evolution of the mean age, $\mgh$, and $\mgfe$ of stellar populations. In Section \ref{sec:observations} we compare our model predictions to the observed trends. Our models struggle to produce $\mgfe$ values as high as the observed values, so we also consider variant models with reduced SNIa Fe production or a terminating burst of star formation. Section \ref{sec:discussion} provides further discussion of nucleosynthetic yields and identifies directions of future investigation. We summarize our conclusions in Section \ref{sec:conclusions}. Appendix \ref{appx:code} gives instructions for accessing \texttt{fanCE}; Appendix \ref{appx:burst} gives details of our treatment of terminating bursts of star formation; and Appendix \ref{appx:mock} tests the relation between SSP fits and the light-weighted average quantities of composite stellar populations. 

\section{Observational data}
\label{sec:data}

\begin{deluxetable}{ccccccc}
\tabletypesize{\footnotesize}\tablecolumns{8}
\tablecaption{Stellar age and abundance results from stacked SDSS ($z\sim0$) and LEGA-C ($z\sim0.7$) spectra in bins of velocity dispersion. All values are from \citet{Beverage2023}. \label{tab:alf-results}}
\tablehead{\colhead{$\sigma$ ($\kms$)} & \colhead{Age (Gyr)} & \colhead{[Fe/H]} & \colhead{[Mg/H]} & \colhead{[Mg/Fe]}}
\startdata
 & & \textbf{SDSS} & & &\vspace{0.1cm}\\
87 & $5.5^{+0.1}_{-0.1}$ & $-0.138^{+0.005}_{-0.005}$ & $0.007^{+0.005}_{-0.005}$ & $0.145^{+0.004}_{-0.004}$\\ 
112 & $6.0^{+0.1}_{-0.1}$ & $-0.109^{+0.005}_{-0.005}$ & $0.065^{+0.005}_{-0.005}$ & $0.174^{+0.004}_{-0.004}$\\ 
138 & $6.7^{+0.6}_{-0.1}$ & $-0.085^{+0.008}_{-0.020}$ & $0.106^{+0.007}_{-0.019}$ & $0.191^{+0.005}_{-0.004}$\\ 
170 & $7.9^{+0.1}_{-0.1}$ & $-0.072^{+0.005}_{-0.006}$ & $0.155^{+0.006}_{-0.005}$ & $0.227^{+0.004}_{-0.004}$\\ 
200 & $9.4^{+0.1}_{-0.1}$ & $-0.064^{+0.004}_{-0.005}$ & $0.188^{+0.005}_{-0.005}$ & $0.252^{+0.004}_{-0.004}$\\ 
246 & $10.3^{+0.1}_{-0.1}$ & $-0.059^{+0.005}_{-0.005}$ & $0.240^{+0.005}_{-0.005}$ & $0.299^{+0.004}_{-0.004}$\\ 
299 & $10.7^{+0.2}_{-0.2}$ & $-0.012^{+0.007}_{-0.007}$ & $0.292^{+0.007}_{-0.007}$ & $0.305^{+0.005}_{-0.005}$\\ 
\hline
 & & \textbf{LEGA-C} & & &\vspace{0.1cm}\\
147 & $3.0^{+0.2}_{-0.2}$ & $-0.10^{+0.03}_{-0.03}$ & $0.18^{+0.03}_{-0.03}$ & $0.28^{+0.03}_{-0.03}$\\ 
176 & $3.8^{+0.3}_{-0.3}$ & $-0.09^{+0.03}_{-0.03}$ & $0.21^{+0.02}_{-0.03}$ & $0.30^{+0.03}_{-0.03}$\\ 
207 & $3.6^{+0.4}_{-0.3}$ & $-0.05^{+0.03}_{-0.03}$ & $0.22^{+0.03}_{-0.03}$ & $0.27^{+0.03}_{-0.02}$\\ 
247 & $4.4^{+0.3}_{-0.3}$ & $-0.03^{+0.02}_{-0.02}$ & $0.28^{+0.03}_{-0.02}$ & $0.31^{+0.03}_{-0.02}$\\ 
\enddata
\end{deluxetable}

For our observational comparisons in Section \ref{sec:observations} we use the stellar ages, [Mg/H], and [Mg/Fe] reported by \cite{Beverage2023}, who measure these quantities for massive quiescent galaxies in bins of velocity dispersion using full-spectrum absorption line fitting. \cite{Beverage2023} examine two galaxy samples, one at $z\sim0$ and one at $z\sim0.7$. 

The $z\sim0$ sample was originally selected by \cite{Conroy2014} from SDSS Data Release 7 \citep{Abazajian2009}. Galaxies were chosen from a narrow redshift interval $0.025<z<0.06$ and identified as quiescent using emission line fluxes. Structural properties were also used to identify galaxies with early-type morphologies. The sample was split into seven evenly spaced bins of velocity dispersion centered at $\sigma=87-299 \kms$.

The $z\sim0.7$ sample was selected by \cite{Beverage2023} from the LEGA-C survey. LEGA-C is a deep spectroscopic survey of massive galaxies at intermediate redshifts ($0.5\lesssim z\lesssim1.0$) conducted with VIMOS on the VLT \citep{vanderWel2016, Straatman2018, vanderWel2021}. Galaxies were identified as quiescent using their rest-frame $UVJ$ colors \citep{Wuyts2007,Muzzin2013a}. The sample was limited to galaxies with a rest-frame wavelength coverage of key absorption features (H$\beta$, Mg\textsc{I}, Fe\textsc{I}), translating to a redshift selection of $z\leq0.75$. The sample was further limited to only include spectra with S/N$>$15 to ensure that the elemental abundances were accurately recovered. The typical S/N of a galaxy in the LEGA-C sample is 55. The $z\sim0.7$ galaxies were split equally among four bins of velocity dispersion centered at $\sigma=147-247 \kms$.

The average stellar population properties of the $z\sim0$ and $z\sim0.7$ galaxies in bins of velocity dispersion were derived by \cite{Beverage2023} using two distinct methods. For the $z\sim0$ sample, \cite{Beverage2023} utilized the same stacked SDSS spectra as \cite{Conroy2014}. In contrast, for the $z\sim0.7$ galaxies, they employed a hierarchical Bayesian framework to effectively ``stack’’ the spectra. Specifically, they divided the LEGA-C galaxies into four velocity dispersion bins and fit all spectra within each bin simultaneously. This approach allowed them to constrain both the individual galaxy properties and the overall distribution of these properties for galaxies in each bin. These population-level properties were parameterized as a normal distribution, meaning that the derived LEGA-C properties and uncertainties reflect the mean of this distribution. While this method avoids traditional stacking, the resulting measurements can still be thought of as analogous to those from a stacked spectrum. In fact, \cite{Beverage2023} directly compared the two methods and found good agreement between them. Therefore, the ``average’’ galaxy properties of the $z\sim0.7$ and $z\sim0$ samples, as presented in \cite{Beverage2023} and utilized throughout this work, are directly comparable despite the differing methods used to derive them.

\subsection{Stellar Population Fitting}
The stellar population fitting is accomplished using the full-spectrum absorption line fitting code, \texttt{alf} (Absorption-Line Fitter) \citep{Conroy2012, Conroy2018}. \texttt{alf} generates simple stellar population (SSP) models with variable elemental abundances for the optical-NIR. The code employs an MCMC wrapper \citep{Foreman-Mackey2013} to estimate the posteriors of the stellar population properties and elemental abundances. 

The SSP model ingredients include metallicity-dependent MIST isochrones \citep{choi2016} and the MILES and Extended-IRTF empirical spectral libraries \citep{Sanchez-Blazquez2006, Villaume2017}. To alter the abundances of the 19 individual elements in these models, \texttt{alf} uses metallicity- and age-dependent synthetic response functions to determine the fractional change in spectra due to the variation of each element. \texttt{alf} models the continuum-normalized spectrum, so it does not incorporate constraints from the overall shape of the spectral energy distribution (SED). 

\cite{Conroy2014} also used \texttt{alf} to measure the stellar population properties of the SDSS stacked spectra used here. However, since the publication of \cite{Conroy2014}, the SSP models have been improved, with metallicity- and age-dependent response functions, updated isochrones, and improved interpolation of the empirical spectral libraries \citep{Villaume2017, Conroy2018}. Therefore in this analysis, we adopt the updated SDSS measurements from \cite{Beverage2023}. 

\begin{figure*}[!]
    \includegraphics[width = 1.0\textwidth]{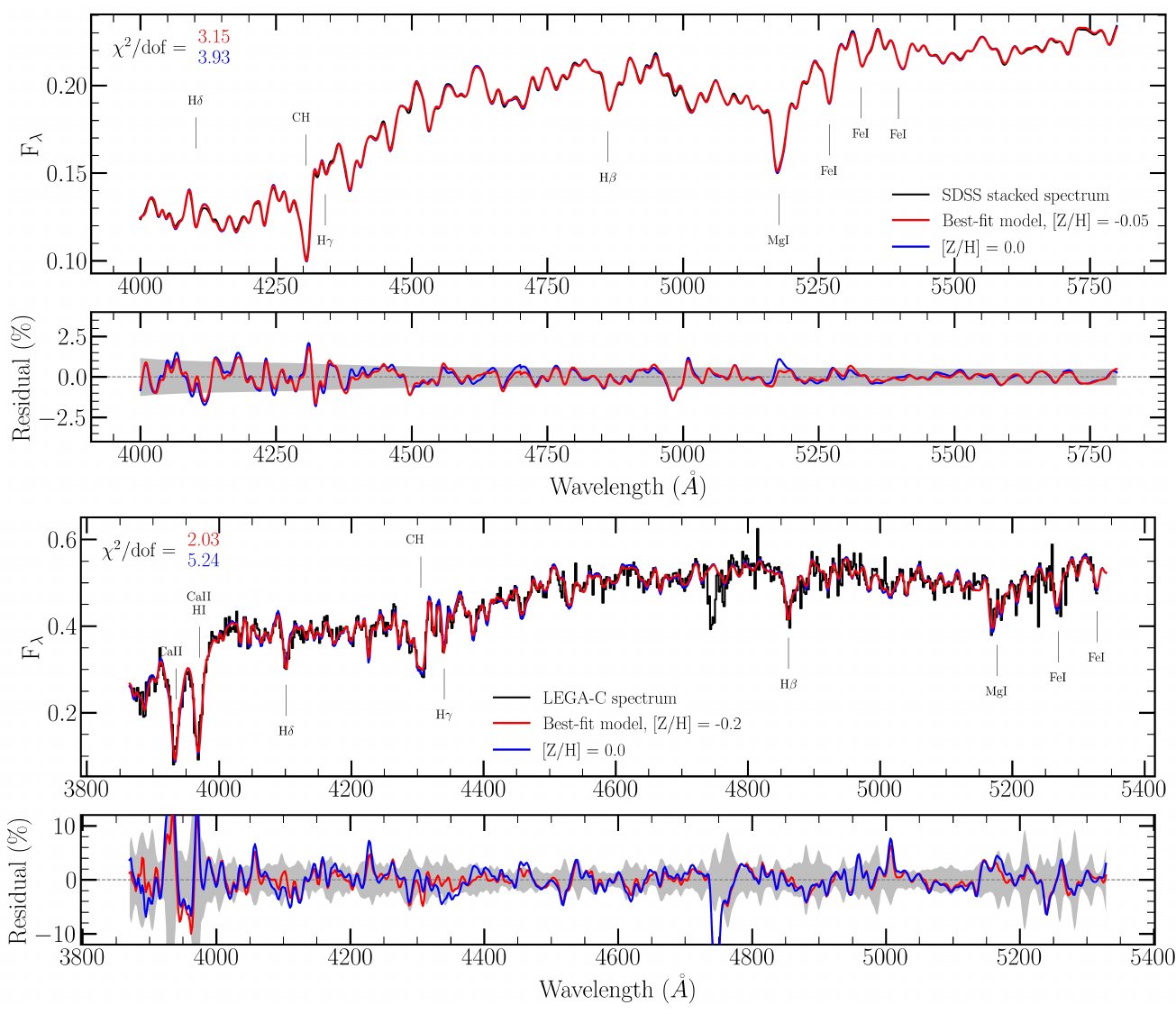}
    \label{subfig:a}
    \caption{(Top) Stacked SDSS spectrum ($z\sim0$) of a velocity dispersion bin (black) and the best-fit SSP spectrum (red) computed with \texttt{alf}. The blue curve shows the \texttt{alf} model with $\feh$ raised from its best-fit value of $\sim0.05$ to 0.0, a change of $3\times$ the fit uncertainty in $\feh$. Other parameters are held fixed. The lower panel plots the fractional residuals between the two SSP models and the observed spectrum. (Bottom) similar to the top, but for an individual LEGA-C galaxy spectrum at $z\sim0.7$. The blue curve again shows a model with $\feh$ perturbed by $3\sigma$ from the best-fit value. Gray bands show the statistical noise per spectral pixel. In the top panels, spectra are binned such that one pixel is $\simeq1$ \AA\ in the rest-frame for both SDSS and LEGA-C.}
    \label{fig:spectra}
\end{figure*}

% \begin{figure*}[!]
%     \begin{subfigure}{\textwidth}
%     \centering
%     \includegraphics[width = 0.9\textwidth]{SDSS_bestfit.png}
%     \label{subfig:a}
%     \end{subfigure}
%     \begin{subfigure}{\textwidth}
%     \centering
%     \includegraphics[width = 0.93\textwidth]{LEGAC_bestfit.png}
%     \label{subfig:b}
%     \end{subfigure}
%     \vspace{-0.5cm}
%     \caption{(Top) Stacked SDSS spectrum ($z\sim0$) of a velocity dispersion bin (black) and the best-fit SSP spectrum (red) computed with \texttt{alf}. The blue curve shows the \texttt{alf} model with $\feh$ raised from its best-fit value of $\sim0.05$ to 0.0, a change of $3\times$ the fit uncertainty in $\feh$. Other parameters are held fixed. The lower panel plots the fractional residuals between the two SSP models and the observed spectrum. (Bottom) similar to the top, but for an individual LEGA-C galaxy spectrum at $z\sim0.7$. The blue curve again shows a model with $\feh$ perturbed by $3\sigma$ from the best-fit value. Gray bands show the statistical noise per spectral pixel. In the top panels, spectra are binned such that one pixel is $\simeq1$ \AA\ in the rest-frame for both SDSS and LEGA-C.}
%     \label{fig:spectra}
% \end{figure*}

The stellar population properties for the $z\sim0$ and $z\sim0.7$ populations from \cite{Beverage2023} are tabulated in Table \ref{tab:alf-results}. In the top panel of Figure \ref{fig:spectra}, we show the observed spectrum and best-fit model spectrum for one velocity dispersion stack from the SDSS. Since the LEGA-C spectra were all fit simultaneously using a hierarchical Bayesian model, we show an example individual spectrum in the bottom panel and its corresponding best-fit model. In both panels, we show the spectrum for a model in which we increase the total metallicity, [Z/H] by 3$\sigma$ while keeping other parameters fixed. The $\chi^2/{\rm dof}$ for each model is shown in the upper left. The residuals and $\chi^2/{\rm dof}$ for the perturbed model are larger, as expected, but the differences are subtle. The response of the predicted spectrum to changes in $\feh$ or $\mgh$ is spread over much of the spectrum, as shown by \cite{Conroy2014, Conroy2018}, so a high S/N allows precise constraints even though the changes are small. While this translates to substantial uncertainties for individual LEGA-C galaxies, the hierarchical approach employed by \cite{Beverage2023} provides a much more precise measurement than what is found for an individual LEGA-C spectrum shown here.

\subsection{SSP Fits vs. Light-Weighted Averages}
\label{sec:sspassumption}
The age and abundance measurements in Table \ref{tab:alf-results}  are derived by modeling the observed spectrum as produced by a single stellar population, even though the light is a composite of populations with different ages and abundances. This approach is a long-standing practice in the study of elliptical galaxies \citep[e.g.,][]{worthey_comprehensive_1994, thomas_stellar_2003, vazdekis_evolutionary_2010, Conroy2014}. With the flexible GCE models used in this paper, it would be possible to compute composite spectra for each model and compare them to data directly in spectral space. However, in this paper we adopt the simpler procedure of computing the light-weighted average quantities from models and comparing them to the empirical values from SSP fits. 

\begin{figure}[!]
\centering
\includegraphics[width = \columnwidth]{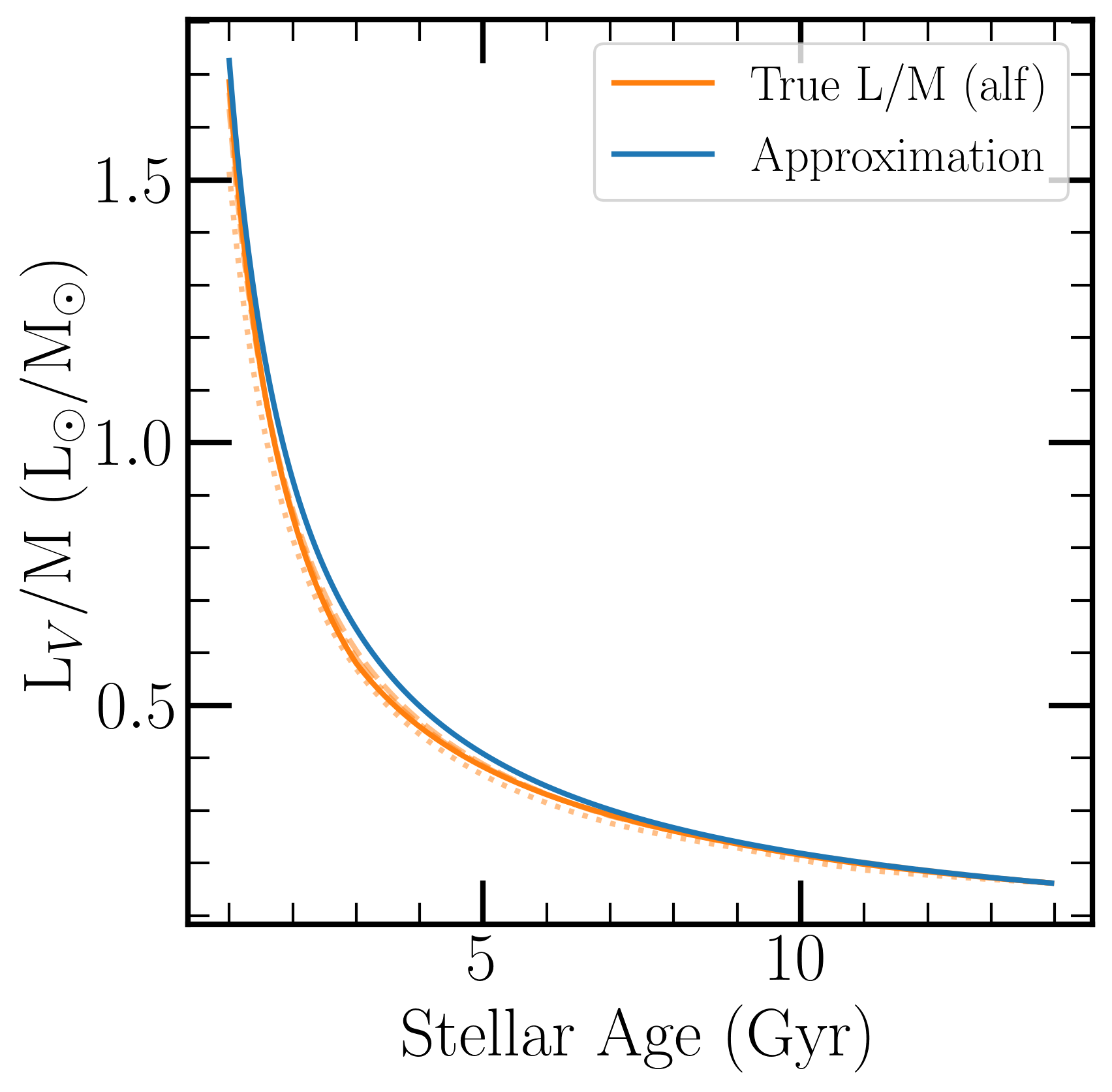}
\vspace{-5mm}
\caption{$L_\mathrm{V}/M$ as a function of stellar population age. The V-band luminosity is used to compute light-weighted quantities from the SFH+GCE models. The solid orange line shows the true $L_\mathrm{V}/M$, as measured from the [Fe/H]$=0$} \citet{Conroy2014} SSP models, while the blue line shows the adopted approximation (Eq. \ref{eqn:weighing}). The dashed, dotted, and dash-dotted transparent orange lines correspond to SSP models with varying [Fe/H] = (-1, -0.5, -0.2), respectively. The adopted light-weighing function appropriately captures the true $L_\mathrm{V}/M$ and is robust to variations in [Fe/H].
\label{fig: lightweighing}
\end{figure}

Younger stellar populations have higher luminosity per unit stellar mass, so to compute model averages that can be compared to SSP fits we must include the impact of varying $L/M$ ratio. In Figure \ref{fig: lightweighing}, orange curves show the V-band $L_\mathrm{V}/M$ predicted by \texttt{alf} as a function of population age for three different $\feh$ values. The blue curve shows an approximation that we use to weight stellar populations in our models, 
\begin{equation}
w(t)_{\mathrm{light}} = L_\mathrm{V}/M = 1.5\left(\frac{t_{\mathrm{obs}}-t}{\mathrm{Gyr}} + 0.1\right)^{-\beta} \frac{L_{\odot}}{M_{\odot}},
\label{eqn:weighing}
\end{equation}
with $\beta=0.90$. Here, $t_{\mathrm{obs}}$ is the age of the universe at the galaxy's observed redshift ($t_{\mathrm{obs}}=7$ and 14 Gyr for $z=0.7$ and $z=0$, respectively), and $t_{\mathrm{obs}}-t$ is the age of the stellar population formed at time $t$. We add 0.1 to avoid divergence at $t=t_{\mathrm{obs}}$, but our model predictions are insensitive to this choice, and the youngest stellar populations in the models we compare to data in Section \ref{sec:data} have ages of 1-2 Gyr. The metallicity dependence of $L_\mathrm{V}/M$ is small compared to the age dependence, so we ignore it in our light-weighting. We have checked that our predictions are insensitive to the precise value of $\beta$, though ignoring light-weighting entirely ($\beta=0$) gives biased results. 

We investigate the accuracy of our light-weighted average method by performing a series of simple tests similar to those of \citet{serra_interpretation_2007}. We describe these tests and their results in Appendix \ref{appx:mock}. In brief, we use \texttt{alf} to compute the spectrum of a composite stellar population (CSP) comprised of two SSPs that have different $\mgh$, $\mgfe$, or age. We then fit this composite spectrum with \texttt{alf} assuming a single population, and we compare the SSP parameters to the light-weighted average of the two input populations. We find excellent agreement between the averaged and SSP-fit values of $\mgh$ and $\mgfe$, and our tests show that it is more accurate to average these logarithmic quantities than to average (Mg/H) and (Mg/Fe) before taking logarithms. Similarly, we find that the mean log(age) agrees with the SSP value, and that linear age averaging is less accurate. For large age differences between the two populations, the log(age) average slightly overestimates the SSP value, e.g., by 0.1 dex for populations that differ by 0.7 dex. Our findings are consistent with those of \citet{serra_interpretation_2007}.

In addition to these two-population tests, we conduct an end-to-end test using continuous GCE models in Section \ref{sec:mock} (see Figure \ref{fig: mock}). These tests confirm the viability of our approach, with the caveat that SSP ages are slightly younger ($\sim1$ Gyr) than the light-weighted mean ages for the oldest galaxies ($\meanage>10$ Gyr) at $z\sim0$.

\section{Galactic chemical evolution models}
\label{sec:models}

\subsection{Star Formation History}
\label{sec:models_sfh}

To interpret the observational data described in Section \ref{sec:data}, we use one-zone models with accretion and outflows treated using the analytic solutions of WAF. We adopt a ``rise-fall" star formation history \citep{Johnson2021}, 
\begin{equation}
\rm{SFH} \equiv \dot{M}_{\star}(t)\propto (1-e^{-t/\tau_{1}})\cdot e^{-t/\tau_{2}},
\label{eqn:SFH}
\end{equation}
with parameters $\tau_1$ and $\tau_2$ describing the rising and falling SFH timescales, respectively. For modeling elliptical galaxies, we consider this form superior to the one-parameter exponential SFH, in which $\dot{M}_{\star}(t)$ begins immediately at its maximum value, or the linear-exponential (a.k.a. ``delayed tau") model, in which the timescale of the rising phase is locked to that of the declining phase. The application of WAF solutions to this SFH is described in the Appendix of \cite{Weinberg2024}, hereafter \citetalias{Weinberg2024}. Although the $\tau_1$ parameter provides flexibility in the early SFH, we do not want to require that the initially linear form already applies at $t=0$ (i.e., at extremely high redshift). We therefore allow $t_{\mathrm{start}}$ to be another adjustable parameter, with $t_{\mathrm{start}}=0.5$ Gyr as our reference choice and $t_{\mathrm{start}}=$ 0 and 1 Gyr as alternatives. We take $t_\mathrm{obs}=14$ Gyr and $t_\mathrm{obs}=7$ Gyr as the age of the Universe at $z=0$ and $z=0.7$, respectively. 

\begin{figure}
\centering
\includegraphics[width=\columnwidth]{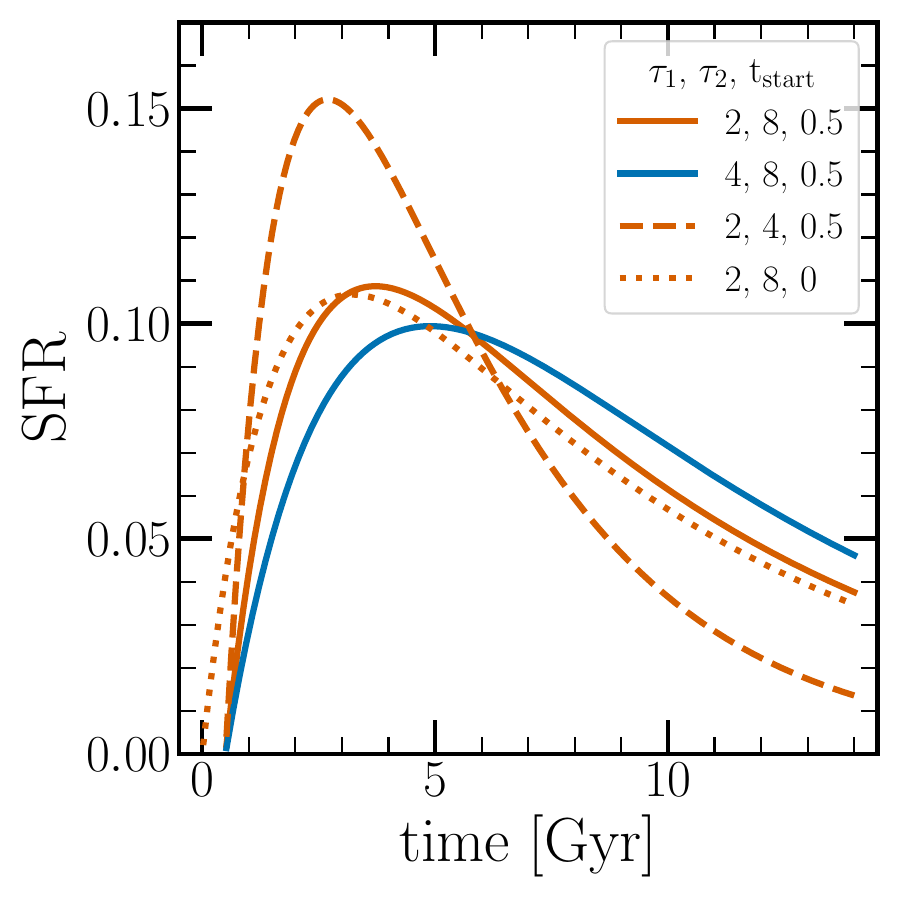}
\caption{Examples of star formation histories for different $\tau_1$, $\tau_2$, $t_{\mathrm{start}}$ (see Equation \ref{eqn:SFH}). The solid, orange curve represents our reference SFH with $\tau_1=2$ Gyr, $\tau_2=8$ Gyr, and $t_{\mathrm{start}}=0.5$ Gyr. The solid blue and dashed orange curves show the impact of longer rising timescales ($\tau_1=4$ Gyr) and shorter declining timescales ($\tau_2=4$ Gyr), respectively. The dotted orange curve shows the reference SFH with an earlier start of star formation ($t_{\mathrm{start}}=0$ Gyr). The SFR normalization gives the fraction of the galaxy's stellar mass formed per Gyr. Our elliptical galaxy SFH models are parameterized by $\tau_1$, $\tau_2$, $\tstart$, and the time $\tcut$ at which star formation cuts off abruptly.} 
\label{fig:sfh}
\end{figure}

Figure \ref{fig:sfh} illustrates SFHs for a variety of parameter choices. For most calculations in this section, we will consider variations about a reference model with $t_{\mathrm{start}}=0.5$ Gyr, $\tau_1=2$ Gyr, and $\tau_2=8$ Gyr (solid orange curve). For $\tau_1 \ll \tau_2$, the SFH reaches a maximum at $t \approx t_{\mathrm{start}} + \tau_1$, so 2.5 Gyr ($z \approx$ 2.6) in our reference case, before settling into exponential decline at $t \gg \tau_1$. Changing $\tau_2$ to 4 Gyr leads to a sharper maximum and a more rapid decline, while changing $\tau_1$ to 4 Gyr leads to a slower rise and a later, broader maximum. Changing $t_{\mathrm{start}}$ to 0 simply translates the reference SFH in time. For our elliptical models, \textit{we also assume that the SFH truncates sharply at a time $t_{\mathrm{cut}}$.}  Our model SFHs have four adjustable timescale parameters --- $t_{\mathrm{start}}$, $\tau_1$, $\tau_2$, and $t_{\mathrm{cut}}$ --- and $t_{\mathrm{cut}}$ is the one with the largest impact on the predicted mean stellar age and abundances. The form of Equation \ref{eqn:SFH} allows a much wider range of behavior than shown in Figure~\ref{fig:sfh}, including a linearly rising SFH, a pure exponential, and a rise followed by a constant SFR plateau. The parameter choices here represent values we consider plausible for massive ellipticals. 

\subsection{Yields}
\label{sec:models_yields}

Nucleosynthetic yields play a critical role in any GCE calculation. Theoretically predicted yields have uncertainties associated with supernova models, black hole formation, and the stellar IMF. We instead base our choice of yields on the empirical analysis of \citetalias{Weinberg2024}, who anchor the overall scale of yields to the \cite{Rodriguez2023} determination of the mean Fe yield of CCSN. This determination implies (with some uncertainties described in \citetalias{Weinberg2024}) an IMF-averaged CCSN yield 
\begin{equation}
\frac{y^{\rm{cc}}_{\rm{Fe}}}{Z_{\rm{Fe,\odot}}}=0.345,
\label{eqn:ccsn Fe yield}
\end{equation}
where $Z_{\rm{Fe,\odot}}=0.00137$ is the solar Fe mass fraction and $y^{\rm{cc}}_{\rm{Fe}}$ is the mass of iron produced per unit mass of star formation (and is therefore dimensionless). For the IMF-averaged Mg yield we take 
\begin{equation}
\frac{y^{\rm{cc}}_{\rm{Mg}}}{Z_{\rm{Mg,\odot}}}=0.973 \cdot 10^{\mgfecc-0.45}.
\label{eqn:ccsn Mg yield}
\end{equation}
In combination with Equation~\ref{eqn:ccsn Fe yield}, Equation~\ref{eqn:ccsn Mg yield} implies that pure CCSN enrichment produces a Mg/Fe ratio that corresponds to $\mgfe=\mgfecc$.
Following \citetalias{Weinberg2024}, we adopt $\mgfecc=0.45$ as our reference value based on the abundance plateau observed for low metallicity Milky Way disk and halo stars.  We also consider a higher CCSN yield ratio, corresponding to $\mgfecc=0.55$, which implicitly assumes that stars with $\mgfe=0.45$ already have a significant fractional Fe contribution from SNIa \citep{Maoz2017,Conroy2022,Mason2023}. We assume that the IMF-averaged Mg and Fe yields are metallicity-independent. 

For the delay time distribution (DTD) of SNIa we adopt a sum of two exponentials that accurately approximates the $t^{-1.1}$ power-law form advocated by \cite{Maoz2017}. For the minimum delay time we take $t_{\mathrm{d,min}}=0.15$ Gyr as our reference value and consider $t_{\mathrm{d,min}}=0.05$ Gyr as an alternative that increases the amount of early SNIa enrichment. The choice of the exponential parameters is described by WAF. The most important SNIa parameter is the time-integrated Fe yield, which we set with the condition
\begin{equation}
    \yfeIa = \yfecc \left(10^{\Delta\afe}-1\right) \mu^{-1}~,
    \label{eqn:yfeIa}
\end{equation}
where $\Delta\afe$ is the drop between the CCSN plateau and the late-time equilibrium $\afe$, and $\mu \approx 1$ is a parameter that accounts for the impact of SFH on this equilibrium.  Following \citetalias{Weinberg2024} (see their Section 3.2), we set $\Delta\afe=\mgfecc$ and $\mu=1.1$, so that a GCE model with a slowly declining SFH evolves to $\mgfe\approx 0$ at late times. The implied $\yfeIa$ is also consistent with empirical estimates of the SNIa rate and a mean Fe yield of $\approx 0.7$ $M_{\odot}$ per supernova. 

Combining Equations~\ref{eqn:ccsn Fe yield}-\ref{eqn:yfeIa} with our adopted solar abundances gives
\begin{equation}
 \{\ymgcc,\yfecc,\yfeIa\} = 
    \{6.52,4.73,7.82\} \times 10^{-4}
    \label{eqn:yvals}
\end{equation}
for models with $\mgfecc=0.45$ and
\begin{equation}
 \{\ymgcc,\yfecc,\yfeIa\} = 
    \{8.21,4.73,10.95\} \times 10^{-4}
    \label{eqn:yvals2}
\end{equation}
for models with $\mgfecc=0.55$.

Details aside, the core assumption of our models is that we can take yields motivated by empirical studies of the Milky Way and general observations of supernova rates and galaxy properties, and apply them to the specific case of elliptical galaxy evolution. This assumption is reasonable if the stellar scale astrophysics that governs properties such as the stellar initial mass function (IMF) and supernova yields is decoupled from the galactic scale astrophysics that governs star formation history and morphology. However, it is not guaranteed to hold. We return to this issue in Section $\ref{sec:discussion}$.

\subsection{Star formation efficiency and outflows}
\label{sec:models_sfe}

The other parameters of our GCE models are the SFE timescale, $\taustar \equiv M_g / \dot{M}_{\star}$, and the outflow mass loading factor, $\eta \equiv \dot{M}_{\mathrm{out}}/\dot{M}_{\star}$. The SFE timescale has an important impact on the rate of metal enrichment at early times, and $\eta$ governs the metallicity at late times by regulating the degree of metal loss from the ISM. In the molecular gas of the present-day star-forming galaxies, a typical value of $\taustar$ is 2 Gyr, with substantial variations \citep{Leroy2008, Sun2023}. Studies of higher redshift galaxies suggest that this timescale decreases as $(1+z)^{-0.5}$ \citep{Tacconi2008}, and the SFE may be higher than the average in galaxies that evolve to become ellipticals. We therefore adopt $\taustar=1$ Gyr as our reference value and consider values as short as 0.5 Gyr. 

It has long been conjectured that the galaxy mass-metallicity relation is driven primarily by a mass dependence of $\eta$ \citep{Finlator2008, Peeples2011, Dave2012, Zahid2012, Chartab2023}, with lower values of $\eta$ at higher stellar and halo mass. There is a degeneracy between the overall scale of yields and the values of $\eta$ required to reproduce the observed mass-metallicity relation (see discussions by \citetalias{Weinberg2024}). For our adopted yields, the implied values of $\eta$ are relatively low. We take $\eta=0.3$ as a reference value, but in line with previous work, we will assume that $\eta$ decreases with increasing $M_{\star}$ to reproduce the observed mass-metallicity trend. 

The WAF analytic solutions assume metallicity-independent yields, time-independent values of $\taustar$ and $\eta$, instantaneous enrichment of CCSN products, and an SNIa DTD that is exponential or (in our case) a sum of exponentials. These solutions also assume that a fraction $r$ of a stellar population's mass is immediately returned to the ISM with its birth metallicity, where the recycling fraction $r \approx 0.4$ for a \cite{Kroupa2001} IMF. We have used the numerical GCE code \texttt{VICE} \citep{Johnson2020} to compare results for instantaneous recycling and full, time-dependent recycling for our typical parameter choices, and we find that changes to the abundance evolution relevant to this paper are negligible, consistent with the small impact found by WAF.

For a single exponential SFH, with $\tau_1=0$ Gyr and $\Mdotstar \propto e^{-t/\tau_2}$, the time evolution of the ISM mass fraction of a pure CCSN element such as Mg follows
\begin{equation}
\Zmg = {\ymgcc \over 1+\eta-r-\taustar/\tau_2}\left(1-e^{-t/\taubar}\right)
\label{eqn:Zmg}
\end{equation}
with
\begin{equation}
\taubar = {\taustar \over 1+\eta-r-\taustar/\tau_2}~.
\label{eqn:taubar}
\end{equation}
For our more general SFH, the abundance still approaches the same late-time equilibrium
\begin{equation}
\Zmgeq = {\ymgcc \over 1+\eta-r-\taustar/\tau_2}~,
\label{eqn:Zmgeq}
\end{equation}
but the time dependence is more complex. From these equations, it is clear that $\eta$ plays a key role in governing late-time metallicity. 

At early times, $t \ll \Bar{\tau}$, Equation \ref{eqn:Zmg} gives $Z_{\mathrm{Mg}} = \ymgcc (t/\taustar)$, independent of $\eta$. A longer $\taustar$ leads to slower metallicity evolution. A shorter $\tau_2$ at fixed $\taustar$ leads to a higher equilibrium abundance but a slower approach to that equilibrium because of the $-\taustar/\tau_2$ term in the denominators of Equations \ref{eqn:Zmg}-\ref{eqn:Zmgeq}. Because our observable quantities are calibrated to solar abundance, e.g. $\mgh = \log(\Zmg/\Zmgsun)$, it is the scaled yields $y_{\mathrm{Mg}} / \Zmgsun$ and $y_{\mathrm{Fe}} / \Zfesun$ that matter, which is why we specify these scaled values in Equations \ref{eqn:ccsn Fe yield} and \ref{eqn:ccsn Mg yield}.

Table \ref{tab:parameters} summarizes our model parameters and reference choices of these parameters. 

\begin{table*}[ht]
\caption{GCE Model Parameters}
\centering
\begin{tabular}{l ll}
\hline\hline
Parameter & Description & Reference Value \\ [1ex]
\hline\hline
\textit{Smooth SFH} & & \\
\hline
$\taustar$ & =$M_g$/$\dot{M}_{\star}$, (SFE) timescale & 1 Gyr\\
$\eta$ & =$\dot{M}_{out}/\dot{M}_{\star}$, mass loading factor & 0.3 \\ 
$\tstart$ & star formation start time  & 0.5 Gyr \\
$\tau_{1}$ & SFH rise timescale & 2 Gyr \\
$\tau_{2}$ & SFH decline timescale & 8 Gyr \\
$t_{\mathrm{cut}}$ & star formation cutoff time & ...\\
$\ymgcc/Z_{\mathrm{Mg},\odot}$ & IMF-integrated CCSN Mg yield & 0.973  \\ 
$\yfecc/Z_{\mathrm{Fe},\odot}$ & IMF-integrated CCSN Fe yield & 0.345 \\ 
$\yfeIa/Z_{\mathrm{Fe},\odot}$ & time-integrated SNIa Fe yield & 0.571  \\ 
$\tdmin$ & minimum delay time for SNIa & 0.15 Gyr\\
$r$ & mass recycling parameter & 0.4 \\ 
\hline
\textit{Terminating Burst Models} & & \\
\hline
$F_D$ & accreted gas/existing gas & 0, 3 \\
$f_g$ & fraction of gas consumed in burst & 0.05, 0.3 \\
$\eta'$ & burst mass loading factor & 0 \\ 
$r'$ & burst mass recycling parameter & 0.3 \\  
%$\ymgcc$ & IMF-integrated CCSN Mg yield & $6.53 \times 10^{-4}$  \\ 
%$\yfecc$ & IMF-integrated CCSN Fe yield & $4.73 \times 10^{-4}$ \\ 
%$\yfeIa$ & time-integrated SNIa Fe yield & $7.15 \times 10^{-4}$  \\ 
%$Z_{\mathrm{Mg},\odot}$ & solar magnesium abundance by mass & $6.71\times 10^{-4}$ \\
%$Z_{\mathrm{Fe},\odot}$ & solar iron abundance by mass & $1.37 \times 10^{-3}$\\
\hline\hline
\end{tabular}
\caption*{Our GCE models utilize a smooth SFH defined by four parameters: $\tstart$, $\tau_1$, $\tau_2$, and $\tcut$. Other parameters of our GCE models describe the star formation efficiency timescale $\taustar$, outflows $\eta$, and supernova yields. In addition, our burst model adds a terminating burst of star formation to a smooth SFH at $t=\tcut$ and contains parameters that describe gas accretion $F_D$ and the fraction of gas consumed in the burst $f_g$. }
\label{tab:parameters}
\end{table*}

\subsection{Light vs. Mass Weighting}
\label{sec:weighting}

As discussed in Section \ref{sec:data}, our tests (Figure \ref{fig: mock}, Appendix \ref{appx:mock}) show that the inferred stellar population properties and elemental abundances from SSP-equivalent measurements are similar to the light-weighted, log-averaged abundances and ages in our GCE models. The log-averaged abundance ratio $\langle[\mathrm{X/H}]\rangle$ of element $X$ is defined as 
\begin{equation}
\langle[\mathrm{X/H}]\rangle = \frac{\int_{0}^{t_{\mathrm{cut}}} \dot{M}_{\star}(t) \cdot [\mathrm{X/H}](t) \cdot w(t)\,dt }{\int_{0}^{t_{\mathrm{cut}}} \dot{M}_{\star}(t) \cdot w(t) \,dt },
\label{eqn:mean abun}
\end{equation}
where $\dot{M}_{\star}(t)$ is the SFR, $\xh(t)$ is the gas-phase abundance, and $w(t)$ is the light-weighting function described by Equation \ref{eqn:weighing}. This function assigns more weight to recently formed stars to account for the fact that younger stars are more luminous. For the mass-weighted quantities shown in some plots, we set $w(t)=1$.

From $\meanmgh$ and $\meanfeh$ we compute $\meanmgfe=\meanmgh-\meanfeh$. We determine the light-weighted, log-averaged age from
\begin{equation}
\langle\log_{10}(\mathrm{age})\rangle = \frac{\int_{0}^{t_{\mathrm{cut}}} \dot{M}_{\star}(t) \cdot \log_{10}(t_{\mathrm{obs}}-t) \cdot w(t)\,dt }{\int_{0}^{t_{\mathrm{cut}}} \dot{M}_{\star}(t) \cdot w(t) \,dt }.
\label{eqn:mean age}
\end{equation}
For brevity, we use the notation $\meanage \equiv 10^{\langle\log_{10}(\mathrm{age})\rangle}$ throughout this work. 

\begin{figure}[]
\centering
\includegraphics[width = \columnwidth]{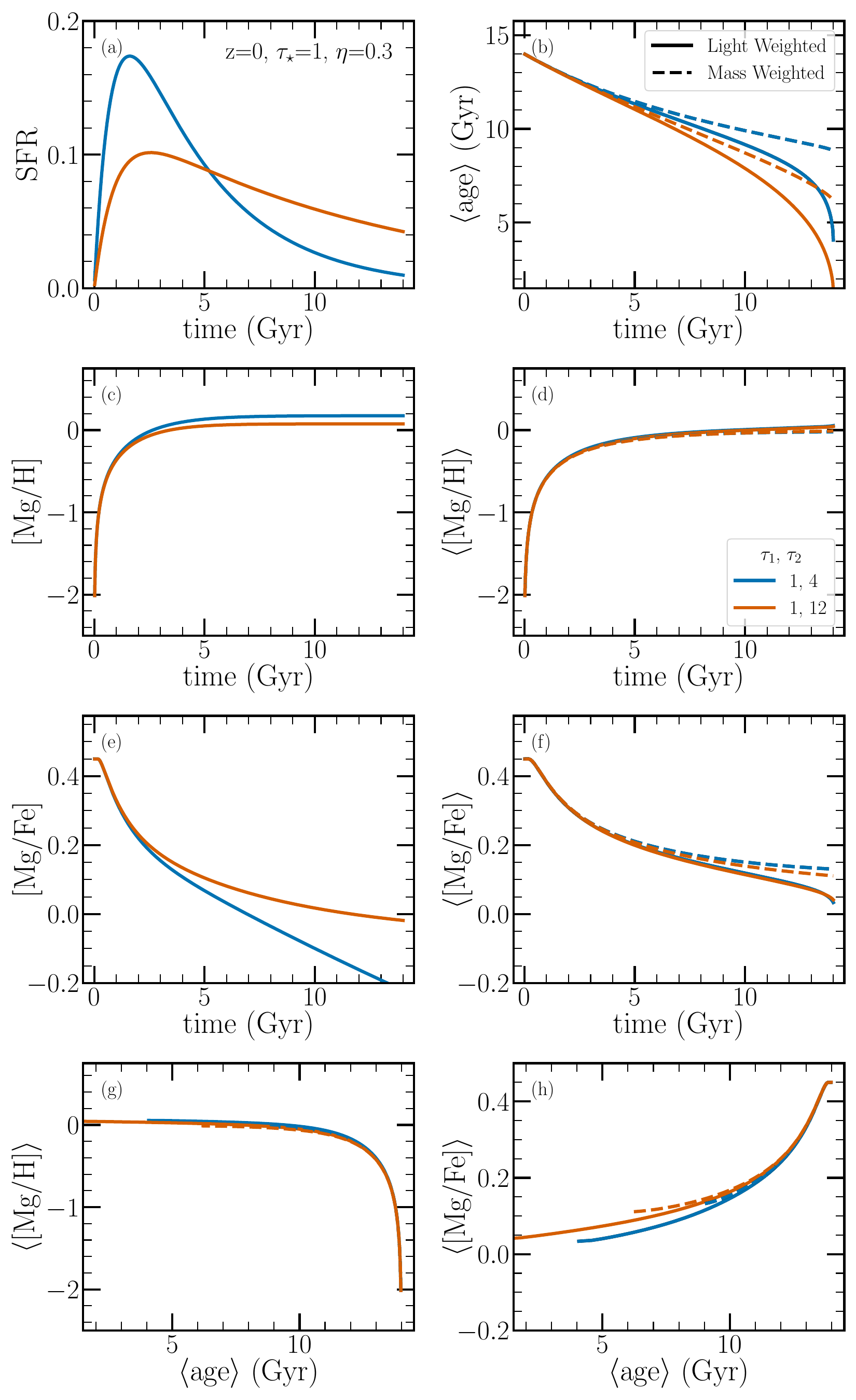}
\caption{Influence of light-weighting on averaged stellar quantities for two example SFHs, shown by the blue and orange curves in panel (a). The legends apply to all panels: all models have $\tau_1=1$ Gyr, $\taustar=1$ Gyr, and $\eta=0.3$, blue and orange curves show models with $\tau_2=$ 4 and 12 Gyr, respectively, and solid and dashed curves (sometimes indistinguishably) show the light- and mass-weighted quantities, respectively. Panel (b) shows the light- and mass-weighted $\meanage$ vs. time. Panels (c) and (e) show the evolution of gas phase $\mgh$ and $\mgfe$. Panels (d) and (f) show the corresponding mean stellar quantities, with light-weighting producing slightly lower $\meanmgfe$ at late times. Panels (g) and (h) plot $\meanmgh$ and $\meanmgfe$ against $\meanage$, which nearly cancels out the impact of light-weighting on $\meanmgfe$.}
\label{fig: Mass vs Light Weighting}
\end{figure}

Figure \ref{fig: Mass vs Light Weighting} shows the influence of light-weighting for two example SFH models. We consider models with a quickly declining SFH ($\tau_2=4$ Gyr, shown in blue) and a slowly declining SFH ($\tau_2=12$ Gyr, shown in orange) for fixed $\tau_1=1$ Gyr, $\taustar=1$ Gyr, and $\eta=0.3$. The normalized SFR in Figure \ref{fig: Mass vs Light Weighting}a gives the fraction of the galaxy's stellar mass formed per Gyr. A galaxy with a quickly declining SFH forms most of its stars early on, while a galaxy with a slowly declining SFH forms stars more steadily over time. As a result, the stellar $\meanage$ at the time of observation is younger for galaxies with a slowly declining SFH versus a quickly declining SFH, as shown by the difference between the blue and orange dashed curves in Figure \ref{fig: Mass vs Light Weighting}b. Light-weighting, represented by the solid curves, gives more weight to younger stars, lowering the $\meanage$ for both models by a couple of Gyr at late times, with deviations from the mass-weighted $\meanage$ rapidly increasing at times close to $t_{\mathrm{obs}}$. 

Figures \ref{fig: Mass vs Light Weighting}c and \ref{fig: Mass vs Light Weighting}e show the evolution of gas-phase $\mgh$ and $\mgfe$ for each model. $\mgh$ initially increases over time as Mg is produced by CCSN, then plateaus as the production of Mg reaches an equilibrium with the removal of Mg by stellar lockup and outflows. $\mgfe$ starts out at the CCSN yield ratio, here $\mgfecc=0.45$, then decreases as SNIa begin contributing Fe. $\mgfe$ is lower for a galaxy with a rapidly declining SFR because the ratio of CCSN to delayed SNIa is lower at late times. Figures \ref{fig: Mass vs Light Weighting}d and \ref{fig: Mass vs Light Weighting}f show the corresponding mean stellar quantities, which have smaller differences between models. Light-weighting has minimal impact on $\meanmgh$ because the gas-phase $\mgh$ is flat at late times. However, light-weighting results in lower $\meanmgfe$ at late times because it gives higher weight to the populations formed with the lowest $\mgfe$. Figures \ref{fig: Mass vs Light Weighting}g and \ref{fig: Mass vs Light Weighting}h plot $\meanmgh$ and $\meanmgfe$ versus $\meanage$. Somewhat coincidentally, the impact of light-weighting on $\meanage$ and on $\meanmgfe$ nearly cancels, so mass-weighted and light-weighted tracks are nearly identical. 

\begin{figure}[]
\centering
\includegraphics[width = \columnwidth]{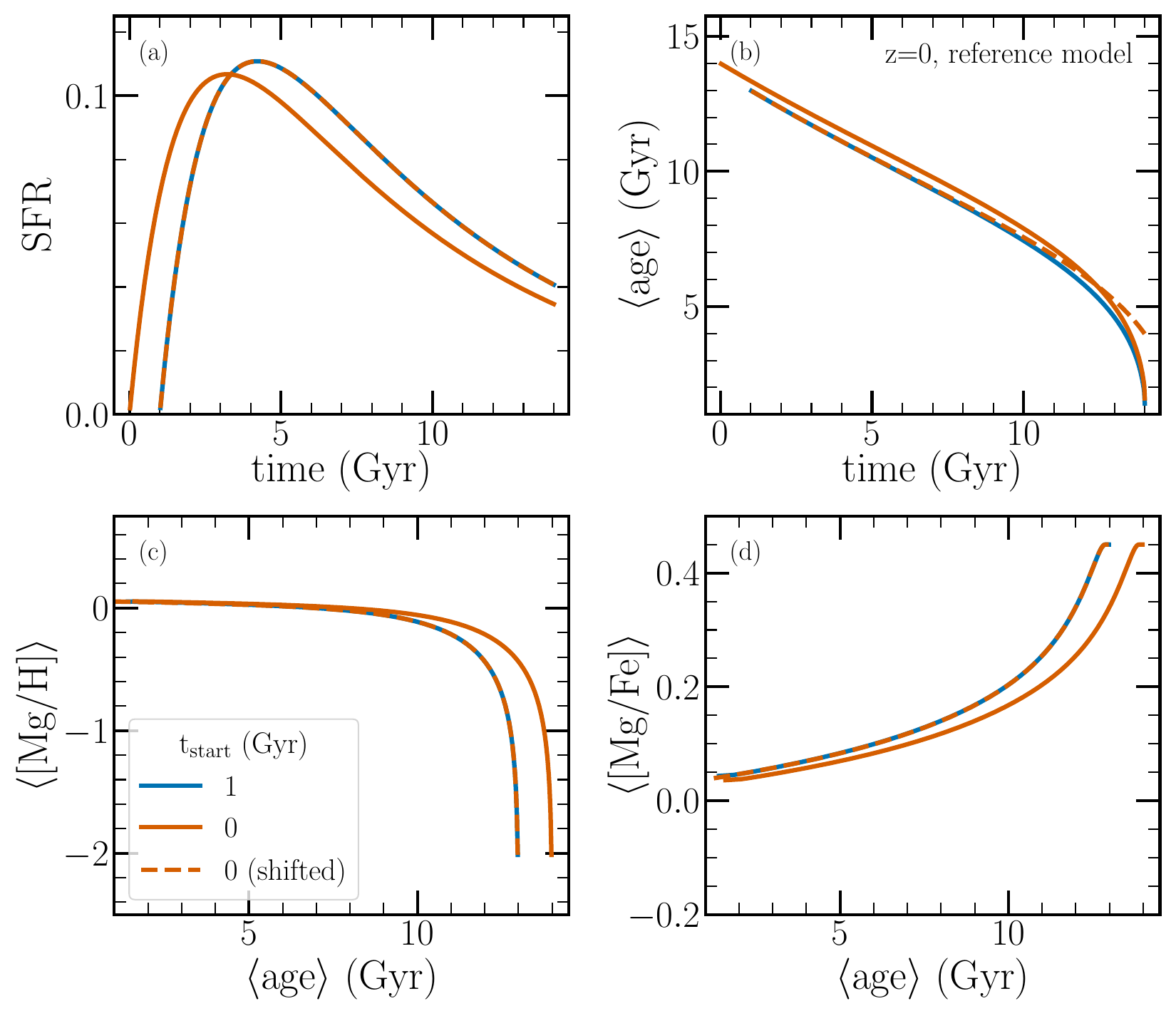}
\caption{Influence of start time on evolution for our reference model with $\tau_1=2$ Gyr, $\tau_2=8$ Gyr, $\taustar=1$ Gyr, and $\eta=0.3$. The solid orange and blue curves in each panel show models with $t_{\mathrm{start}}$ = 0 and 1 Gyr, respectively. Orange dashed curves, nearly superposed on the blue curves, show the $t_{\mathrm{start}}=0$ curves shifted by 1 Gyr to later $t$ or lower $\meanage$. The shifted SFR is renormalized to sum to 1. This 1 Gyr translation captures the impact of $t_{\mathrm{start}}=1$ almost exactly, despite the effects of light-weighting.}
\label{fig: tstart}
\end{figure}

In Figure \ref{fig: tstart} we explore the influence of start time on evolution for our reference model with $\tau_1=2$ Gyr, $\tau_2=8$ Gyr, $\taustar=1$ Gyr, and $\eta=0.3$. Incorporating a nonzero $t_{\mathrm{start}}$ is nearly identical to simply shifting the $t_{\mathrm{start}}=0$ curve to later $t$ or younger $\meanage$ by $t_{\mathrm{start}}$, despite the effects of light-weighting. This near-perfect equivalence is convenient as it allows us to treat different $t_{\mathrm{start}}$ values as simple horizontal translations along the time or $\meanage$ axis, without recomputing the models. 

\subsection{Impact of Parameters on Observables}
\label{sec:observables}

\begin{figure}[h]
\centering
 \includegraphics[width = \columnwidth]{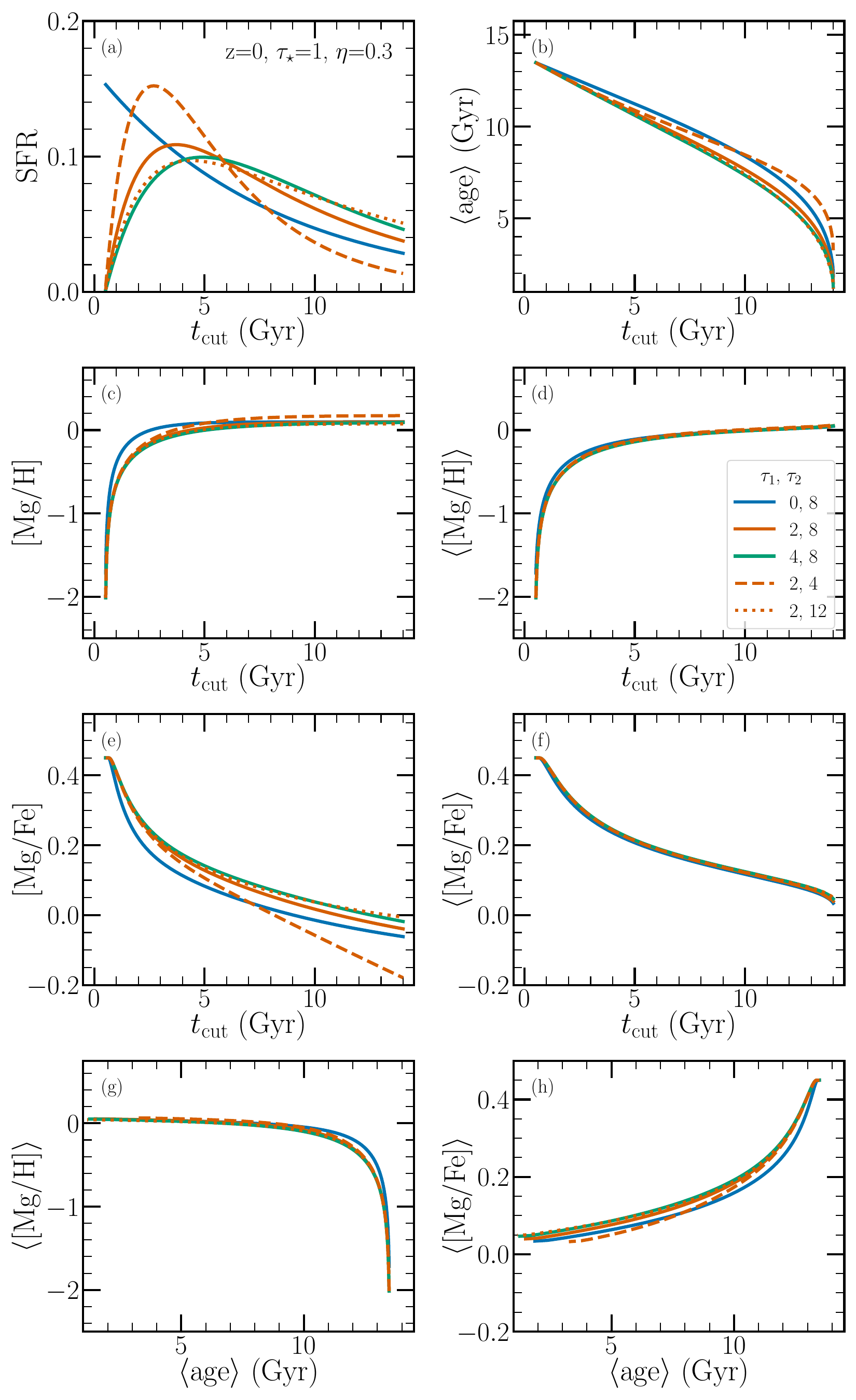}
\caption{Influence of star formation history on the evolution of gas phase ([Mg/H], [Mg/Fe]) and stellar ($\meanmgh$, $\meanmgfe$) abundances and the light-weighted mean stellar age, $\meanage$. The panel layout is the same as in Figure \ref{fig: Mass vs Light Weighting}. Solid blue, orange, and green curves show models with $\tau_2=8$ Gyr and $\tau_1=$ 0, 2, and 4 Gyr, respectively. Dashed and dotted orange curves show models with $\tau_1=2$ Gyr and $\tau_2=$ 4 and 12 Gyr, respectively. An early start of star formation (blue) or a rapid exponential decline (dashed orange) leads to higher $\meanage$ at a given $t$ and slightly lower $\meanmgfe$ at a given $\meanage$. All models have $\taustar=1$ Gyr, $\eta=0.3$, and $t_{\mathrm{start}}=0.5$ Gyr.}
\label{fig: tau1 tau2 z=0}
\end{figure}

We now investigate the impact of model parameters on the observables $\meanmgh$, $\meanmgfe$, and $\meanage$ at different epochs. We begin our analysis with the influence of star formation history on age and abundance evolution for several SFH models. Our results are presented in Figure \ref{fig: tau1 tau2 z=0} with the same panel layout as Figure \ref{fig: Mass vs Light Weighting}. However, we have changed the $x$-axis label in the upper panels from time to $\tcut$ to emphasize the role of this parameter in our predictions: we simply halt model evolution at $t=\tcut$ and read the values of the observables there.  Evolving a model to $t=14\Gyr$ allows us to make predictions for all possible choices of $\tcut$.

Figure \ref{fig: tau1 tau2 z=0}a shows the selected SFH models, with the reference model represented by the solid orange curve. These SFHs are identical to those in Figure \ref{fig:sfh} but with the addition of a slowly declining SFR ($\tau_2=12$ Gyr) and a strictly declining SFR with high initial star formation ($\tau_1=0$ Gyr). As expected, we find that models with higher SFR at later times, corresponding to longer $\tau_1$ and $\tau_2$, have younger, light-weighted $\meanage$ at $z=0$. 

Figure \ref{fig: tau1 tau2 z=0}c shows the time evolution of the gas $\mgh$. The different SFH models attain almost the same equilibrium $\mgh$; however, the pure exponential model reaches equilibrium sooner because of its higher early star formation. The time evolution of the gas $\mgfe$, shown in Figure \ref{fig: tau1 tau2 z=0}e, has a stronger dependence on the SFH. As already seen in Figure \ref{fig: Mass vs Light Weighting}, a rapidly declining SFH leads to lower $\mgfe$ at late times. Curves for the other models are similar, but the pure exponential again approaches equilibrium more quickly. Figures \ref{fig: tau1 tau2 z=0}d and \ref{fig: tau1 tau2 z=0}f show the time evolution of the stellar $\meanmgh$ and $\meanmgfe$, respectively. Averaging over the stellar population greatly reduces the model differences; in particular, while a sharply declining SFH produces lower $\mgfe$ at late times, there are few stars formed at these abundances. Overall, the impact of SFH on the predicted time evolution of $\meanmgh$ and $\meanmgfe$ is remarkably small, at least over the range of histories considered here. 

%The cutoff time $\tcut$ is an additional SFH parameter, and its implementation in our models is trivial: $\tcut$ specifies the time at which one reads the observables $\meanmgh$, $\meanmgfe$, and $\meanage$ from the model curves. 
Figures \ref{fig: tau1 tau2 z=0}g and \ref{fig: tau1 tau2 z=0}h plot the mean stellar abundance ratios against $\meanage$, yielding model predictions that can be compared to the SSP observables discussed in Section \ref{sec:data}. The predicted $\meanmgh$-$\meanage$ tracks are insensitive to the SFH, except that the exponential SFH, which has its maximum SFR at $t=t_{\mathrm{start}}$, achieves higher $\meanmgh$ at the oldest ages. For $\meanmgfe$ the model curves are more separated as a function of $\meanage$ than they were as a function of $\tcut$ because both the $\tau_1=0$ and $\tau_2=4$ Gyr models have older ages at a given time (Figure \ref{fig: tau1 tau2 z=0}b). Nonetheless, the influence of SFH on the $\meanmgfe$-$\meanage$ trajectory remains small. 

\begin{figure}[]
\centering
\includegraphics[width = \columnwidth]{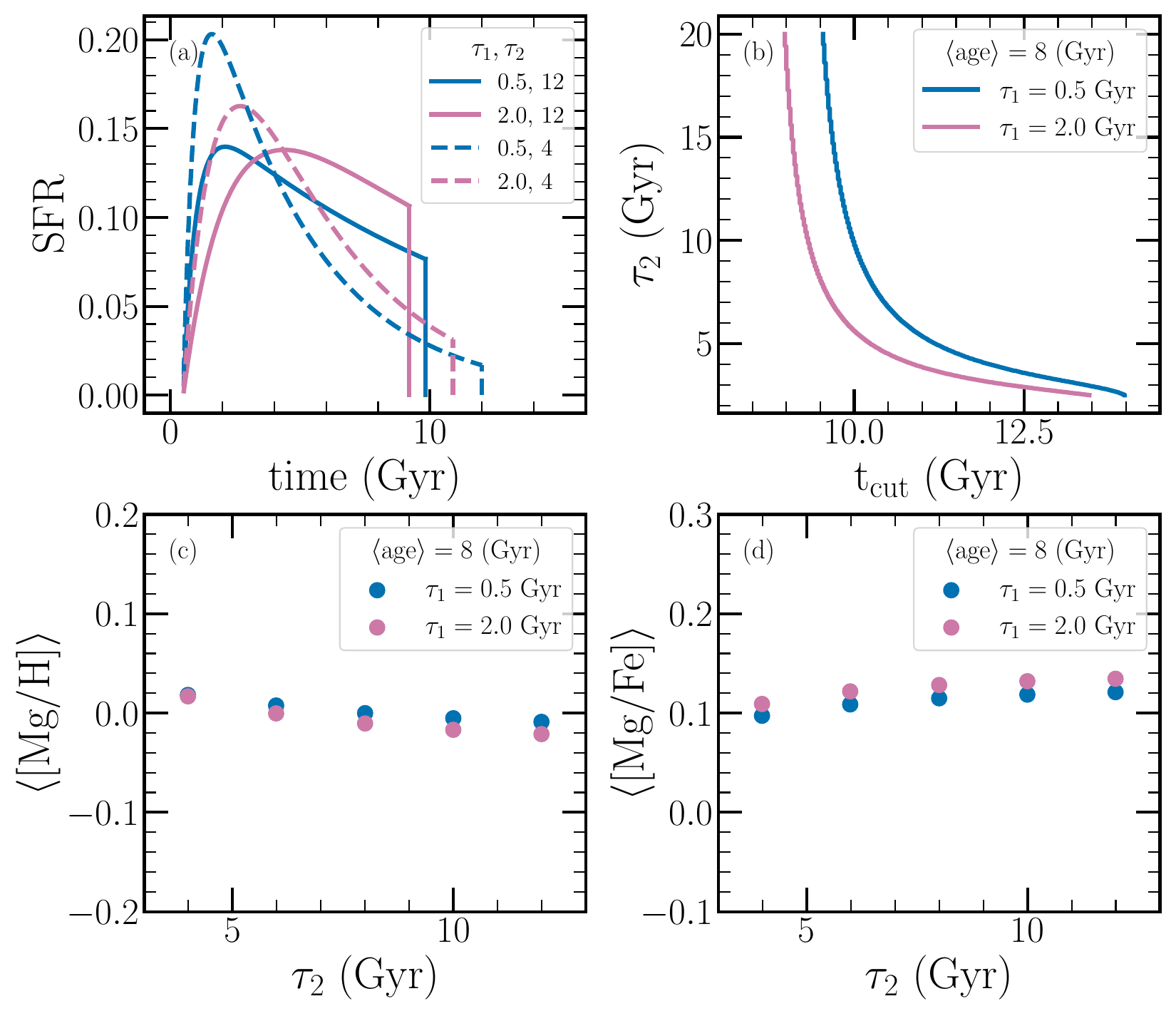}
\caption{Insensitivity of $\meanmgh$ and $\meanmgfe$ to the SFH. Panel (a) presents SFHs of models with four different combinations of long and short ($\tau_1$,$\tau_2$) provided in the figure legend, with $\tcut$ chosen to give $\meanage=8$ Gyr. All other parameters are described by their reference values. Panel (b) fixes $\tau_1=0.5$ Gyr (blue curve) and $\tau_1=2$ Gyr (pink curve) and shows different combinations of $\tau_2$ and $\tcut$ that have $\meanage=8$ Gyr. 
%Models with a higher late-time SFR (higher $\tau_2$) require an earlier $\tcut$ to obtain the same $\meanage$. 
Panels (c) and (d) take five combinations of $\tau_2$ and $\tcut$ for each fixed value of $\tau_1$ from Panel (b) and plot $\meanmgh$ and $\meanmgfe$ versus $\tau_2$, respectively. Combinations of $\tau_1$, $\tau_2$, and $\tcut$ that give the same $\meanage$ also give nearly the same $\meanmgh$ and $\meanmgfe$.}
\label{fig: tau2 tcut}
\end{figure}

We further illustrate the insensitivity of $\meanmgh$ and $\meanmgfe$ to the SFH in Figure \ref{fig: tau2 tcut}. Figure \ref{fig: tau2 tcut}a presents the SFHs of models with four different combinations of long and short $\tau_1$ (0.5, 2 Gyr) and $\tau_2$ (4, 12 Gyr), with $\tcut$ chosen to give $\meanage=8$ Gyr in each case. All other parameters are described by their reference values. Models with a higher late-time SFR ($\tau_2=12$ Gyr, solid curves) require an earlier $\tcut$ to obtain the same $\meanage$. Figure \ref{fig: tau2 tcut}b further shows this trade-off between $\tau_2$ and $\tcut$ by presenting different combinations of $\tau_2$ and $\tcut$ that obtain $\meanage=8$ Gyr for fixed $\tau_1=0.5$ Gyr (blue curve) and $\tau_1=2$ Gyr (pink curve). As shown in Figure \ref{fig: tau2 tcut}a, a higher $\tau_2$ requires an earlier $\tcut$ to produce the same $\meanage$. Panels (c) and (d) take five combinations of $\tau_2$ and $\tcut$ for each fixed value of $\tau_1$ from Panel (b) and plot $\meanmgh$ and $\meanmgfe$ versus $\tau_2$, respectively. We find that $\meanmgh$ and $\meanmgfe$ are insensitive to $\tau_1$, $\tau_2$, and $\tcut$ for a given $\meanage$. In other words, while $\tau_1$, $\tau_2$, and $\tcut$ all influence $\meanage$, combinations of these parameters that give the same $\meanage$ also give nearly the same $\meanmgh$ and $\meanmgfe$.

\begin{figure}[]
\centering
\includegraphics[width = \columnwidth]{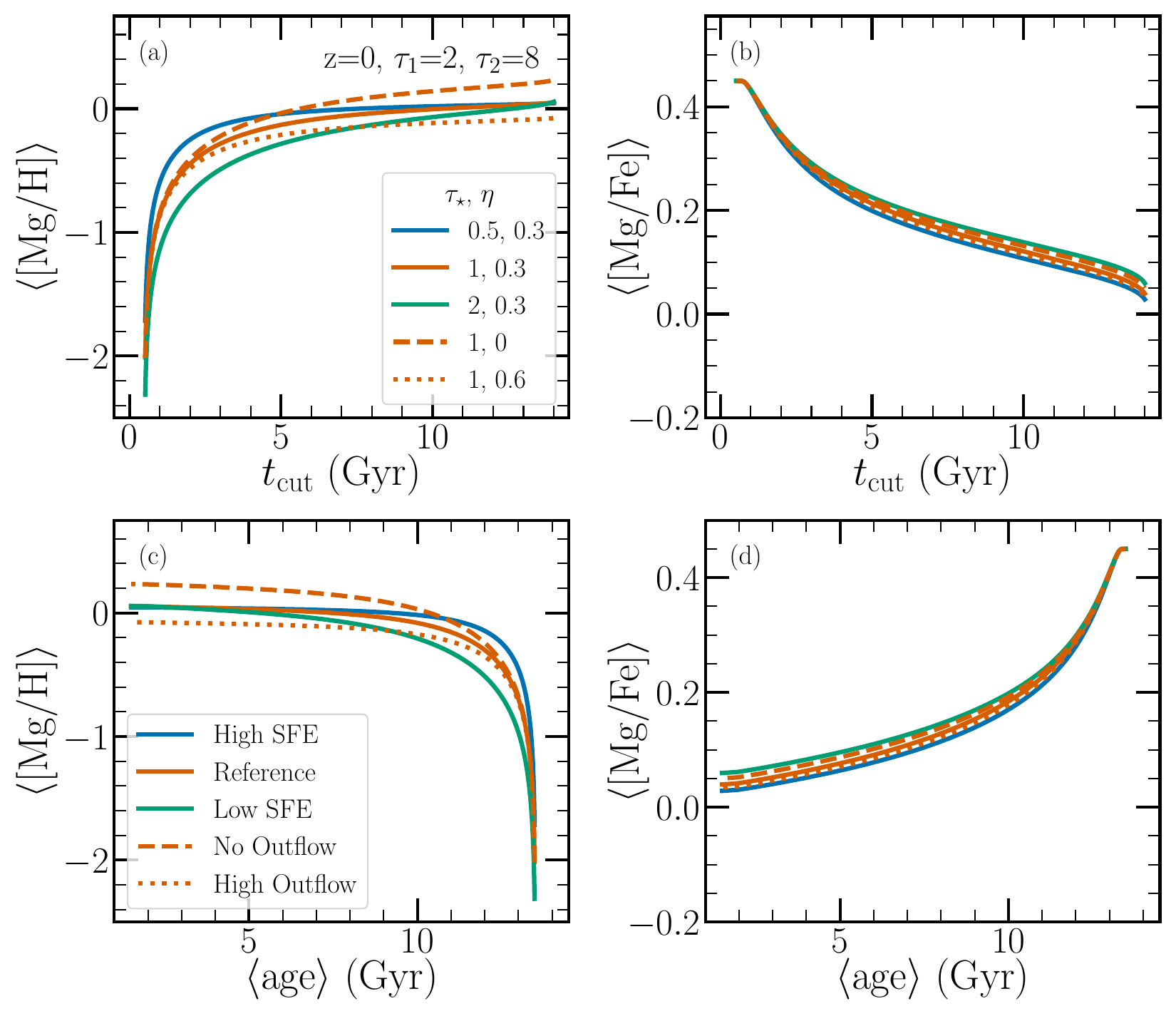}
\caption{Influence of SFE and outflows on mean abundance evolution for models with the same SFH ($\tau_1=2$ Gyr and $\tau_2=8$ Gyr, solid orange curve in Figure \ref{fig: tau1 tau2 z=0}a). The solid orange curve shows our reference choices of $\taustar=1$ Gyr and $\eta=0.3$. Solid blue and green curves show models with high and low SFE, respectively, while dashed and dotted curves show models with no outflows and high outflows, respectively. Corresponding ($\taustar$, $\eta$) values are given in the legend. Changing $\taustar$ affects the early evolution of $\meanmgh$, while changing $\eta$ alters the late evolution. Neither parameter has much impact on $\meanmgfe$ evolution.}
\label{fig: taustar eta z=0}
\end{figure}

Figure \ref{fig: taustar eta z=0} shows the influence of SFE and outflows on the mean abundance evolution for models with the reference SFH (solid orange curves in Figure \ref{fig: tau1 tau2 z=0}a and \ref{fig: tau1 tau2 z=0}b). Reducing (increasing) $\taustar$ raises (lowers) $\meanmgh$ at early times and thus in galaxies with high $\meanage$ caused by early truncation of star formation. Conversely, reducing (increasing) $\eta$ raises (lowers) $\meanmgh$ at late times and thus in galaxies with younger $\meanage$, while having little impact at early times. This behavior is consistent with our previous discussion: high SFE (short $\taustar$) leads to more rapid growth of metallicity, but the late time equilibrium is governed by the degree of metal loss in outflows. In contrast to $\meanmgh$, the influence of $\taustar$ and $\eta$ on the $\meanmgfe$-$\tcut$ and $\meanmgfe$-$\meanage$ trajectories is barely discernible. Although SFE and outflows have a well-known impact on evolutionary tracks in [$\alpha$/Fe] - [Fe/H] (see, e.g., Figure 3 of \cite{Andrews2017}), the evolution of [$\alpha$/Fe] with time is determined mainly by yields and by the SNIa DTD. With these quantities specified, the prediction of $\meanmgfe$ vs. $\meanage$ is remarkably robust within our family of models, with little sensitivity to the SFH parameters (Figure \ref{fig: tau1 tau2 z=0}) and even less to $\taustar$ and $\eta$.

\begin{figure}[]
\centering
\includegraphics[width = \columnwidth]{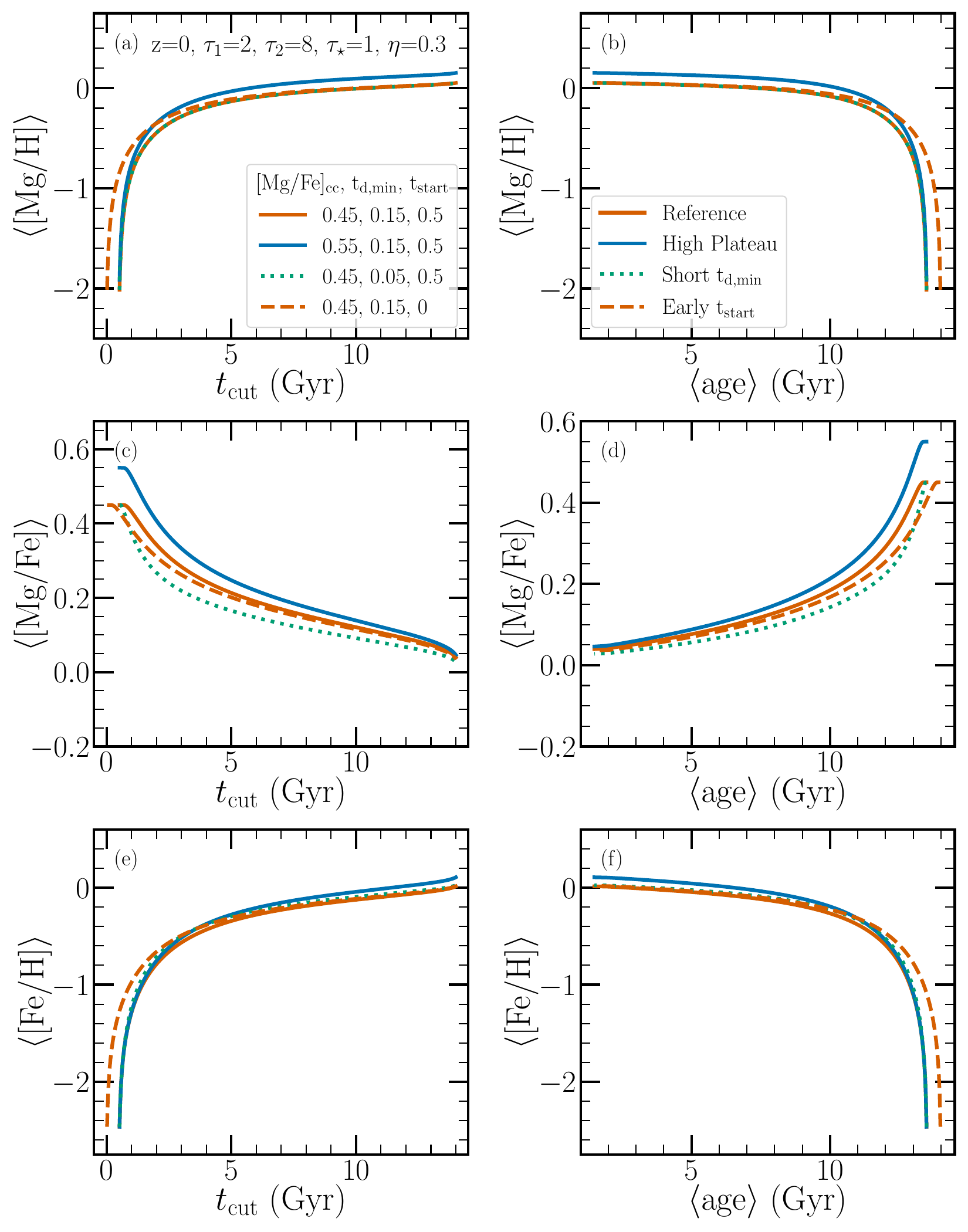}
\caption{Influence of the $\mgfecc$, $\tstart$, and $\tauIa$ on models with the reference values of SFH, SFE, and outflow parameters ($\tau_1=2$ Gyr, $\tau_2=8$ Gyr, $\taustar=1$ Gyr and $\eta=0.3$). Solid orange curves, identical to those in Figure \ref{fig: taustar eta z=0}, adopt our reference choices of $t_{\mathrm{d,min}}=0.15$ Gyr, $t_{\mathrm{start}}=0.5$ Gyr, and $\ymgcc/\Zmgsun = 0.973$, corresponding to $\mgfecc=0.45$ in Equation \ref{eqn:ccsn Mg yield}. The solid blue curve shows a model with boosted $\ymgcc$, corresponding to  $\mgfecc=0.55$. The dotted green and dashed orange curves show models with short $t_{\mathrm{d,min}}$ and early $t_{\mathrm{start}}$, respectively, with parameter values given in the legend. Boosting the Mg yield (high $\mgfecc$) raises $\meanmgh$ and $\meanfeh$ by 0.1 dex at all times and boosts $\meanmgfe$ by 0.1 dex at early times. Changing $t_{\mathrm{d,min}}$ only affects the Fe abundance, slightly raising $\meanfeh$ above the reference value at all times and altering $\meanmgfe$ significantly at early and intermediate times, but having little impact on the late evolution. Changing $t_{\mathrm{start}}$ affects the early evolution of $\meanmgh$, $\meanfeh$, and $\meanmgfe$ by shifting the curves in $t$ and $\meanage$ (see Figure \ref{fig: tstart}), but it has little impact on the late evolution.}
\label{fig: tdmin tstart z=0}
\end{figure}

Figure \ref{fig: tdmin tstart z=0} explores the impact of the Mg yield, the SNIa minimum time delay, and the star formation start time, with other parameters fixed to their reference values. As discussed in Section \ref{sec:models_yields}, we adopt an empirically estimated $\yfecc$ and compute $\ymgcc$ based on the plateau value of $\mgfecc$ taken to represent pure CCSN enrichment. Our reference choice of $\ymgcc = 0.973$ $\Zmgsun$ corresponds to $\mgfecc=0.45$. Boosting $\ymgcc$ by 0.1 dex, with $\mgfecc=0.55$, raises $\meanmgh$ by 0.1 dex at all times. It also boosts $\meanmgfe$ by 0.1 dex at early times, as expected. However, our empirical normalization of $\yfeIa$ is designed to produce $\mgfe \approx 0$ at late times for a non-truncated SFH, and this boosted $\yfeIa$ largely erases the impact of boosted $\ymgcc$ on $\meanmgfe$ for $t > 6$ Gyr and $\meanage <10$ Gyr.

By definition, $t_{\mathrm{d,min}}$ has no impact on $\meanmgh$. However, reducing $t_{\mathrm{d,min}}$ from 0.15 Gyr to 0.05 Gyr depresses $\meanmgfe$ noticeably ($\sim 0.05$ dex) over a wide range of time and $\meanage$. Although we hold the time-integrated $\yfeIa$  fixed, a shorter $t_{\mathrm{d,min}}$ with a $t^{-1.1}$ DTD leads to substantially more rapid Fe enrichment, and the signature of this faster enrichment persists for several Gyr. As previously shown (Figure \ref{fig: tstart}), changing $t_{\mathrm{start}}$ is effectively equivalent to shifting $\meanmgh$ and $\meanmgfe$ horizontally along the time and $\meanage$ axes. Because these trajectories are steep at early times and flat at late times, the impact on the predicted abundances is significant at the oldest ages but small at $\meanage< 10$ Gyr.  The evolution of $\meanfeh$ follows directly from $\meanmgh$ and $\meanmgfe$.  In this figure only, we show $\meanfeh$ evolution in separate panels.  While the tracks resemble those of $\meanmgh$, the $\meanfeh$ tracks are less flat at late times and young ages because of the Fe contribution from delayed SNIa.

\begin{figure}[]
\centering
\includegraphics[width = \columnwidth]{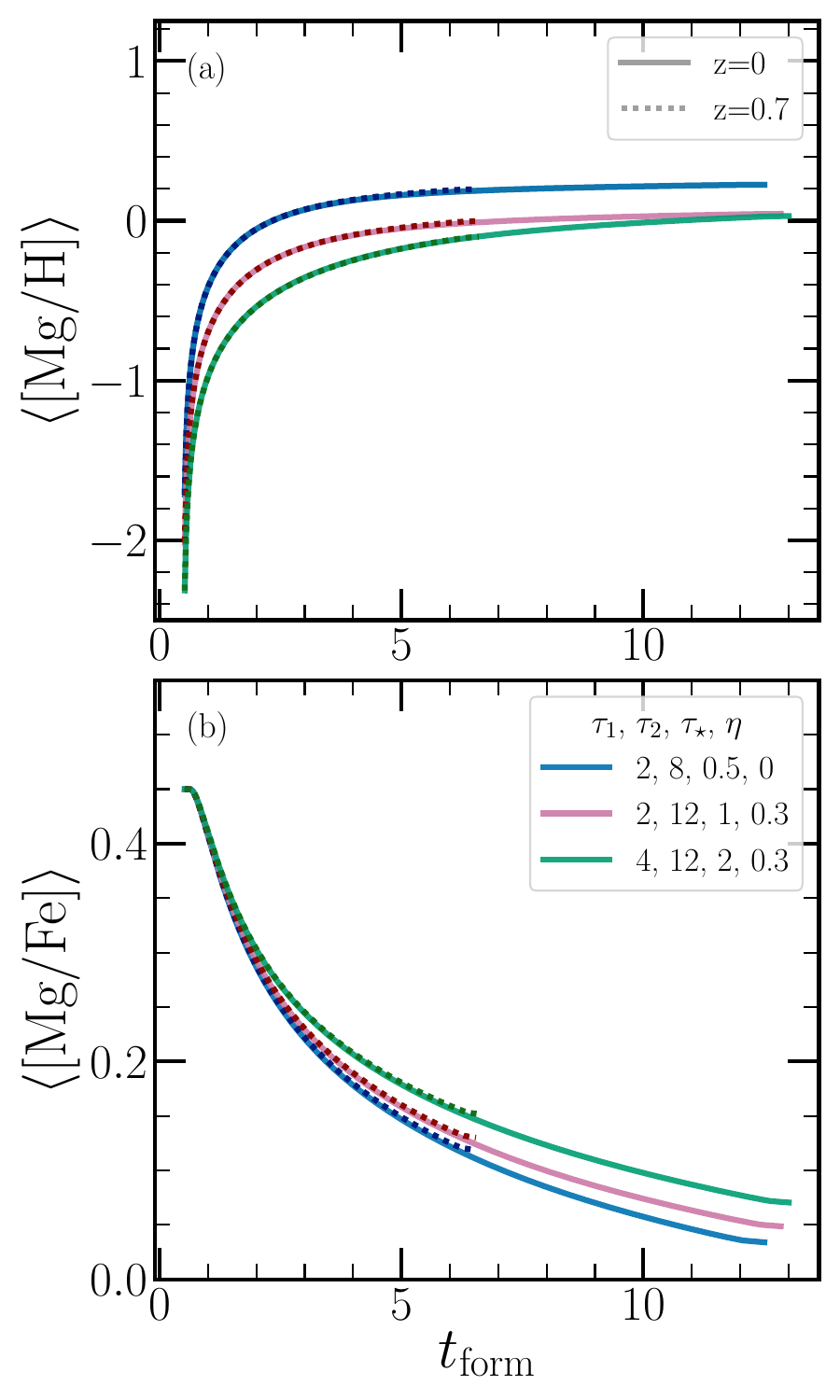}
\caption{Abundances as a function of $\tform\equiv\tobs-\meanage$ for three example models at $z=0$ and $z=0.7$.  In SSP fits, $\tobs-\mathrm{age}$ is referred to as $\tform$ because it represents the time (relative to the big bang) at which model stars are assumed to form.  As a function of $\tform$, the predicted light-weighted abundances are nearly identical for models evolved to $z=0$ (solid curves) or to $z=0.7$ (dotted curves).}
\label{fig: z=0.7 shift}
\end{figure}

A similar analysis can be done for galaxies at $z=0.7$ by setting $t_{\mathrm{obs}}=7$ Gyr in Equations \ref{eqn:weighing} and \ref{eqn:mean age}. Figure \ref{fig: z=0.7 shift} compares the $\meanmgh$ and $\meanmgfe$ trajectories at $z=0$ (solid curves) and $z=0.7$ (dashed curves) for three example models.  For this figure only, we plot the model abundances as a function of $\tform \equiv \tobs-\meanage$ rather than $\meanage$ itself.  For a stellar population that forms instantaneously at a single epoch, as assumed in an SSP fit, $\tform$ is simply the time at which the stars formed, relative to the big bang at $t=0$.  With linear, mass-weighted age averaging, predictions for $z=0$ and $z=0.7$ galaxies with the same model parameters would be identical as a function of $\tform$ because the evolutionary tracks are the same as a function of time.  Figure~\ref{fig: z=0.7 shift} shows that predictions are nearly identical as a function of $\tform$ even with logarithmic, light-weighted averaging.  Equivalently, the $\meanmgh-\meanage$ and $\meanmgfe-\meanage$ predictions for galaxies at $z=0$ and $z=0.7$ are simply shifted in age by $\tobs(z=0)-\tobs(z=0.7)=7\Gyr$, if we adopt the same model parameters.  

\begin{figure*}[]
\centering
\includegraphics[width = \textwidth]{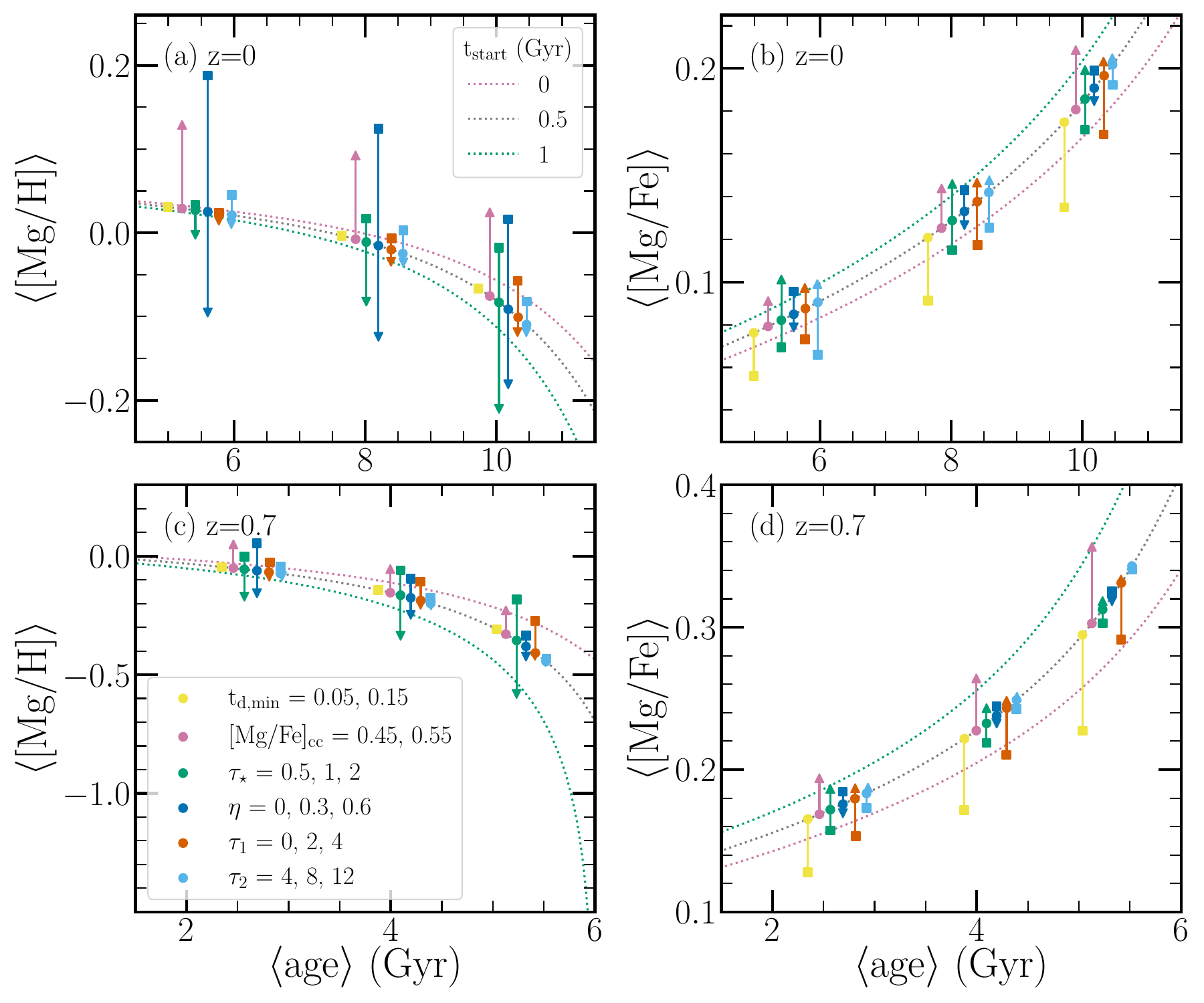}
\vspace{-5mm}
\caption{Summary of parameter impact on reference model abundances at different age cuts. The red, black, and green, dotted curves show the reference model with t$_{\mathrm{start}}=$ 0, 0.5, and 1 Gyr, respectively. The vertical lines show the possible abundance range from changing one parameter in the t$_{\mathrm{start}}=0.5$ Gyr reference model at a given age. Parameters are represented by different colors and are provided in the legend. Circles mark the reference values, while squares and triangles represent lower and higher parameter values, respectively. (One can read the points as an ``arrow" that follows the order of parameter values in the legend.) While some parameters have a significant impact at some epochs, none of these parameters boost $\meanmgfe$ substantially above the reference model curve.}
\label{fig: summary}
\end{figure*}

Figure \ref{fig: summary} summarizes the impact of changing each parameter in the reference model for galaxies at $z=0$ and $z=0.7$. The red, black, and green dotted curves show the reference model with $t_{\mathrm{start}}=$ 0, 0.5, and 1 Gyr, respectively. The vertical lines show the abundance shifts from changing one parameter in the $t_{\mathrm{start}}=0.5$ Gyr reference model at a given $\meanage$. Parameters are represented by different colors and are provided in the legend. Circles mark the reference values, while the squares and triangles represent the lowest and highest parameter values, respectively. 

The physics underlying the changes shown in Figure \ref{fig: summary} has already been discussed above, but the presentation here helps compare their magnitude. For $\meanmgh$, the values of $\eta$ and the CCSN Mg yield have the largest impact. The SFE timescale $\taustar$ has an important impact at the oldest ages but a smaller impact at young ages. In the context of our comparison to observed abundance trends in Section \ref{sec:observations}, the most important implication of Figure \ref{fig: summary} is that none of the changes we consider substantially boost $\meanmgfe$ relative to the reference model. Increasing $\mgfecc$ boosts $\meanmgfe$ for the oldest galaxies, but the effect diminishes at younger ages. Several parameter changes can decrease $\meanmgfe$ relative to the reference model: a shorter $\tdmin$, a shorter rise time $\tau_1$, or a shorter decline time $\tau_2$. Shifting $\tstart$ from 0.5 Gyr to 1 Gyr boosts $\meanmgfe$ at a given $\meanage$ by roughly the same amount as a 0.1-dex bump in $\mgfecc$. Increasing $\tstart$ reduces the $\meanmgh$ of the oldest galaxies, as there is less time for enrichment. For the $z=0$ galaxies, changing $\tstart$ from 0.5 Gyr to 1 Gyr is qualitatively similar to extending the rise time $\tau_1$ from 2 Gyr to 4 Gyr. However, for the $z=0.7$ galaxies the longer rise time produces a minimal change in $\meanmgfe$ at a given $\meanage$ while shifting $\tstart$ has a substantial impact.

\section{Modeling observed elliptical populations}
\label{sec:observations}

\subsection{Representative Models}
\label{sec:representative}

\begin{figure}
\centering
\includegraphics[width=\columnwidth]{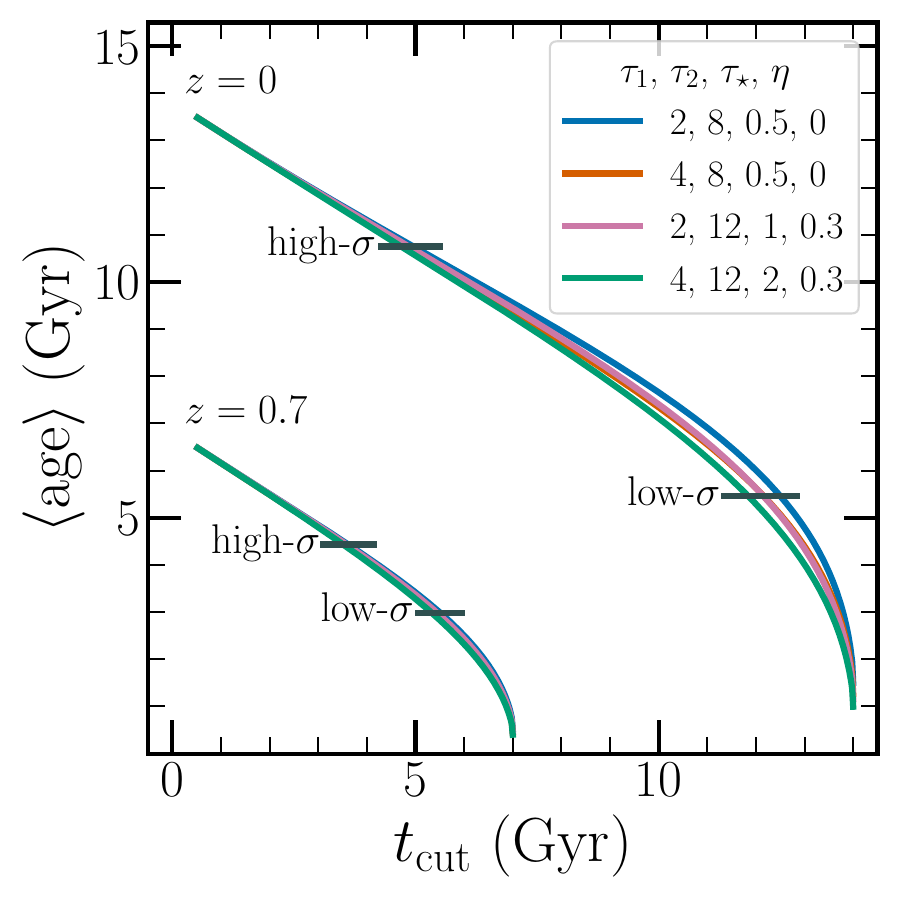}
\caption{Relation between $\meanage$ and $t_{\mathrm{cut}}$ for representative models (see Figure \ref{fig: best fits}) at redshift $z=0$ and $z=0.7$. The $z=0.7$ models display overlapping tracks in $\meanage$-$\tcut$ while the $z=0$ models have distinguishable tracks for $\tcut \gtrsim 4$ Gyr. However, the pink and orange models at $z=0$ are nearly identical. The horizontal lines mark the observed $\meanage$ for the lowest (youngest) and highest (oldest) velocity dispersion bins at each redshift. The intersection of the horizontal lines and model curves indicate the value of $\tcut$ required to obtain the observed $\meanage$.} 
\label{fig:tcut}
\end{figure}

\begin{figure*}[]
\centering
\includegraphics[width = \textwidth]{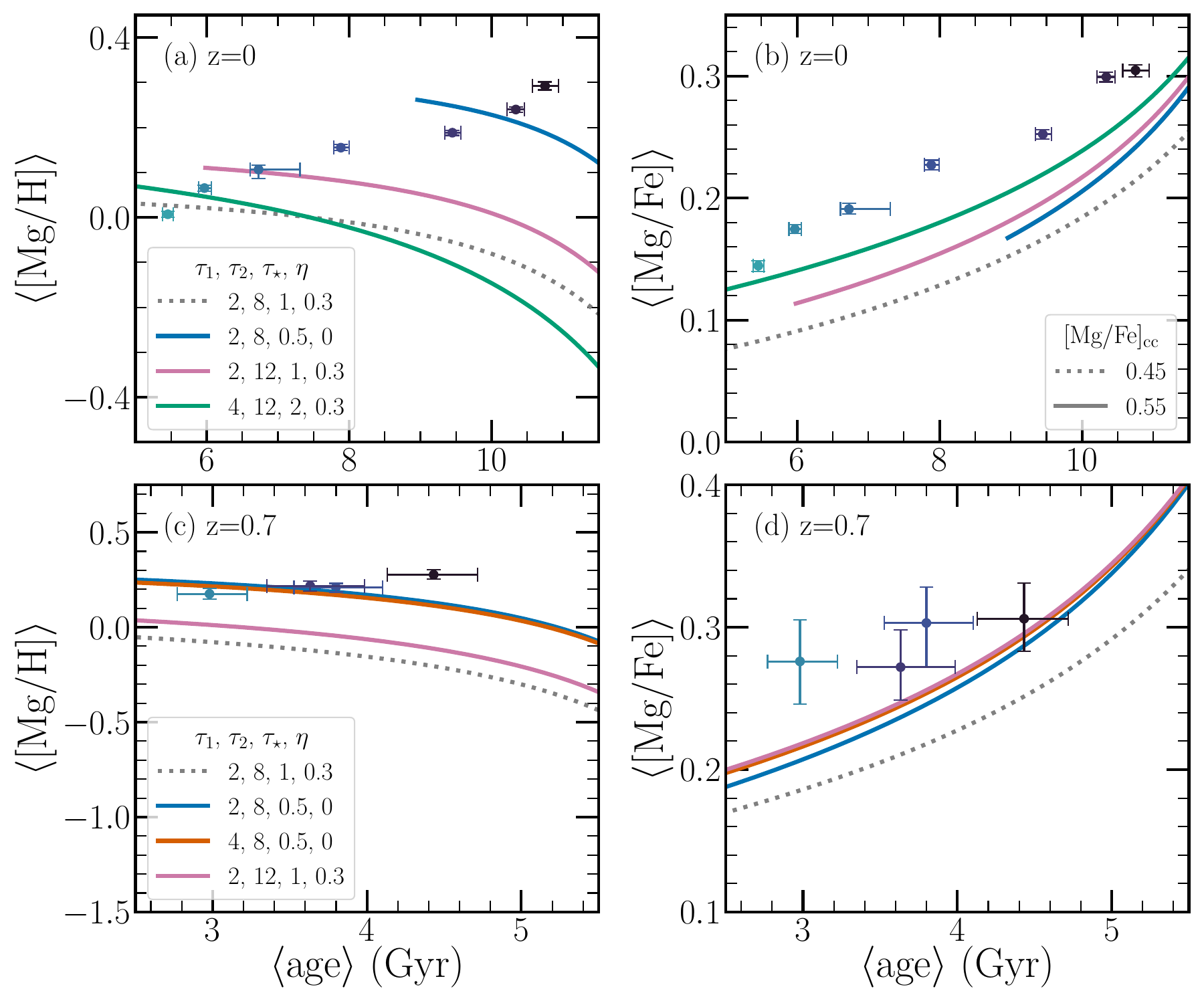}
\vspace{-5mm}
\caption{Comparison between representative models and the observed ages and abundances of galaxies stacked in bins of velocity dispersion, $\sigma$. Observed properties are represented by blue points and contain SDSS \citep{Conroy2014} and LEGA-C \citep{Beverage2023} data for galaxies at $z=0$ and $z=0.7$, respectively. The points increase in $\sigma$ moving from light to dark blue. The solid curves show representative models with parameters chosen from values presented in Figure \ref{fig: summary} to reproduce the observed $\meanmgh$ and $\meanmgfe$ of low, medium, and high velocity dispersion galaxies at $z=0$ and galaxies at $z=0.7$. All of these models have $\tstart=0.5$ Gyr. Reproducing the $\meanage$ of high-$\sigma$ ellipticals requires $\eta \approx 0$ and high SFE, while low-$\sigma$ ellipticals can accommodate moderate outflows and lower SFE. At $z=0.7$, the high $\meanmgfe$ requires $\eta \approx 0$ and high SFE for all $\sigma$-bins. All representative models have a high Mg yield, with $\mgfecc=0.55$, but even so, they under-predict the observed $\meanmgfe$ at $z=0$ for all but the lowest-$\sigma$ galaxies. The dotted gray curve shows the reference model displayed by the solid orange curves in Figures \ref{fig: tau1 tau2 z=0}-\ref{fig: tdmin tstart z=0} and by the central dotted curves in Figure \ref{fig: summary}. The reference model has $\mgfecc=0.45$, which reduces the predicted $\meanmgh$ and $\meanmgfe$ and worsens the agreement with observations. }
\label{fig: best fits}
\end{figure*}

We use our results from Figure \ref{fig: summary} to select representative models that reproduce the observed $\meanmgh$ and $\meanmgfe$ of low, medium, and high velocity dispersion galaxies at $z=0$ and galaxies at $z=0.7$. These models are described by star formation histories with rising timescales of $\tau_1=2$ and 4 Gyr, declining timescales of $\tau_2=8$ and 12 Gyr, and a star formation start time of $\tstart=0.5$ Gyr. The parameter that has the strongest impact on $\meanage$ is $\tcut$, the time at which star formation rapidly truncates. Figure \ref{fig:tcut} illustrates the relationship between $\tcut$ and the predicted $\meanage$ for the representative models (see Figure \ref{fig: best fits}) at $z=0$ and $z=0.7$. The $z=0.7$ models display nearly identical tracks in $\meanage$-$\tcut$. However, there are differences as large as $\sim1$ Gyr in $\meanage$ between the $z=0$ models for a given $\tcut$ when $\tcut\gtrsim4.5$ Gyr. Here, models having a higher star formation rate at late times (longer $\tau_1$ and/or longer $\tau_2$) have younger $\meanage$ at fixed $\tcut$. The horizontal lines mark the observed $\meanage$ for the lowest (youngest) and highest (oldest) velocity dispersion bins at each redshift (see Figure \ref{fig: best fits}). The intersections of the horizontal lines and model curves indicate the truncation time required for each model to obtain the observed $\meanage$. More massive, high-$\sigma$ ellipticals are older than less massive, low-$\sigma$ ellipticals and therefore cease star formation at an earlier $\tcut$. 

Figure \ref{fig: best fits} compares our representative models to the observed ages and abundances of galaxies stacked in bins of velocity dispersion, $\sigma$. Figure \ref{fig: summary} shows that once the yields are fixed, there is not much leverage to boost $\meanmgfe$. With empirical yields described in Section \ref{sec:models_yields}, there are no combinations of $\tau_1$, $\tau_2$, $\taustar$, and $\eta$ that can reproduce the observations. Because our models generally underproduce the observed $\meanmgfe$, we adopt $\mgfecc=0.55$ as our reference value in this section rather than 0.45 as in Section \ref{sec:models}. While Milky Way stellar observations generally favor a lower $\mgfe$ plateau (see \citetalias{Weinberg2024}), the value of $\mgfecc=0.55$ remains possible, especially if one allows for a significant SNIa Fe contribution to stars with $-1.5<\feh<-0.5$ as argued by \cite{Maoz2017}, \cite{Conroy2022}, and \cite{Mason2023}. The dotted gray curves in Figure \ref{fig: best fits} show the reference model that is the central curve from Figure \ref{fig: summary}; the most important difference from the models in this section is the lower $\mgfecc=0.45$.

The blue points in Figure \ref{fig: best fits}a show the $\meanmgh$ and $\meanage$ of the observed SDSS galaxy bins at $z=0$.\footnote{More precisely, values shown by blue points are the SSP fits, which we refer to with the same notation $\meanage$, $\meanmgh$, and $\meanmgfe$ that we use for light-weighted model averages.} Velocity dispersion increases when moving from the light to dark blue points, corresponding to $\meanmgh$ and $\meanage$ increasing with increasing $\sigma$ \citep{Conroy2014}. The $\meanmgh$ and $\meanage$ of the low-$\sigma$ bins are best represented by a slowly rising and declining SFH ($\tau_1=4$ and $\tau_2=12$ Gyr) with mild outflows ($\eta=0.3$) and SFE typical of local molecular gas disks ($\taustar=2$ Gyr) (green curve). Longer $\tau_1$, $\tau_2$, and $\taustar$ allow for higher late-time star formation that results in young $\meanage$. Mild outflows balance the late Mg production and push $\meanmgh$ closer to Solar. 

The representative model for the mid-$\sigma$ bins (pink curve) is similar to the low-$\sigma$ model but with a more rapid SFR rise ($\tau_1=2$ Gyr) and higher SFE ($\taustar=1$ Gyr) that produces more stars at earlier times. This increases the $\meanage$ and raises the $\meanmgh$ by producing more Mg earlier and quenching before all the SNIa explode. The high-$\sigma$ bins (blue curve) are best represented by a rapidly rising and declining SFH ($\tau_1=2\Gyr$ and $\tau_2=8\Gyr$) with high SFE ($\taustar=0.5$ Gyr) and no outflows ($\eta=0$). Shorter $\tau_1$, $\tau_2$, and $\taustar$ result in the majority of stars being formed at early times, increasing the $\meanage$ and producing significant amounts of Mg before early quenching. No outflows results in all the Mg being retained, driving $\meanmgh$ to $\sim 0.25$ dex above Solar. The high SFE is needed to achieve the high observed $\meanmgh$ at such old ages.

The parameter choices for these three representative models are not unique, but they reflect our expectation that the oldest and highest-$\sigma$ galaxies have high star formation efficiency and rapid early star formation, and they ascribe the trend of $\meanmgh$ with $\sigma$ to the lower outflow efficiency in deeper potential wells. The mid-$\sigma$ model is similar to the reference model from Section \ref{sec:models} (gray dotted curve), but it predicts higher $\meanmgh$ because $\mgfecc=0.55$ implies a higher Mg yield $\ymgcc$.

The blue points in Figure \ref{fig: best fits}c show the $\meanmgh$ and $\meanage$ of the observed LEGA-C galaxy bins at $z=0.7$, with velocity dispersion increasing from the light blue to the dark blue points. At this redshift, $\meanmgh$ is nearly constant across all bins of $\sigma$. The blue and pink representative models at $z=0.7$ have the same parameters as the blue and pink representative models at $z=0$. The orange model is similar to the blue model but with a longer rise timescale $\tau_1=4$ Gyr, chosen because it slightly boosts $\meanmgfe$ for the youngest galaxies at this redshift. As shown in Figure \ref{fig: summary}, $\meanmgh$ at $z=0.7$ is insensitive to $\tau_1$ for all $\meanage$, so the blue and orange models have nearly identical $\meanmgh$-$\meanage$ trajectories. The blue and orange models retain all of the Mg they produce ($\eta=0$) and can match the high observed  $\meanmgh$ at $z=0.7$. Outflows with $\eta=0.3$ cause the pink model to undershoot the observed $\meanmgh$ by $\sim0.2$ dex. 

Figures Figure \ref{fig: best fits}b and Figure \ref{fig: best fits}d compare the representative models to the observed $\meanmgfe$ and $\meanage$ of the SDSS and LEGA-C bins, respectively. At $z=0$, the green model predicts the observed $\meanmgfe$ of the low-$\sigma$ galaxies and nearly reproduces the observed values at older ages. The pink and blue models, which reproduce the $\meanmgh$ of mid- and high-$\sigma$ galaxies, respectively, underpredict the observed $\meanmgfe$ for all ages. At $z=0.7$, the blue, orange, and pink models make nearly identical predictions and are reasonably consistent with the observed $\meanmgfe$ of the LEGA-C galaxies within the estimated uncertainties.

Figure \ref{fig: best fits} demonstrates the difficulty of reproducing observed abundance trends in ellipticals with smooth star formation histories and yields that are empirically motivated by abundance trends in the Milky Way. Our reference model with $\mgfecc=0.45$ underpredicts the observed $\meanmgfe$ values by $\sim0.08$ dex at $z=0$ and $\sim0.05$ dex at $z=0.7$. For $\mgfecc=0.55$, we find models that come close to the observed $\meanmgh$-$\meanage$ and $\meanmgfe$-$\meanage$ trends at $z=0.7$ (blue and orange curves in panels c and d) and underpredict the observed $\meanmgfe$ by $\sim0.02-0.05$ dex for low-$\sigma$ ellipticals at $z=0$ (green curves in panels a and b). For the higher-$\sigma$ galaxies, achieving super-solar $\meanmgh$ at older ages requires higher SFE, which leads to a more rapid drop in $\meanmgfe$. For the mid-$\sigma$ galaxies, the gap between our representative model (pink curve) and the observed $\meanmgfe$ is $\sim0.06-0.08$ dex, and for the oldest, high-$\sigma$ ellipticals (blue curve) it is $\sim0.07$ dex. We have made parameter choices that tend to maximize $\meanmgfe$ at a given $\meanage$ subject to the constraints imposed by $\meanmgh$, and one can see from Figure \ref{fig: summary} that many alternative choices would worsen the agreement with the observed abundance trends.

Gaps of $\sim0.05$ dex are small enough that they might be explained by systematic errors in the SSP fits themselves, e.g., because of imperfections in the stellar libraries used for the spectral synthesis. However, since we have already pushed our parameter choices to reduce discrepancies to this level, we will examine possible physical resolutions, arising from lower SNIa yields (Section \ref{sec:SNIa}) or from terminating bursts of star formation (Section \ref{sec:bursts}). Even if systematic errors can adequately explain the gaps between our model predictions and the binned-average data, other mechanisms may be needed to explain individual galaxies that lie above the mean. Before considering these mechanisms, we re-examine the relationship between SSP fits and full evolutionary models, now guided by the models that come closest to matching the data. We return to a broader discussion of our results, including the choice of CCSN yields and the effects of the stellar IMF, in Section \ref{sec:discussion}.

\subsection{Test of SSP Fits}
\label{sec:mock}

\begin{figure*}[!]
\centering
\includegraphics[width = \textwidth]{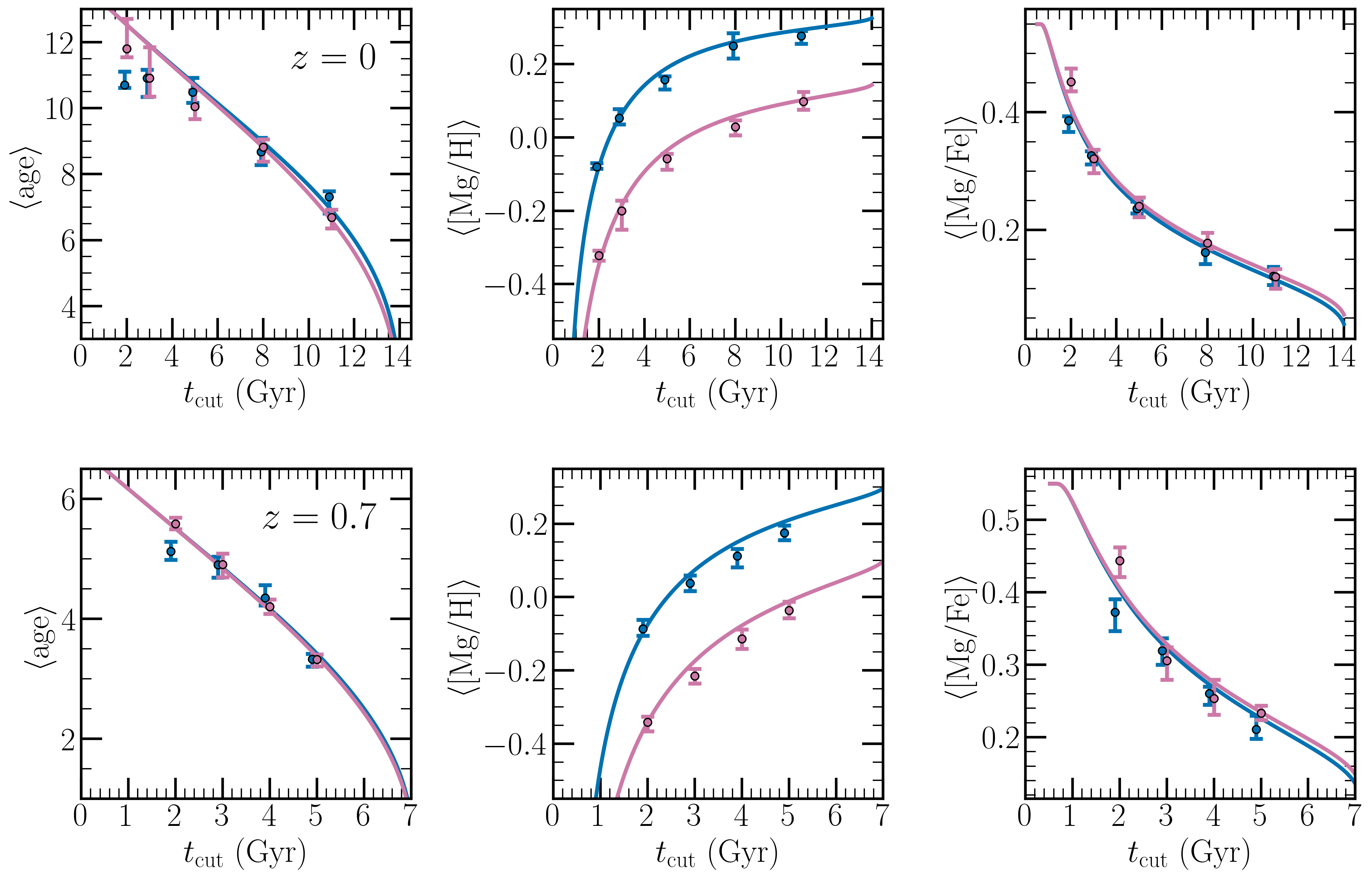}
\vspace{-5mm}
\caption{Mock-recovery test results comparing the light-weighted model predictions (lines) to the best-fit SSP-equivalent quantities derived from mock CSP spectra (points). CSPs were generated using the pink and blue GCE models from Figure \ref{fig: best fits} for $z=0$ (top row) and $z=0.7$ (bottom row). At various cutoff times, mock spectra with 20 noise realizations were fit using the method described in Section~\ref{sec:data}. Error bars represent the 16th and 84th percentiles of the best-fit SSP-equivalent quantities for the 20 realizations of each CSP. This test shows that SSP-equivalent quantities can be well-approximated by light-weighted values, to within 0.02 dex for $\meanmgh$ and $\meanmgfe$, with a small bias in $\meanage$ for the oldest models.  }
\label{fig: mock}
\end{figure*}

Our two-population tests in Appendix \ref{appx:mock}, summarized in Section \ref{sec:data}, show that an SSP fit recovers the light-weighted, logarithmically averaged age, $\mgh$, and $\mgfe$ of a composite stellar population. Now that we have guidance on the model parameters relevant to observed ellipticals, we test the consistency between SSP fits and light-weighted averages for models with extended SFHs and continuously evolving metal content. 

We construct composite stellar population spectra corresponding to the pink and blue GCE models presented in Figure~\ref{fig: best fits}. To do so we compute SSP spectra at each time step based on the evolving gas-phase [Mg/H] and [Mg/Fe] and the look-back time $t_{\mathrm{obs}}-t$. For each choice of truncation time $t_{\rm cut}$, we combine all SSPs up to $t_{\rm cut}$, weighting each by the mass fraction formed in each time step.  The appropriate light weighting at each wavelength follows directly from the spectral synthesis.
%normalizing them by their $L_\mathrm{V}/M$ (Eq. \ref{eqn:weighing}) and by the mass-fraction formed at each time step.

We simulate observational data from these CSP spectra by adding Gaussian noise corresponding to S/N = 40\,\AA$^{-1}$, which is typical of individual LEGA-C and SDSS spectra. These mock observations are then fit using the same full-spectrum SSP fitting method applied to the observations as described in Section~\ref{sec:data}. 

In Figure~\ref{fig: mock}, smooth curves show the models' light-weighted $\meanage$, $\meanmgh$, and $\meanmgfe$, as a function of $\tcut$, with $t_{\mathrm{obs}}=14$ Gyr ($z=0$, top row) and $t_{\mathrm{obs}}=7$ Gyr ($z=0.7$, bottom row). Points show the mean SSP-fitting results from 20 mock spectra noise realizations at selected values of $\tcut$. Error bars show 16th-84th percentile range of the 20 realizations.  The uncertainty in the mean prediction is smaller by $\approx \sqrt{20}$, and the statistical errors on the observational data points in Figure~\ref{fig: best fits} are often smaller than this because more than 20 galaxies contribute to an individual $\sigma$-bin.

Figure~\ref{fig: mock} demonstrates good agreement between light-weighted model quantities and SSP fits for realistic SFH and enrichment histories, as expected from the two-population tests. The model averages typically overpredict the SSP $\meanmgh$, but by $\lesssim 0.02$ dex. Agreement in $\meanmgfe$ is similarly at the 0.01-0.02 dex level in most cases.
The logarithmically averaged $\meanage$ typically agrees with the SSP values to $\lesssim 0.5$ Gyr. However, for the oldest models (earliest $\tcut$) at $z=0$, the difference rises to 1-1.5 Gyr. Accounting for this effect would noticeably improve the agreement between models and data for the most massive $z=0$ galaxies, as one can see by mentally translating the blue and pink curves in Figure \ref{fig: best fits} leftward by 1-1.5 Gyr. The differences for the oldest galaxies may be a consequence of incorporating CSPs at $\feh < -0.7$ that rely on empirical stellar libraries in a poorly constrained metallicity regime \citep{villaume_extended_2017}.  More accurate libraries in this regime might reduce the difference between light-weighted evolutionary histories and SSP fits.
%When we limit the metallicity of the input SSPs to [Fe/H]$>-0.7$, the age underestimation is only $\lesssim0.5$\;Gyr.

\subsection{SNIa Yields}
\label{sec:SNIa}

\begin{figure}[!]
\centering
\includegraphics[width = \columnwidth]{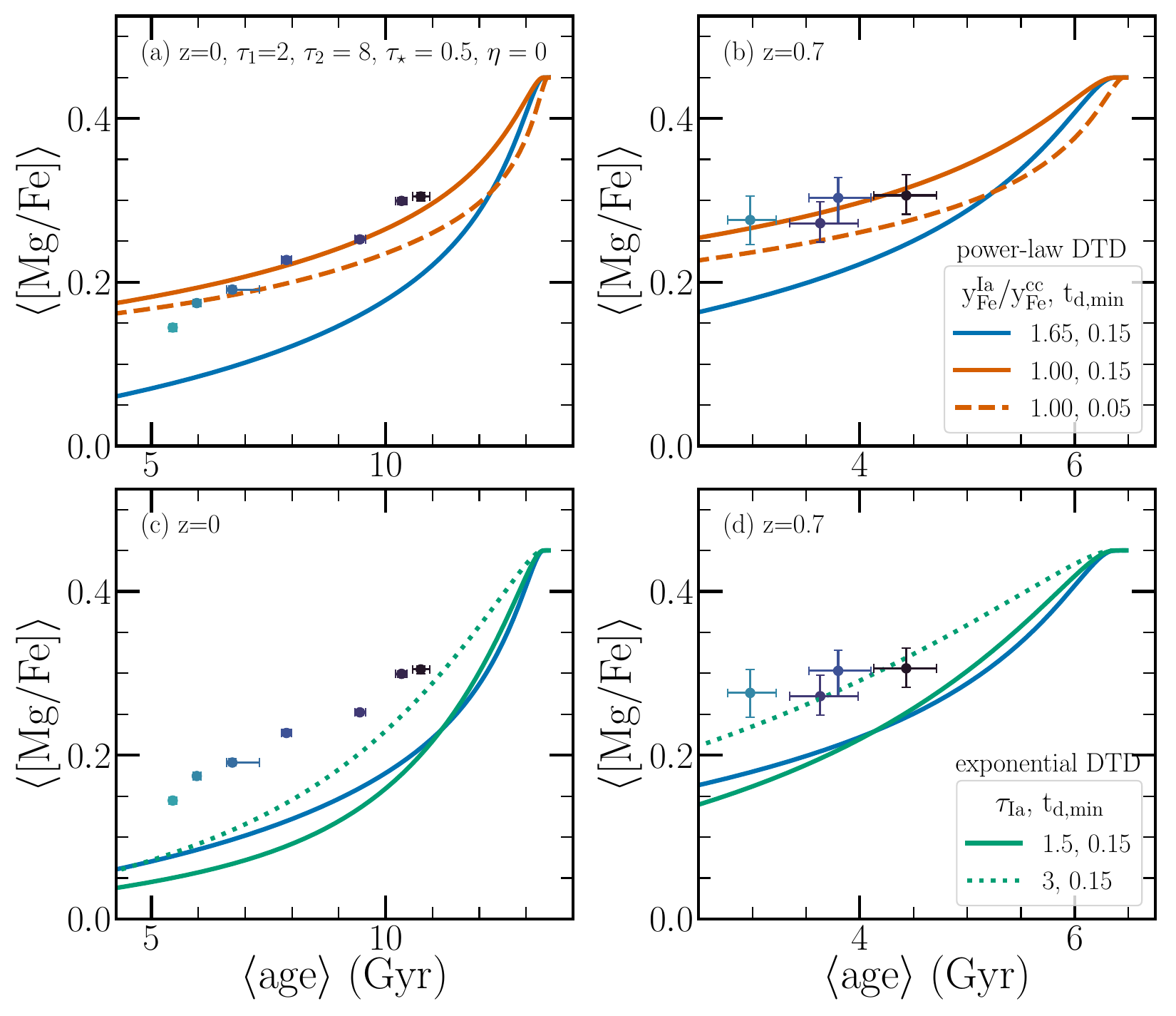}
\vspace{-5mm}
\caption{Influence of SNIa Fe yields. {\it (Top panels)} The solid blue and orange curves compare models with the reference SNIa Fe yield ($\ironratio=1.65$) and a reduced SNIa Fe yield ($\ironratio=1.00$), respectively. Shortening $t_{\mathrm{d,min}}$ for the reduced yield model (dashed orange curve) results in lower $\meanmgfe$ at early times but similar $\meanmgfe$ at the ages of the observed galaxies. {\it (Bottom panels)} The blue and green curves compare models with a $t^{-1.1}$ power-law DTD and a single exponential DTD, respectively. The solid green curve has $\tauIa=1.5$ Gyr while the dotted green curve has $\tauIa=3$ Gyr. All models are based on the representative model shown by the blue curve in Figure \ref{fig: best fits} ($\tau_1=2$ Gyr, $\tau_2=8$ Gyr, $\taustar=0.5$ Gyr and $\eta=0$), but with the reference [Mg/Fe]$_{\mathrm{cc}}=0.45$. A lower SNIa Fe yield or preferential loss of SNIa Fe in galactic outflows would enable our elliptical galaxy models to reproduce observed $\meanmgfe$ ratios.}
\label{fig: SNIa}
\end{figure}

Although we can obtain marginal agreement with the elliptical galaxy observables by adopting a high $\mgfecc=0.55$, most Milky Way observations favor $\mgfecc \approx 0.45$ or (in the case of APOGEE) even lower.  Another parameter that we have not yet explored is the SNIa Fe yield.  Our normalization condition (Equation~\ref{eqn:yfeIa}) is chosen to produce near-solar $\mgfe$ in the ISM at late times, but galaxies that become massive ellipticals could have a different $\yfeIa$ from disk galaxies like the Milky Way.

For $\mgfecc=0.45$, Equation~\ref{eqn:yfeIa} gives $\yfeIa/\yfecc=1.65$. 
The blue solid curves in Figure \ref{fig: SNIa} show $\meanmgfe$ vs. $\meanage$ in a model with these yields and other parameters ($\tau_1=2$ Gyr, $\tau_2=8$ Gyr, $\taustar=0.5$ Gyr and $\eta=0$) matched to the blue-curve model of Figure~\ref{fig: best fits}.  The solid orange curves show the impact of reducing the SNIa yield to $\yfeIa=\yfecc$, with other yields unchanged.  This reduced $\yfeIa$ model is remarkably successful at reproducing the observed $\meanmgfe$-$\meanage$ relations at both $z=0$ and $z=0.7$.  If we also reduce $\tdmin$ from 0.15 Gyr to 0.05 Gyr (dashed orange curves), the agreement with the data is somewhat worse, but still better than that of the original reference model.

Reducing $\yfeIa$ by a factor of 1.65 is equivalent to reducing the normalization of the SNIa DTD in ellipticals by the same factor.  However, observations suggest that the SNIa rate in galaxy clusters, which have a high fraction of elliptical and lenticular galaxies, is higher than the rate in star-forming field galaxies, not lower \citep{Maoz2017}. Because the stellar populations of ellipticals are old, observed rates probe the late phases of the delay time distribution, so this inference of a higher time-integrated rate depends on assuming that the DTD form is the same in cluster and field galaxies. It is also possible that the $\ironratio$ in ellipticals is the same as that in our reference model, or higher, but that ellipticals preferentially lose SNIa products in galactic winds relative to CCSN products, even at epochs when they are actively forming stars and enriching their ISM. Such preferential ejection could arise if CCSN ejecta are trapped in dense star-forming gas while SNIa ejecta are released in lower-density gas that is more easily vented from the galaxy. Nonetheless, while these scenarios are possible, there seems to be no good empirical or theoretical argument for reduced $\ironratio$ in elliptical galaxies {\it other} than the high $\meanmgfe$ ratios. 

\cite{Dubay2024} find that GCE models are more successful at reproducing Milky Way observations if the SNIa DTD has fewer short-delay supernovae than predicted by the $t^{-1.1}$ power-law advocated by, e.g., \cite{Maoz2017} and \cite{Wiseman2021}.  \cite{Strolger2020} argue that an exponential-like DTD fits observed SNIa rates as well as the power-law model (see also \citealt{Palicio2024}).  Dubay et al.\ show (in their Appendix B) that an exponential DTD with a timescale $\tauIa \approx 1.5\Gyr$ is similar to the theoretically motivated single-degenerate DTD of \cite{Greggio2005}, and a number of GCE studies have adopted this exponential form for the DTD (e.g., \citealt{Schoenrich2009,Andrews2017}; WAF).  Relative to the $t^{-1.1}$ power-law with $\tdmin=0.15\Gyr$, the $\tauIa=1.5\Gyr$ exponential DTD has a similar median delay time ($\approx 1\Gyr$) but fewer SNIa at $t \ll 1\Gyr$.  
%DW: My calculations give median of 0.98 Gyr for power-law with td = 0.15-10 Gyr and median of 0.91 Gyr for the exponential with tauIa=1.5.
The solid green curve in Figure~\ref{fig: SNIa} adopts the same parameters and $\yfeIa$ as the blue curve but employs a single exponential DTD with $\tauIa=1.5\Gyr$.  For the oldest ages, the predicted $\meanmgfe$ is slightly higher than that of the reference model, which uses a sum of two exponentials to approximate a $t^{-1.1}$ power-law.  However, at younger ages the predicted $\meanmgfe$ is slightly lower because the exponential DTD has more intermediate-time SNIa than the long-tailed, power-law DTD.  Thus, simply changing the form of the DTD to a single exponential does not improve agreement with the elliptical galaxy observations.

\cite{Dubay2024} also consider an exponential DTD with $\tauIa=3\Gyr$.  We have repeated our calculations for this DTD (dotted green curve), and in this case, we get much better agreement with the data at $z=0.7$ and $z=0$. Like the findings of \cite{Dubay2024}, Figure~\ref{fig: SNIa} highlights the importance of pinning down both the form and the normalization of the SNIa DTD, and testing its universality across different types of galaxies.

\subsection{Terminating Bursts of Star Formation}
\label{sec:bursts}

Our models assume flexible but smooth star formation histories.  Bursts of star formation produce temporary boosts in [Mg/Fe] because the rate of $\alpha$-element production from massive stars increases during the burst (WAF; \citealt{Johnson2020}).  If a burst occurs at the beginning or middle of the SFH, then the Fe from SNIa progenitors formed in the burst will eventually be incorporated into stars, so we expect the impact on $\meanmgfe$ to be small.  A burst that {\it terminates} the galaxy's SFH is different because the newly produced Mg can be incorporated into stars during the burst (self-enrichment), but the SNIa Fe production occurs after star formation has ceased.  It has long been known that some galaxies, variously referred to as ``E+A'', ``K+A'', or ``post-starburst'', undergo bursts of star formation before abruptly transitioning to quiescence \citep{Dressler1983,Zabludoff1996,Balogh1999,Leung2023,Park2023}.  However, it is unclear whether {\it most} early type galaxies end their SFH with a final burst. 

To estimate the impact of a terminating burst, we compute the mean metallicity of stars formed in the burst using a ``leaky box'' chemical evolution model \citep{Hartwick1976}, which extends ``closed box'' evolution \citep{Talbot1971} by allowing outflow.  We implicitly assume that the burst duration is long compared to the $\sim 30\Myr$ lifetime of massive stars, so that newly produced CCSN elements are incorporated into later generations within the burst, but short compared to the $\sim 1\Gyr$ timescale over which substantial SNIa Fe production would occur.  A terminating burst could be triggered by a galaxy merger that brings a substantial amount of fresh gas, which dilutes the metallicity of the ISM but increases the fuel available to make stars.  Our use of the leaky box model implies that any further gas accretion {\it during} the burst is negligible compared to the initial supply.  While our calculation is idealized, it captures the distinctive ways that a terminating burst affects the mean stellar abundances of a galaxy and gives an approximate value of the quantitative impact on observables.

We give details of this calculation in Appendix~\ref{appx:burst}.  The parameters of a burst in our model are $f_g$, the ratio of gas remaining at the end of the burst to gas at the start of the burst, and $F_D$, the ratio of accreted gas to gas present in the galaxy at the end of smooth evolution.  For representative values of $f_g$ we consider $f_g=0.3$ and $f_g=0.05$, corresponding to bursts that consume 70\% and 95\% of the initial gas supply.  We assume that outflows during the burst are negligible, regardless of the value of $\eta$ assumed during smooth evolution.  This choice maximizes the impact of the burst because all metals formed during the burst are retained.

For the dilution factor we consider $F_D=0$, corresponding to a burst that only consumes the pre-existing gas in the ISM, and $F_D=3$, corresponding to an accretion event that increases the gas supply by a factor $1+F_D=4$.  For example, $F_D=3$ could describe a galaxy with $10^{11}\Msun$ of stars and $10^{10}\Msun$ of remaining gas that accretes a gas rich companion with $3\times 10^{10}\Msun$ of gas, triggering a burst.  We ignore the impact of accreted stars.  We assume that accretion lowers the ISM abundance of both Mg and Fe by a factor $(1+F_D)$, which is a reasonable approximation for a minor merger in which the accreted system has much lower metallicity than the main galaxy.  The mass and abundances of stars formed during the burst are computed from \Eqref{mburst} and \Eqref{Zxburst}, respectively.  

The impact of a burst on observables is enhanced by light-weighting, as the stellar population formed during the burst is younger and brighter than the stellar populations formed during smooth evolution.  We assume that the burst duration is long enough for self-enrichment but short compared to other GCE timescales.  We therefore assign all stellar mass from the burst to the timestep ($\Delta t = 0.02\Gyr$) that immediately follows $t_{\rm burst}=\tcut$.  Our procedure should be accurate for burst durations up to $\sim 0.3\Gyr$, but for longer burst durations one could not ignore SNIa enrichment.

\begin{figure}[!]
\centering
\includegraphics[width = \columnwidth]{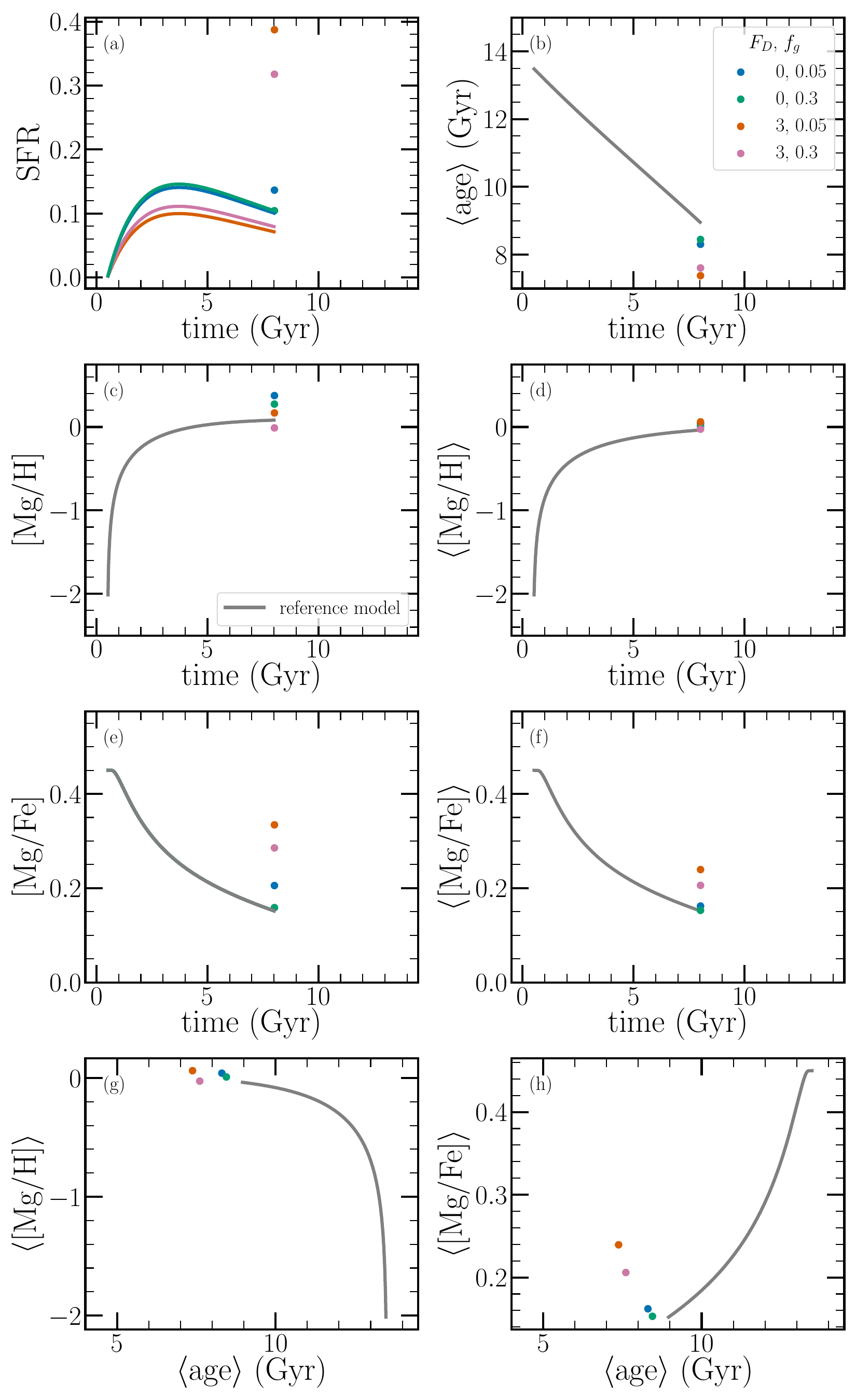}
\vspace{-5mm}
\caption{Impact of a burst of star formation at $t_{\mathrm{burst}}=\tcut=8$ Gyr on age and abundance evolution. The reference model ($\tau_1=2$ Gyr, $\tau_2=8$ Gyr, $\taustar=1$ Gyr, $\eta=0.3$, solid orange curve in Figure \ref{fig: Mass vs Light Weighting}) before the burst is shown in gray. The colored points in panel (a) represent the fraction of total stars formed in the burst, ranging from $\sim10\%$ ($F_D=0$, $f_g=0.3$) to $\sim40\%$ ($F_D=3$, $f_g=0.05$); the colored curves differ only in normalization, such that the area under the curve plus the final point sum to one. Points in the other panels show the post-burst age or abundance. The colors indicate different starburst models, with corresponding ($F_D$, $f_g$) values given in the legend. A burst of star formation decreases $\meanage$ while increasing $\meanmgfe$.}
\label{fig: example burst}
\end{figure}

Figure \ref{fig: example burst} shows the effects of a final burst of star formation at $t_{\mathrm{burst}}=8$ Gyr on the reference model for different burst parameters. The curves in Figure \ref{fig: example burst}a show the smooth SFH as before, while points at 8 Gyr show the fraction of the galaxy's stars that form in the final burst. Burst models that have accretion of pristine gas ($F_D=3$) and use a large fraction of gas in the burst ($f_g=0.05$) form a higher fraction of stars ($\sim$40\%) during the burst. In comparison, burst models with no gas accretion ($F_D=0$) and a lower fraction of gas used in the burst ($f_g=0.3$) form $\sim$10\% of their total stars during the burst. Light-weighting makes the final burst have a significant impact on $\meanage$, especially for late $t_{\mathrm{burst}}$. The gray curve in Figure \ref{fig: example burst}b shows age evolution for the reference model before the burst, with the colored points indicating the post-burst values. A final burst of star formation lowers $\meanage$ by as much as 1 Gyr from the pre-burst value, with models that form a higher fraction of stars during the burst having younger, final $\meanage$. 

Points in Figures \ref{fig: example burst}c and \ref{fig: example burst}e show the $\mgh$ and $\mgfe$ of stars formed during the burst, computed as described in Appendix \ref{appx:burst}. If there is no accretion ($F_D=0$), $\mgh$ increases due to self-enrichment from CCSN producing Mg during the burst. Models that convert a higher fraction of gas into stars during the burst ($f_g=0.05$) produce more CCSN and therefore show a larger increase in $\mgh$. If the ISM is diluted by pristine gas immediately before the burst ($F_D=3$), then the $\mgh$ of burst stars is lower, and it even lies below the $\mgh$ at the end of the smooth evolution if only a small fraction of gas is converted into stars ($f_g=0.3$).

We assume that the final bursts are long enough to incorporate CCSN products produced by stars formed in the burst, but short enough that there is no Fe enrichment from SNIa. As a result, all burst models increase $\mgfe$. Models with larger burst fractions show larger $\mgfe$ increases, so in contrast to $\mgh$ the $F_D=3$ models have the largest boost of $\mgfe$. Figures \ref{fig: example burst}d and \ref{fig: example burst}f show the final light-weighted $\meanmgh$ and $\meanmgfe$ of the entire galaxy. Relative to Figures \ref{fig: example burst}c and \ref{fig: example burst}e, abundance changes are reduced because the burst comprises only a fraction of the galaxy's stars. Figures \ref{fig: example burst}g and \ref{fig: example burst}h plot $\meanmgh$ and $\meanmgfe$ against $\meanage$. Overall, adding a final burst of star formation increases $\meanmgfe$ and decreases $\meanage$, with both changes moving the models in the right direction to reproduce the observed data.

\begin{figure*}[!]
\centering
\includegraphics[width = \textwidth]{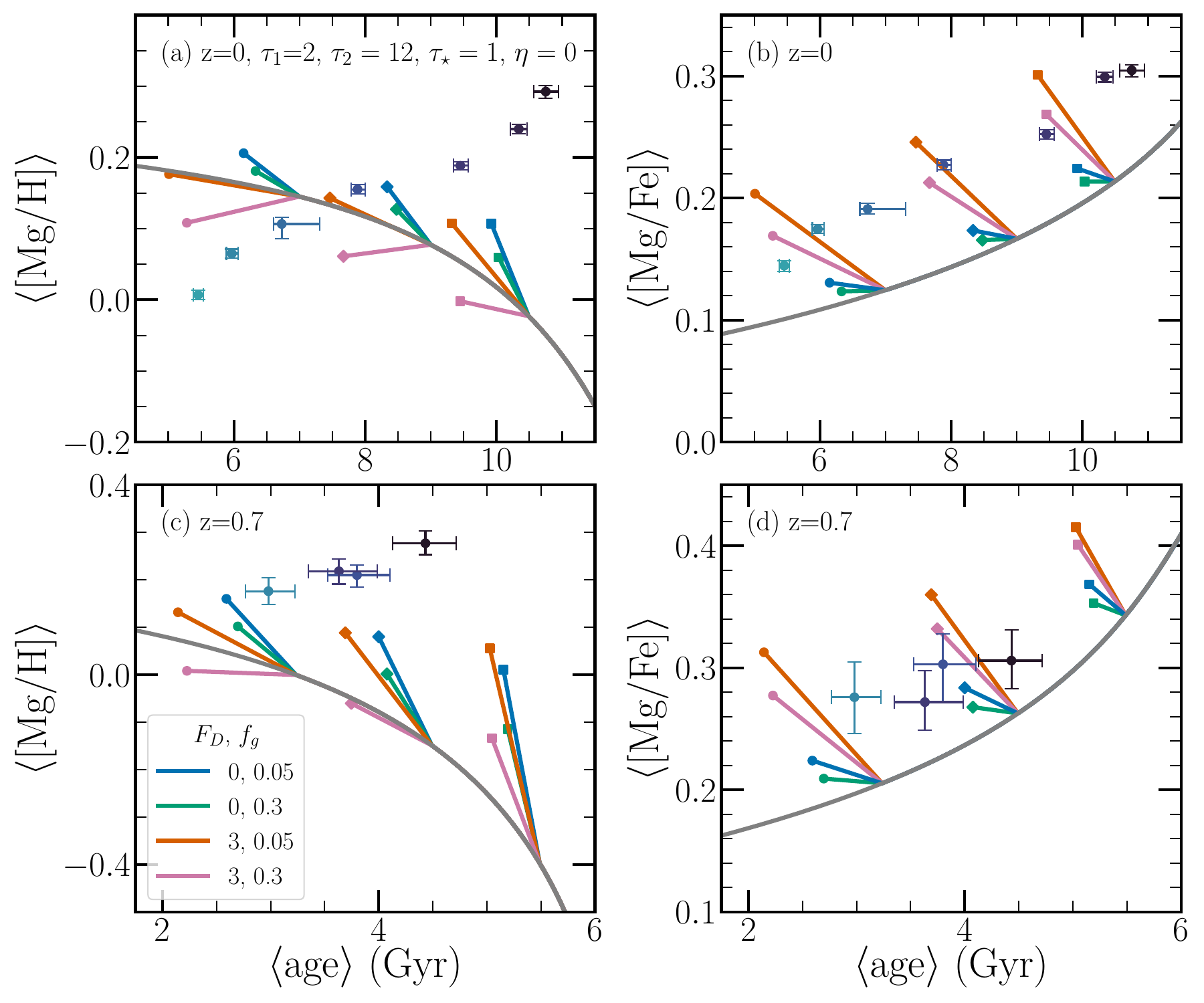}
\vspace{-5mm}
\caption{Comparison between representative starburst models and observed ages and abundances for galaxies stacked in bins of velocity dispersion. The gray curve is similar to the pink, representative model shown in Figure \ref{fig: best fits} ($\tau_1=2$ Gyr, $\tau_2=12$ Gyr, $\taustar=1$ Gyr), but with a lower Mg yield ($\mgfecc = 0.45$) and with no outflows ($\eta=0$). The colored lines show the effects of a burst of star formation occurring at $t_{\mathrm{burst}}=5.30,$ 7.70, and 10.52 Gyr at $z=0$ and $t_{\mathrm{burst}}=2.02,$ 3.50, and 5.16 Gyr at $z=0.7$, with endpoints marking the post-burst value. The colors indicate different starburst models, with corresponding ($F_D$, $f_g$) values given in the legend. Incorporating a substantial burst of star formation right before quenching can reproduce the observed, elevated $\meanmgfe$ ratios. The bursts typically boost $\meanmgh$ as well, but the effect is reduced if the burst is fueled by freshly accreted, low-metallicity gas (high $F_D$).}
\label{fig: burst}
\end{figure*}

Figure \ref{fig: burst} provides a comparison between representative starburst models and the observed ages and abundances for galaxies stacked in bins of velocity dispersion, $\sigma$. We introduce bursts of star formation at $t_{\mathrm{burst}}=5.30,$ 7.70, and 10.52 Gyr for the $z=0$ comparison and $t_{\mathrm{burst}}=2.04,$ 3.50, and 5.16 Gyr for the $z=0.7$ comparison. As a baseline smooth model, we adopt parameters similar to those of the central model (pink curve) shown in Figure \ref{fig: best fits} ($\tau_1=2$ Gyr, $\tau_2=12$ Gyr, $\taustar=1$ Gyr), but we adopt the lower Mg yield favored by Milky Way disk data ($\mgfecc = 0.45$), and we set outflows to $\eta=0$ to compensate for this lower yield. The colored lines show the effects of the burst models at each $t_{\mathrm{burst}}$, with the endpoints marking the post-burst value. 

Figure \ref{fig: burst} shows that adding a substantial final burst of star formation to a reasonable smooth star formation history can bridge the gap between predicted and observed $\meanmgfe$ ratios, bringing models with $\mgfecc=0.45$ within the range of the data. The effects of the burst on $\meanage$ and on $\meanmgfe$ are about equally important. While a burst that consumes the existing gas (e.g., $F_D=0$, $f_g=0.05$)  can resolve the tension with data at $z=0.7$, fully resolving the tension at $z=0$ requires a larger burst with accreted gas. In these $F_D=3$ models, the $\meanmgh$ at a given age is similar to or even below that of the smooth model, as dilution outweighs the effect of self-enrichment. The models shown in Figure \ref{fig: burst} fall short of the observed $\meanmgh$ of the high-$\sigma$ galaxies. However, this discrepancy could be resolved by adopting a shorter $\taustar$ and $\tau_2$ for this population and, if necessary, boosting the Mg yield above the level assumed here (see Figure \ref{fig: best fits}).

If self-enrichment during terminating bursts is indeed the explanation for the high observed $\meanmgfe$ relative to our model predictions, it implies that these large final bursts of star formation are the norm in elliptical galaxy evolution, not the exception. Population synthesis models that isolate late star formation can test this conjecture. Furthermore, since not all galaxies will experience bursts of the same magnitude, we expect correlations between the fraction of stars formed immediately before quiescence and the mean abundances of the stellar population. 

To illustrate this point, we construct models with varying values of $F_D$ and $f_g$, and $t_{\mathrm{burst}}$ chosen to produce a specified $\meanage$, so we can predict correlations that might be seen within ellipticals of a given age or velocity dispersion. The smooth models are the blue, pink, and green representative models presented in Figure \ref{fig: best fits} but with lower Mg yields, $\mgfecc=0.45$, and with outflows set to $\eta=0$ for the pink model to boost $\meanmgh$. In all cases, the burst is assumed to terminate the galaxy's SFH at $t_{\mathrm{burst}}=\tcut$. We choose $F_D=0$, 1, 2, or 3, and for each value, we generate models with randomly chosen values of $f_g$ in the range 0.05-0.5 and randomly chosen values of $t_{\mathrm{burst}}$. We then select $(f_g,t_{\mathrm{burst}})$ combinations that give $\meanage$ in a narrow range, corresponding to low-$\sigma$, mid-$\sigma$, and high-$\sigma$ ellipticals at $z=0$ and to mid-$\sigma$ ellipticals at $z=0.7$. 

\begin{figure*}[!]
\centering
\includegraphics[width = \textwidth]{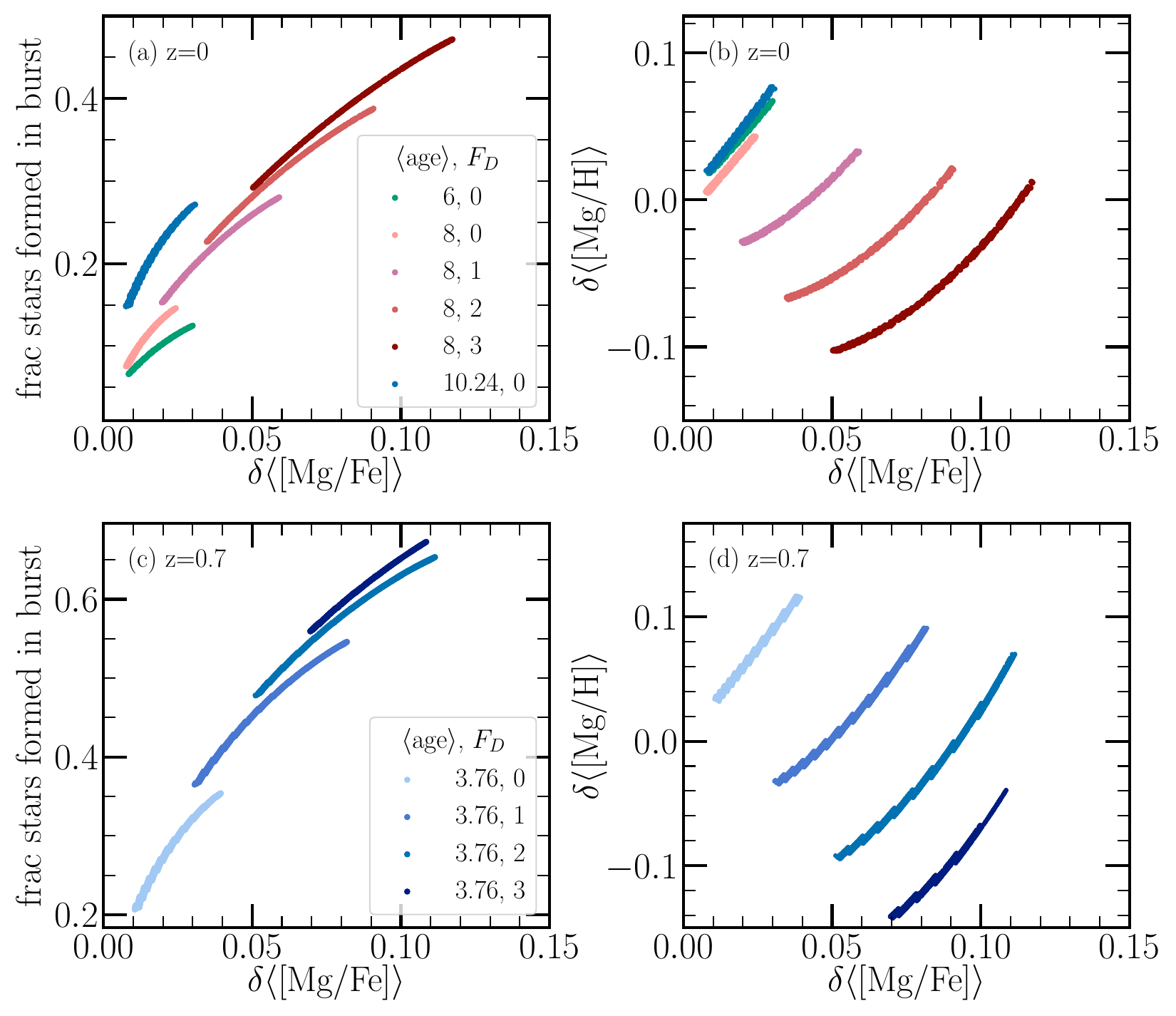}
\vspace{-5mm}
\caption{Correlation of burst stellar fraction and abundance differences for models with terminating bursts of star formation at $z=0$ (top) and $z=0.7$ (bottom). The quantities $\delta\meanmgfe$ and $\delta\meanmgh$ are the changes in average abundances relative to the model with no burst. We consider models with dilution factor $F_D=0$, 1, 2, and 3, and for each $F_D$ the points trace out a range of final gas fraction from $f_g=0.5$ (smallest $\delta\meanmgfe$) to $f_g=0.05$ (largest  $\delta\meanmgfe$). The smooth model parameters of the blue, pink, and green sequences correspond to those of the blue, pink, and green models in Figure \ref{fig: best fits}, except here we adopt $\mgfecc=0.45$ and set $\eta=0$ for the pink model. In each sequence we choose $t_{\mathrm{burst}}$ to give $\meanage\approx10.24$ Gyr (blue), 8 Gyr (pink), and 6 Gyr (green) at $z=0$ or $\meanage\approx3.76$ Gyr at $z=0.7$. }
\label{fig: gal sample}
\end{figure*}

Figures \ref{fig: gal sample}a and \ref{fig: gal sample}c correlate the fraction of stars formed during the burst with $\delta\meanmgfe$, the deviation of the final $\meanmgfe$ from the smooth model $\meanmgfe$ at the corresponding $\meanage$, for the model elliptical populations at $z=0$ and $z=0.7$, respectively. Each curve shows the results for the different-age populations at a fixed $F_D$, with a lower final gas fraction $f_g$ corresponding to a higher burst stellar fraction and larger $\delta\meanmgfe$. For $F_D=0$, even the $f_g=0.05$ models have only small increments of $\delta\meanmgfe<0.035$ dex, at either redshift. However, burst models with substantial accreted gas fractions (large $F_D$) can have larger $\delta\meanmgfe$, with large fractions of stars formed during the burst.

Figures \ref{fig: gal sample}b and \ref{fig: gal sample}d compare $\delta\meanmgh$ to $\delta\meanmgfe$ for the model elliptical populations at $z=0$ and $z=0.7$, respectively. For $F_D=0$, models that consume more gas (lower $f_g$) have both higher $\meanmgfe$ and higher $\meanmgh$, with a similar correlation for the three population ages. However, this relationship is more complicated for $F_D>0$ because of the dilution effect on $\meanmgh$. At fixed $\delta\meanmgfe$, ellipticals with larger $F_D$ (darker blue curves) have more negative $\delta\meanmgh$ than other ellipticals of similar $\meanage$. 

Figure \ref{fig: gal sample} shows that raising $\meanmgfe$ as much as 0.05 dex above the smooth model prediction requires a substantial fraction of stars to be formed in a terminating burst of star formation. This can be achieved by adding a significant amount of fresh gas during the burst through a major accretion event. Smaller changes in $\meanmgfe$ can be obtained by forming 10-20\% of stars in a terminating burst with no gas accretion. A test of this terminating burst hypothesis is that galaxies with higher $\meanmgfe$ at a given $\meanage$ or velocity dispersion should show signatures of substantial late star formation. However, there is no clear predicted correlation between deviations in $\meanmgh$ and $\meanmgfe$ because of the impact of dilution on $\meanmgh$. 

\section{Discussion}
\label{sec:discussion}

\subsection{Yields and IMF}
\label{sec:IMF}

Our calculations show that the nucleosynthetic yields and SNIa DTD are the key factors governing the predicted $\meanmgh$-$\meanage$ and $\meanmgfe$-$\meanage$ relations of ellipticals. Once we choose the yields and DTD, the SFH has only a small impact on the predicted relations (see Figures \ref{fig: tau1 tau2 z=0}gh and \ref{fig: summary}), within the fairly wide range of behavior allowed by our smooth SFH parameters. The choices of SFE timescale $\taustar$ and outflow efficiency $\eta$ have an important impact on the early and late evolution, respectively, of $\meanmgh$, but they have limited impact on $\meanmgfe$ (Figure \ref{fig: taustar eta z=0}). The high SFE needed to get early enrichment in $\meanmgh$ tends to drive down $\meanmgfe$, making it difficult to explain the high observed values of both quantities in the oldest, high-$\sigma$ ellipticals (Figure \ref{fig: summary}, Figure \ref{fig: best fits}).

For our fiducial choice of yields, Equation~\ref{eqn:yvals}, we are unable to reproduce the high observed $\meanmgfe$ ratios in our $z\sim0$ and $z\sim0.7$ samples, falling short by 0.05-0.1 dex (dotted gray curve in Figure \ref{fig: best fits}). As discussed in Section \ref{sec:models}, the overall normalization of these yields is motivated by the \cite{Rodriguez2023} determination of the mean Fe yield of CCSN; the ratio $\ymgcc/\yfecc$ corresponding to $\mgfecc=0.45$ is motivated by the observed plateau in $\mgfe$ for low metallicity stars in the Milky Way disk and halo; and the ratio $\yfeIa/\yfecc$ is motivated by the expectation that models evolve to $\mgfe\approx0$ at late times to explain solar abundances in MW disk stars. 

Nearly all stars in APOGEE with $\feh>-1.5$ have $\mgfe<0.45$ (see Figure 1 of \citetalias{Weinberg2024}), and a plateau at $\mgfecc\leq0.45$ is also supported by most of the data compiled by \cite{Kobayashi2020} (see their Figure 12) and by the Sgr dwarf and Milky Way halo stars shown by \cite{Sestito2024} (see their Figure 6). However, based on a sample of halo stars from the H3 survey extending to $\feh=-3$, \cite{Conroy2022} argue that the true core collapse plateau lies at $\mgfecc\approx0.6$, and that the roughly flat trend for stars with $-1.5\lesssim\feh\lesssim-0.5$ already includes a substantial SNIa contribution (see also \citealt{Maoz2017}). With our higher choice of $\mgfecc=0.55$, giving the yields in Equation~\ref{eqn:yvals2}, our smooth SFH models still underproduce the observed $\meanmgfe$, but at a level that is possibly within the systematic uncertainties of the SSP abundance measurements (Figure \ref{fig: best fits}). Unfortunately, massive star and CCSN models are not robust enough to settle the choice of $\mgfecc$ theoretically, as illustrated by the fact that state-of-the-art models (e.g., \citealt{Sukhbold2016}) miss the solar [O/Mg] ratio by 0.4 dex even though O and Mg are produced in stars of similar mass \citep{Griffith2021snyield}.

The empirical evidence for $\mgfecc\approx0.45$ comes from stars with $\feh<-0.5$. Our models assume $\mgfecc$ is independent of metallicity, which is consistent with theoretical predictions (e.g., \cite{Nomoto2013,Andrews2017}). However, these theoretical predictions could be incorrect, and the yield ratios at the $\mgh\approx0$ metallicities that matter most for our models could be higher than the plateau observed for lower metallicity stars. 

Another possible explanation of the difference between our models and the data is that the IMF-averaged Mg and Fe yields in ellipticals are simply different than those in the Milky Way. For CCSN, the IMF-averaged yields could be different because of a difference in the high-mass slope of the IMF, perhaps connected to star formation in environments of high ambient pressure. A top-heavy IMF would increase the Mg and Fe yields $\ymgcc$ and $\yfecc$, making it easier to reproduce the observed $\meanmgh$ and $\meanmgfe$ in the oldest galaxies and reducing the need for a high SFE to explain this rapid enrichment. Using the CCSN yields of \cite{Chieffi2013}, \cite{Griffith2021bulge} find that changing the high-mass IMF slope from -2.3 \citep{Kroupa2001} to -2.0 slightly decreases the predicted $\mgfecc$, by 0.03 dex. We used \texttt{VICE} \citep{Johnson2020} to perform a similar calculation for yields from \cite{Sukhbold2016} with different choices of explosion landscape.  Here we find that changing the high-mass IMF slope from -2.3 to -2.0 increases $\mgfecc$, but only by 0.02-0.03 dex.  Both of these calculations suggest that the IMF shape has little leverage on the CCSN Mg/Fe ratio.  However, this dependence merits further investigation with other yield models.

Raising the CCSN yields relative to SNIa yields (higher $\yfecc/\yfeIa$) with a top-heavy IMF would help to boost $\meanmgfe$ even if $\mgfecc$ is unchanged, similar to the effect shown in Figure~\ref{fig: SNIa}. Black hole formation can also have a large impact on IMF-averaged yields (\citealt{Griffith2021snyield}; \citetalias{Weinberg2024}), even if the IMF itself is unchanged. However, while black hole formation may well depend on metallicity, there is no obvious reason it should be different in ellipticals vs. the Milky Way at a given metallicity.

Much of the discussion about IMF differences in ellipticals has focused on low-mass behavior, in particular on observations suggesting that high-$\sigma$ ellipticals have more low mass ($M<0.5$ $M_\odot$) stars than predicted by a \cite{Kroupa2001} IMF \citep{vanDokkum2010,vanDokkum2012,Conroy2012IMF,Conroy2017IMF}. Because these star have lifetimes longer than the age of the Universe, the main impact of increasing their relative numbers is to lower all yields --- $\ymgcc$, $\yfecc$, and $\yfeIa$ --- by the same factor. A bottom-heavy IMF thus makes it more difficult to explain rapid enrichment, but it probably has little impact on $\meanmgfe$ predictions. If the increased numbers extend above the $\sim1$ $M_\odot$ turnoff, then the SNIa rate could be boosted by increased white dwarf formation, possibly explaining the high SNIa rate inferred in galaxy clusters \citep{Maoz2017}. An increased $\yfeIa/\yfecc$ would {\it worsen} the discrepancy with observed $\meanmgfe$, opposite to the reduced $\yfeIa/\yfecc$ models considered in Section \ref{sec:SNIa}.

\subsection{Future Directions}
\label{sec:future}

Our results suggest several directions for future investigations. One is to fit model parameters directly to spectra instead of comparing the light-weighted $\meanage$, $\meanmgh$, $\meanmgfe$ to SSP fit values. The tests presented in Figure \ref{fig: mock} indicate that predicting SSP quantities is an effective shortcut, but the residual inaccuracies in this process may still be enough to affect the assessment of models. Fitting directly to observed spectra requires computing model spectra for continuous sequences of stellar populations as we have done in Figure $\ref{fig: mock}$, but this process could be made reasonably efficient. For a smooth SFH and $\mgfecc=0.45$, we would not expect to obtain statistically acceptable fits to the data, since we cannot reproduce the observed $\meanmgfe$-$\meanage$ relation, but inspecting the discrepancies of best-fit models in spectral space might better reveal whether these discrepancies could arise from systematic uncertainties in the spectral modeling. 

Another promising direction is to use broad-band SED measurements (photometry or spectrophotometry) in addition to the continuum-normalized absorption spectra modeled by \texttt{alf}. Codes such as \texttt{Prospector} \citep{BJohnson2021} and \texttt{BagPipes} \citep{Carnall2018} use the broad-band SED to derive SFH constraints that go beyond effective population age defined by SSP fits. We might therefore hope to constrain our SFH parameters with the broad-band SED and allow absorption spectra to better constrain the yields and GCE parameters. In the context of ellipticals, it would be especially valuable to use the SED and spectra combination to probe the cessation of star formation and thus get insights into the physics of quenching. Our models assume a sharp cutoff at $\tcut$, but they allow an exponential decline before that, and the same $\meanage$ can be obtained with a rapid exponential decline and late cutoff or a slow decline and early cutoff. The predicted $\meanmgfe$-$\meanage$ tracks are surprisingly insensitive to the difference (Figure \ref{fig: tau1 tau2 z=0}), but the combination with SED-fitting might distinguish such models. Most importantly, our modeling in Section \ref{sec:bursts} suggests that terminating bursts of star formation may play an important role in boosting $\meanmgfe$ relative to the smooth-SFH predictions. Broad-band SED modeling may be able to confirm or refute the hypothesis that typical ellipticals terminate their star formation with a final burst. 

The combined spectra-SED constraints become still more promising if they can be applied on a galaxy-by-galaxy basis in addition to the high-precision constraints for binned samples. Our preliminary investigations imply that galaxies at fixed $M_\star$ exhibit intrinsic scatter in SSP age, $\mgh$, and $\mgfe$ comparable to the statistical uncertainties for individual SDSS or LEGA-C spectra (A. Beverage et al., in preparation). If terminating bursts are a common phenomenon, we expect correlations between deviations from the mean $\mgfe$ and the fraction of stars formed at the end of the galaxy's SFH, as illustrated in Figure \ref{fig: gal sample}. The correlation of scatter in age, $\mgh$, and $\mgfe$ at fixed $M_\star$ or $\sigma$ can provide further insights into the termination of star formation and the factors that govern abundance ratios. 

HST and JWST studies show that there are already substantial populations of quiescent galaxies present at $z=1-5$. The combination of ground-based and JWST spectroscopy allows the first assessments of stellar metallicities and $\mgfe$ ratios in such systems \citep[e.g.,][]{Beverage2024, Beverage2024Suspense, Kriek2024, Slob2024, Carnall2024}. For massive {\it star-forming} galaxies, one can combine stellar $\feh$ measurements with gas phase [O/H] measurements to infer the $\alpha$-enhancement \citep[e.g.,][]{Steidel2016, Topping2020, Kashino2022}. In a recent study, \cite{Stanton2024} applied this technique to massive star-forming galaxies at $z\approx3.5$, finding [O/Fe]$=0.425\pm0.026$ for the average value of this mixed gas-stellar abundance ratio. These high redshift observations provide another direction to extend our modeling, and probing to early epochs may help break remaining degeneracies between yields and other GCE parameters. Applying the WAF one-zone models to their $z\approx3.5$ data set, \cite{Stanton2024} find that explaining their high observed [O/Fe] ratios requires a high oxygen yield corresponding to [O/Fe]$_{\mathrm{cc}}\approx0.7$, a conclusion similar to but even stronger than the one inferred here from the $\meanmgfe$ ratios of $z\sim0$ and $z\sim0.7$ ellipticals. 

The other natural direction is to extend this work to model other element abundances. With SDSS and LEGA-C samples, we can obtain robust measurements for the mean abundances of up to a dozen elements at $z\sim0$ and six elements at $z\sim0.7$ \citep{Conroy2014, Beverage2023}. These include elements with different ratios of CCSN/SNIa enrichment (e.g., Si, Ca, Ni, Mn) and elements in which the delayed contribution comes from asymptotic giant branch (AGB) stars rather than SNIa (e.g., C, N, Sr, Ba). For all these elements, we would like to know whether their abundances in ellipticals can be explained by adopting stellar yields that reproduce Milky Way data combined with the GCE parameters inferred from Mg and Fe. The differing CCSN/SNIa ratios and the distinct DTD of AGB enrichment can further test the GCE models themselves.

\section{Conclusions}
\label{sec:conclusions}
Massive quiescent galaxy populations exhibit well-defined trends of stellar age, $\mgh$, and $\mgfe$ with galaxy velocity dispersion or stellar mass. We have constructed one-zone GCE models with flexible star formation histories to interpret the trends measured by \cite{Beverage2023}, who fit SSP models to the spectra of galaxies from the SDSS (at $z \approx 0$) and from LEGA-C (at $z \approx 0.7$). We predict SSP quantities from the light-weighted, logarithmically averaged quantities in the GCE model, denoted $\meanage$, $\meanmgh$, and $\meanmgfe$. Our tests with synthetic spectra from composite stellar populations show that this prediction is generally accurate, though the light-weighted $\meanage$ slightly exceeds the SSP age (by $\sim0.5$-2 Gyr) for the oldest galaxies (Figures \ref{fig: mock}, \ref{fig: simple_test}).

The components of our GCE models are the star formation history (SFH), the star formation efficiency timescale $\taustar=\Mgas/\Mdotstar$, the outflow efficiency $\eta=\Mdotout/\Mdotstar$, and the IMF-averaged Mg and Fe yields from CCSN and SNIa. Following \citetalias{Weinberg2024}, we normalize $\yfecc$ using the \cite{Rodriguez2023} determination of the mean observed Fe yield of CCSN, choose $\ymgcc/\yfecc$ based on the plateau of $\mgfe$ seen in metal-poor Milky Way stars, and choose $\yfeIa/\yfecc$ so that GCE models with a continuous, slowly declining SFH evolve to $\mgfe\approx0$ at late times. We adopt $\mgfecc=0.45$ as our default value of the plateau, but we consider models with a higher $\ymgcc/\yfecc$ corresponding to $\mgfecc=0.55$.

With our adopted SFH parameterization (Equation \ref{eqn:SFH}), star formation begins at $\tstart$, rises approximately linearly to a maximum at $t\approx\tau_1$, then declines exponentially on a timescale $\tau_2$ and truncates sharply at a cutoff time $\tcut$, where the galaxy transitions to quiescence. Because our parameterization already allows a variety of rise times $\tau_1$, we fix $\tstart=0.5\pm0.5$ Gyr, assuming that star formation in massive ellipticals commences at $z \gtrsim6$. While all four SFH parameters affect the mean age and abundances, the trajectories of $\meanmgh$ vs. $\meanage$ and $\meanmgfe$ vs. $\meanage$ are not sensitive to the individual values of $\tau_1$, $\tau_2$, and $\tcut$, and the value of $\tstart$ simply translates these trajectories in time (Figure \ref{fig: tstart}). Thus, once the yields, $\taustar$, and $\eta$ are specified, the choice of SFH has surprisingly little leverage on the predicted $\meanmgh$ and $\meanmgfe$ ratios for galaxies of a specified $\meanage$. 

Figure \ref{fig: summary} summarizes the impact of GCE model parameters on $\meanmgh$-$\meanage$ and $\meanmgfe$-$\meanage$ trajectories. Beyond the linear scaling of Mg abundance with $\ymgcc$, the predicted $\meanmgh$ depends mainly on $\taustar$ in old galaxies and on $\eta$ in younger galaxies: higher SFE (lower $\taustar$) leads to more rapid early enrichment, while higher $\eta$ leads to greater metal loss and lower metallicity at later times. Several parameters affect the predicted $\meanmgfe$ at fixed $\meanage$, but only at the $\sim0.02$-dex level, and it is generally easier to decrease the predicted $\meanmgfe$ than to increase it. Our fiducial models adopt a double-exponential DTD for SNIa that approximates a $t^{-1.1}$ power-law with a minimum delay time $\tdmin=150$ Myr. If we reduce $\tdmin$ to 50 Myr, predicted $\meanmgfe$ ratios drop by $\sim0.05$ dex because of the more rapid onset of SNIa Fe enrichment. 

Figure \ref{fig: best fits} compares a representative set of GCE models to the SDSS and LEGA-C data. With reasonable parameter choices, the models can reproduce the observed $\meanmgh$ values, with the SDSS trend of higher $\meanmgh$ at higher $\sigma$ explained mainly by decreasing outflow efficiency (lower $\eta$) in deeper potential wells. Even with $\eta=0$, reproducing the super-solar $\meanmgh$ observed in higher-$\sigma$ SDSS galaxies and all LEGA-C galaxies requires $\ymgcc$ to be $\sim0.1$ dex higher than our default value, and a high SFE ($\taustar\sim0.5$ Gyr) is needed for the oldest SDSS galaxies. The higher $\ymgcc$ is within the uncertainty estimated by \citetalias{Weinberg2024}, so we do not consider the high observed $\meanmgh$ to be problematic for our models. However, a bottom-heavy IMF in the most massive ellipticals \citep{vanDokkum2010, Conroy2017IMF} would lower the $\ymgcc$ in these systems, making their high $\meanmgh$ more difficult to explain; shifting from a \cite{Kroupa2001} IMF to a Salpeter-like IMF lowers $\ymgcc$ by 0.14 dex.

With our fiducial CCSN yield ratio of $\mgfecc=0.45$, our models underpredict the observed $\meanmgfe$ values at a given $\meanage$ by $\sim 0.08$ dex at $z=0$ and $\sim 0.05$ dex at $z=0.7$.  With the higher ratio $\mgfecc=0.55$, the discrepancy with the data is smaller, though once we choose parameters to match $\meanmgh-\meanage$ the models still consistently underpredict the observed $\meanmgfe$ (Figure~\ref{fig: best fits}bd).  At $z=0.7$ the difference is comparable to the $1\sigma$ statistical uncertainties in the observed values, but at $z=0$ it is a statistically significant 0.05-0.07 dex in most velocity dispersion bins.  Spectral fits imply that the observed stellar populations of ellipticals are old, but they are still young enough to allow ample time for SNIa Fe enrichment.  Within the fairly flexible framework of our models, that SNIa enrichment should have depressed $\meanmgfe$ below the observationally inferred values.  

The tension between models and the data is mild enough that it might be explained by systematic uncertainties in the SSP spectral fits, e.g., from imperfect stellar libraries. Reducing the observational estimates of $\meanmgfe$ by $\sim0.05$ dex or increasing the estimated ages by 1-2 Gyr would align the data with our $\mgfecc=0.55$ models. Reducing $\meanmgfe$ by 0.1 dex would remove the conflict with our $\mgfecc=0.45$ models. 

Alternatively, this tension could be a sign that the IMF-averaged yields in elliptical galaxies are systematically different from the empirical estimates of \citetalias{Weinberg2024}, which are motivated by global statistics of CCSN and SNIa rates and by abundance trends in the Milky Way. Agreement with the observed $\mgfe$ would be improved by increasing $\ymgcc/\yfecc$ (above $\mgfecc=0.55$), or by decreasing $\yfeIa/\yfecc$. For example, Figure \ref{fig: SNIa} shows that a model with $\mgfecc=0.45$ and $\yfeIa/\yfecc$ reduced from 1.65 to 1.0 can reproduce the $\meanmgfe$-$\meanage$ trends from SDSS and LEGA-C. However, observations suggest a higher SNIa DTD normalization in galaxy clusters relative to the field \citep{Maoz2017}, which would imply a higher $\yfeIa/\yfecc$ in ellipticals, not lower. 

Agreement between models and data can also be improved by adding a self-enriching burst of star formation to the end of a galaxy's smooth SFH, immediately before the transition to quiescence. This terminating burst reduces the light-weighted age, and it increases both $\meanmgh$ and $\meanmgfe$ because self-enrichment in a short burst is dominated by CCSN. We have developed a simple, approximate model for the impact of these terminating bursts (Figures \ref{fig: example burst} and \ref{fig: burst}, Appendix \ref{appx:burst}). Bursts that consume just the gas that remains after the smooth star formation can boost $\meanmgfe$ at fixed $\meanage$ by $\sim0.03$ dex, while also boosting $\meanmgh$ by $\sim0.02-0.08$ dex. If the burst is fueled by accretion of a metal-poor, gas-rich companion, then the increase of $\meanmgfe$ can be larger, while $\meanmgh$ may increase, stay flat, or even drop because of the dilution of the pre-burst ISM. The importance of terminating bursts could be probed observationally by looking for correlations between residuals from the $\meanmgfe$-$\meanage$ relation and spectral signatures of late star formation, going beyond the SSP-fitting techniques used here. 

Our results suggest several directions for future work. One is to compare our model star formation histories to SED-based observational estimates, including investigating the frequency and magnitude of terminating bursts. A second is to apply our models to emerging observational results at higher redshift, which should include progenitors of some of the oldest low-$z$ quiescent populations. A third is to extend our analysis to other elements whose abundance can be estimated by full spectral fitting, including elements that have substantial contributions from intermediate-mass ABG stars. Our results also highlight the importance of sharpening empirical constraints on IMF-averaged yields, in the Milky Way and in other local galaxies. 

Elliptical galaxies differ from disk galaxies in their star formation histories and potentially in their outflows and other galactic scale phenomena. Stellar yields are governed by stellar-scale astrophysics, including the IMF, the physics of supernova explosions and black hole populations, and binary populations that may affect massive star evolution and the SNIa DTD.
Building on the framework developed here, we hope to learn whether galactic scale effects fully explain the distinctive abundance patterns of ellipticals, or whether these patterns also imply different stellar scale astrophysics in the progenitors of massive quiescent galaxies. 

\section*{Acknowledgments}
We thank Adam Wheeler, Todd Thompson, Liam Dubay, and Jennifer Johnson for helpful discussions.
This work is supported by NSF grant AST-2307621. NMG acknowledges the hospitality of the Kavli Institute for Theoretical Physics (KITP), supported by NSF PHY-2309135, during the completion of this work. This work is dedicated to the loving memory of Marcelina Kroenke. Niech spoczywa w pokoju.

\appendix
\section{\texttt{fanCE}}
\label{appx:code}
\texttt{fanCE} (\textbf{F}lexible \textbf{An}alytic \textbf{C}hemical \textbf{E}volution) \citep{Gountanis2024} is an analytic one-zone chemical evolution code. It implements the Weinberg, Andrews, \& Freudenburg (2017) analytic solution for a constant, exponential declining, or linear-exponential (delayed tau) star formation history. It also implements a flexible, two-parameter star formation history $\propto(1-e^{-t/\tau_1})\cdot e^{-t/\tau_2}$, which rises linearly on a timescale $\tau_1$ and then falls exponentially on a timescale $\tau_2$ for $\tau_2>\tau_1>0$.

We provide {\tt Python} implementations of \texttt{fanCE} via \url{https://github.com/nmgountanis/fanCE} so that others can easily adapt them to interpret other galaxies or stellar populations or explore alternative assumptions. The source code for \texttt{fanCE} can be downloaded from \href{https://github.com/nmgountanis/fanCE}{GitHub} and installed by running: \\

\noindent\texttt{cd <path\_to\_installation>}

\noindent\texttt{git clone https://github.com/nmgountanis/fanCE}

\noindent\texttt{cd fanCE}

\noindent\texttt{pip install -e .} \\

If you make use of this code, please cite \cite{Weinberg2017equ} for the analytic solutions and 
Gountanis et al. (2024, this paper) for the code implementation.  The reference for the default yield choices is \cite{Weinberg2024}.

\section{Mean metallicity of burst stars}
\label{appx:burst}

Our goal is to compute the mean metallicity of stars formed during a burst described by a ``leaky box'' chemical evolution model (\citealt{Hartwick1976}; \citealt{Binney1998}, section 5.3; WAF, section 3.5), which generalizes ``closed box'' evolution \citep{Talbot1971} to allow outflow.  These models assume instantaneous enrichment and instantaneous recycling of stellar mass loss.  For notational simplicity, we will consider a single element of mass fraction $Z$ and yield $y$.  The metallicity evolution of the leaky box model is
\begin{equation}
    Z(t) = {y \over 1+\eta'-r'}\left[-\ln f_g(t)\right]~,
    \label{eqn:leakyZ}
\end{equation}
 where
 \begin{equation}
     f_g(t) = M_g(t)/M_{g,i}
     \label{eqn:fg}
\end{equation}
is the ratio of the gas mass to the initial gas mass.  We have denoted the outflow mass-loading by $\eta'$ to emphasize that it may differ from the value $\eta$ that applies during pre-burst evolution, and we have denoted the recycling factor by $r'$ because a burst may not last long enough for the full recycling factor $r$ to be applicable.  If the initial gas metallicity is non-zero (as it is in our models), then the enrichment described by \Eqref{leakyZ} is added to the initial abundance $Z_i$ at the start of the burst.

Because there is, by assumption, no ongoing gas accretion during the burst, mass conservation implies
\begin{equation}
    M_g(t) + (1-r')\Mstarform(t) + \eta'\Mstarform(t) = M_{g,i}~,
    \label{eqn:mgas}
\end{equation}
where
\begin{equation}
  \Mstarform(t) = \int_0^t \Mdotstar(t') dt'
  \label{eqn:mstarform}
\end{equation}
is the mass of stars formed and multiplying by $(1-r')$ gives the current stellar mass.  Dividing by $M_{g,i}$ and rearranging yields
\begin{equation}
    f_*(t) \equiv {\Mstarform(t) \over M_{g,i}} = {1-f_g(t) \over 1+\eta'-r'}~.
    \label{eqn:fstar}
\end{equation}
For small $\eta'$ and $f_g$, the value of $f_*$ can exceed one because recycled gas can be formed into new stars.  Stars formed at time $t'$ have metallicity $Z(t')$, so the mean metallicity of all stars formed up to time $t$ is
\begin{equation}
    \langle Z_* \rangle = {\int_0^t Z(t') \Mdotstar(t') dt' \over \int_0^t \Mdotstar(t') dt'} =
                           {\int_0^{f_*} Z(f_*')df_*' \over \int_0^{f_*} df_*'} =
                           {\int_1^{f_g} Z(f_g') df_g' \over \int_1^{f_g} df_g'}~,
    \label{eqn:Zbarintegral}
\end{equation}
where the last equality uses the fact that $df_*'/df_g' = -(1+\eta'-r')^{-1}$ is constant and therefore cancels between the numerator and denominator.  Applying \Eqref{leakyZ} and evaluating the integral yields the final result
\begin{equation}
    \langle Z_* \rangle = {y \over 1+\eta'-r'} \left[1+ {f_g \ln f_g \over 1-f_g}\right]~.
    \label{eqn:Zbar}
\end{equation}
The factor $\left[1+ {f_g \ln f_g \over 1-f_g}\right]$ grows from $\frac{1}{2} (1-f_g)$ at early times, when $(1-f_g)\ll 1$, to unity at late times, when $f_g \ll 1$.  

Our terminating burst models are parameterized by a dilution factor $F_D$ and a final gas fraction $f_g$.  (Recall that $f_g$ is the ratio of the gas mass at the end of the burst to the gas mass at the start of the burst, {\it not} to the stellar mass.)  At the end of smooth evolution the gas mass is $\taustar \Mdotstar$, so the initial gas mass for the burst is $M_{g,i} = (1+F_D)\taustar\Mdotstar$.  We assume that the accreted gas is much lower metallicity than that in the parent elliptical, so that the gas phase abundance of element X is reduced by a factor $(1+F_D)$.  We therefore compute the impact of a burst on our elliptical galaxy models by adding, in the final timestep, a mass of stars
\begin{equation}
    M_{\rm burst} = \taustar\Mdotstar (1+F_D) {1-f_g \over 1+\eta'-r'}
    \label{eqn:mburst}
\end{equation}
(cf. \Eqref{fstar}), with abundances
\begin{equation}
    \bar{Z}_{X,{\rm burst}} = {Z_X \over 1+F_D} + {\yxcc \over 1+\eta'-r'}
                               \left[1+ {f_g \ln f_g \over 1-f_g}\right] ~,
    \label{eqn:Zxburst}
\end{equation}
where X = Mg, Fe, and $Z_X$ is the ISM abundance at the last timestep before the burst.  We assume that the burst is too short to allow significant SNIa iron production.  We adopt $\eta'=0$ to maximize the burst impact, and we adopt $r'=0.3$ in place of $r=0.4$ to allow for a short burst duration.  Our assumption that accreted gas has negligibly low metallicity could be adjusted by changing the $1+F_D$ dilution factor in the first term.  For the values $f_g=0.3$ and $f_g=0.05$ used in our illustrative burst models, the factor $\left[1+ {f_g \ln f_g \over 1-f_g}\right]$ is 0.484 and 0.842, respectively.

\section{Testing the SSP assumption against mock observations}
\label{appx:mock}

\begin{figure*}[!]
\centering
\includegraphics[width = \textwidth]{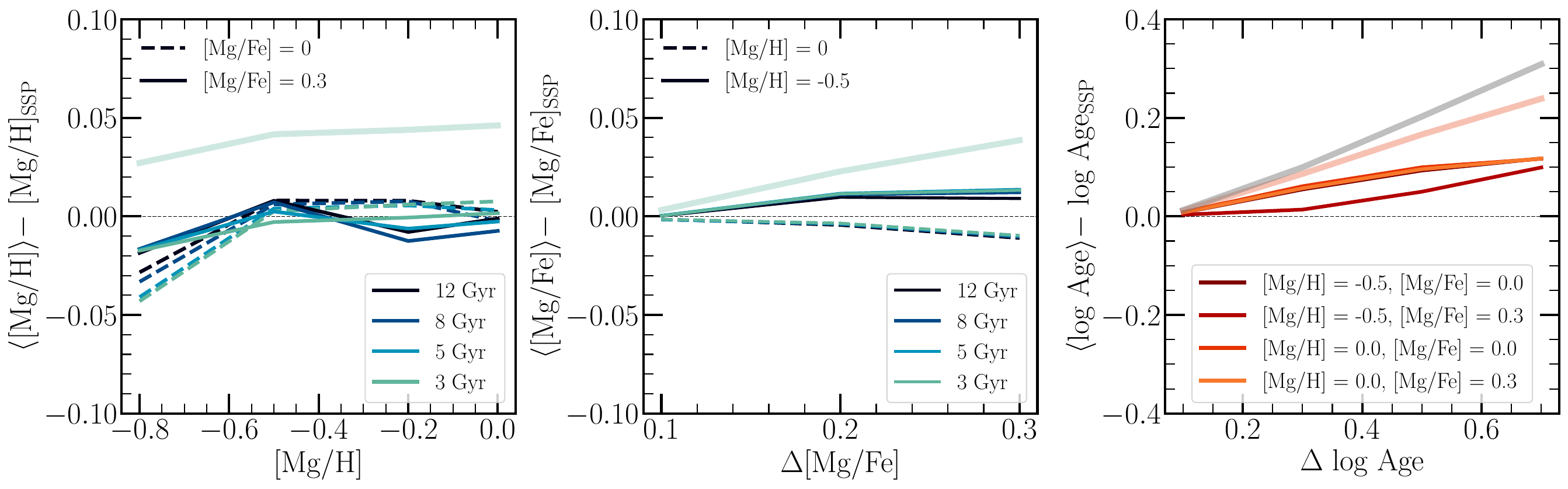}
\vspace{-5mm}
\caption{Results from simple two-population CSP tests. The $y$-axis shows the deviations between the best-fit SSP-equivalent stellar population properties and the adopted logarithmic light-weighted quantities. \textit{Left:} Both populations have the same stellar age and [Mg/Fe], but they differ in [Mg/H] by 0.4\;dex. The $x$-axis represents [Mg/H] of the lower metallicity population. All dashed lines have [Mg/Fe] = 0, while all solid lines have [Mg/Fe] = 0.3. The low transparency solid line repeats the 3 Gyr case but averaging linear (Mg/H) rather than the logarithmic quantity [Mg/H].  \textit{Center:} Both populations have the same stellar age and [Mg/H], but they differ in [Mg/Fe] by the amount listed on the $x$-axis. In all cases, the lower [Mg/Fe] population has [Mg/Fe] = 0. All dashed lines have [Mg/H] = 0, while all solid lines have [Mg/H] = -0.5. The low transparency solid line repeats the 3 Gyr case with linear averaging of (Mg/Fe).   \textit{Right:} Both populations have the same [Mg/H] and [Mg/Fe] but differ in log Age by the amount listed on the x-axis. In all cases, the lowest age is log(age) = 0.4 (2.5\,Gyr). The low transparency line repeats the solar abundance case but with linear rather than logarithmic age averaging.  The gray line shows the same case but for the mass-weighted logarithmic age. }
\label{fig: simple_test}
\end{figure*}

The observational measurements from SDSS and LEGA-C are derived assuming the galaxies form in a single burst of star formation from the gas of a single metallicity and elemental abundance pattern. On the other hand, the chemical evolution models applied to these data have extended star formation and chemical histories. Here, we test whether there is significant systematic bias caused by our comparison of observed SSP-equivalent values and the GCE light-weighted ones.

We generate simple mock CSP spectra by combining two SSPs with different ages, metallicities, and abundance ratios. We then fit the CSPs with the same SSP models used in the LEGA-C and SDSS measurements. Finally, we compare these best-fit SSP-equivalent ages and abundances to the true light-weighted values from the underlying SSPs. 

The results from the mock tests are shown in Figure~\ref{fig: simple_test}. We show the effects of altering only one parameter at a time. In the left panel, we fix the stellar age and [Mg/Fe] of the two populations but shift [Mg/H] by 0.4 dex. The $x$-axis represents [Mg/H] of the lower metallicity population. The solid (dashed) lines show the results when both populations have [Mg/Fe] fixed to 0.0 (0.4). The average $\mgh$ of these two populations matches the $\mgh$ of the SSP fit to within 0.02 dex in most cases, with larger deviations up to 0.05 dex for $\mgh_{\mathrm{lower}}=-0.8$ and $\mgfe=0.0$. Because the two populations have the same age and thus the same $L_{\rm V}/M$, it does not matter whether we weight by mass or by light.  However, it does matter whether we average the logarithmic quantity $\mgh$ or average (Mg/Fe) and then take the logarithm. The former logarithmic average better approximates the SSP result, while the linear average over-predicts the SSP $\mgh$ by $\sim0.04$ dex, as illustrated by the low transparency blue line for the $\mgfe=0.3$, 5 Gyr case. 

In the center panel, we vary the abundance ratio [Mg/Fe], fixing the lowest [Mg/Fe] population to [Mg/Fe]$=0$. Again, we show cases where the population ages and [Mg/H] are fixed to different values. Even for an $\mgfe$ difference of 0.3 dex between the two populations, the mean $\mgfe$ matches the SSP fit to within $\sim0.01$ dex. For $\mgh=-0.5$, linear averaging of (Mg/Fe) leads to a less accurate prediction than directly averaging $\mgfe$, as shown by the low transparency line. For $\mgh=0.0$, the bias of linear averaging is similar in magnitude but opposite in sign to that of logarithmic averaging. 

In the right panel, we vary the ages of the populations but fix [Mg/H] and [Mg/Fe]. In all tests, the lowest age is set to 2.5 Gyr. The light-weighted average of log(age) slightly overpredicts the log(age) of the SSP fit, by up to $\sim0.1$ dex for a $\Delta\mathrm{log(age)}=0.7$ dex between the populations. Linear averaging of ages roughly doubles the error in predicting the SSP age, as illustrated by the low transparency line for the Solar abundance case. The $L_{\rm V}/M$ for the young and old populations differs, and mass-weighted averages give much less accurate predictions of the SSP age as expected (dashed line). 

In other tests, we find that the slight age bias shown in Figure \ref{fig: simple_test} becomes progressively stronger when the youngest SSP has a younger age ($\leq$1 Gyr), because the A-star-dominated SSP has very strong Balmer absorption features that dramatically outshines the older stellar populations. This bias at young ages, however, is not an issue in this study, as the SDSS and LEGA-C populations are selected to be (often much) older than 1 Gyr. 

The results of these tests agree with what has been previously found by \citet{serra_interpretation_2007}; light-weighted abundances and abundance ratios are well approximated by SSP-equivalent measurements, whereas the stellar ages can be biased towards the youngest SSP. These two-population tests support our adopted procedure of predicting light-weighted $\mgh$, $\mgfe$, and log(age) values from our continuous models to compare to SSP observational results. We present an end-to-end test for representative continuous models in Section \ref{sec:mock} (Figure \ref{fig: mock}). 

\bibliography{ellipticalGCE}{}

@article{villaume_extended_2017,
	title = {The {Extended} {IRTF} {Spectral} {Library}: {Expanded} {Coverage} in {Metallicity}, {Temperature}, and {Surface} {Gravity}},
	volume = {230},
	issn = {1538-4365},
	shorttitle = {The {Extended} {IRTF} {Spectral} {Library}},
	url = {https://iopscience.iop.org/article/10.3847/1538-4365/aa72ed},
	doi = {10.3847/1538-4365/aa72ed},
	language = {en},
	number = {2},
	urldate = {2020-12-11},
	journal = {The Astrophysical Journal Supplement Series},
	author = {Villaume, Alexa and Conroy, Charlie and Johnson, Benjamin and Rayner, John and Mann, Andrew W. and van Dokkum, Pieter},
	month = jun,
	year = {2017},
	pages = {23},
}

@article{vazdekis_evolutionary_2010,
	title = {Evolutionary stellar population synthesis with {MILES} - {I}. {The} base models and a new line index system},
	issn = {00358711, 13652966},
	url = {https://academic.oup.com/mnras/article-lookup/doi/10.1111/j.1365-2966.2010.16407.x},
	doi = {10.1111/j.1365-2966.2010.16407.x},
	language = {en},
	urldate = {2024-05-10},
	journal = {Monthly Notices of the Royal Astronomical Society},
	author = {Vazdekis, A. and Sánchez-Blázquez, P. and Falcón-Barroso, J. and Cenarro, A. J. and Beasley, M. A. and Cardiel, N. and Gorgas, J. and Peletier, R. F.},
	month = mar,
	year = {2010},
}

@article{thomas_stellar_2003,
	title = {Stellar population models of {Lick} indices with variable element abundance ratios},
	volume = {339},
	issn = {0035-8711, 1365-2966},
	url = {https://academic.oup.com/mnras/article/339/3/897/972092},
	doi = {10.1046/j.1365-8711.2003.06248.x},
	language = {en},
	number = {3},
	urldate = {2023-07-27},
	journal = {Monthly Notices of the Royal Astronomical Society},
	author = {Thomas, D. and Maraston, C. and Bender, R.},
	month = mar,
	year = {2003},
	pages = {897--911},
}

@article{worthey_comprehensive_1994,
	title = {Comprehensive stellar population models and the disentanglement of age and metallicity effects},
	volume = {95},
	issn = {0067-0049, 1538-4365},
	url = {http://adsabs.harvard.edu/doi/10.1086/192096},
	doi = {10.1086/192096},
	language = {en},
	urldate = {2023-06-28},
	journal = {The Astrophysical Journal Supplement Series},
	author = {Worthey, Guy},
	month = nov,
	year = {1994},
	pages = {107},
}

@article{serra_interpretation_2007,
	title = {On the interpretation of the age and chemical composition of composite stellar populations determined with line-strength indices},
	volume = {374},
	issn = {0035-8711, 1365-2966},
	url = {https://academic.oup.com/mnras/article-lookup/doi/10.1111/j.1365-2966.2006.11188.x},
	doi = {10.1111/j.1365-2966.2006.11188.x},
	language = {en},
	number = {3},
	urldate = {2022-05-06},
	journal = {Monthly Notices of the Royal Astronomical Society},
	author = {Serra, P. and Trager, S. C.},
	month = jan,
	year = {2007},
	pages = {769--774},
}

@ARTICLE{Weinberg2024,
       author = {{Weinberg}, David H. and {Griffith}, Emily J. and {Johnson}, James W. and {Thompson}, Todd A.},
        title = "{The Scale of Stellar Yields: Implications of the Measured Mean Iron Yield of Core Collapse Supernovae}",
      journal = {arXiv e-prints},
         year = 2024,
        month = sep,
          eid = {arXiv:2309.05719},
        pages = {arXiv:2309.05719},
          doi = {10.48550/arXiv.2309.05719},
archivePrefix = {arXiv},
       eprint = {2309.05719},
 primaryClass = {astro-ph.GA},
       adsurl = {https://ui.adsabs.harvard.edu/abs/2023arXiv230905719W}
}

@ARTICLE{Sestito2024,
       author = {{Sestito}, Federico and {Vitali}, Sara and {Jofre}, Paula and {Venn}, Kim A. and {Aguado}, David S. and {Aguilera-G{\'o}mez}, Claudia and {Ardern-Arentsen}, Anke and {de Brito Silva}, Danielle and {Carlberg}, Raymond and {Eldridge}, Camilla J.~L. and {Gran}, Felipe and {Hill}, Vanessa and {Jablonka}, Pascale and {Kordopatis}, Georges and {Martin}, Nicolas F. and {Matsuno}, Tadafumi and {Rusterucci}, Samuel and {Starkenburg}, Else and {Viswanathan}, Akshara},
        title = "{Pristine Inner Galaxy Survey (PIGS) IX. The largest detailed chemical analysis of very metal-poor stars in the Sagittarius dwarf galaxy}",
      journal = {arXiv e-prints},
         year = 2024,
        month = apr,
          eid = {arXiv:2405.00096},
        pages = {arXiv:2405.00096},
          doi = {10.48550/arXiv.2405.00096},
archivePrefix = {arXiv},
       eprint = {2405.00096},
 primaryClass = {astro-ph.GA},
       adsurl = {https://ui.adsabs.harvard.edu/abs/2024arXiv240500096S}
}

@ARTICLE{Tacconi2008,
       author = {{Tacconi}, L.~J. and {Genzel}, R. and {Saintonge}, A. and {Combes}, F. and {Garc{\'\i}a-Burillo}, S. and {Neri}, R. and {Bolatto}, A. and {Contini}, T. and {F{\"o}rster Schreiber}, N.~M. and {Lilly}, S. and {Lutz}, D. and {Wuyts}, S. and {Accurso}, G. and {Boissier}, J. and {Boone}, F. and {Bouch{\'e}}, N. and {Bournaud}, F. and {Burkert}, A. and {Carollo}, M. and {Cooper}, M. and {Cox}, P. and {Feruglio}, C. and {Freundlich}, J. and {Herrera-Camus}, R. and {Juneau}, S. and {Lippa}, M. and {Naab}, T. and {Renzini}, A. and {Salome}, P. and {Sternberg}, A. and {Tadaki}, K. and {{\"U}bler}, H. and {Walter}, F. and {Weiner}, B. and {Weiss}, A.},
        title = "{PHIBSS: Unified Scaling Relations of Gas Depletion Time and Molecular Gas Fractions}",
      journal = {\apj},
         year = 2018,
        month = feb,
       volume = {853},
       number = {2},
          eid = {179},
        pages = {179},
          doi = {10.3847/1538-4357/aaa4b4},
archivePrefix = {arXiv},
       eprint = {1702.01140},
 primaryClass = {astro-ph.GA},
       adsurl = {https://ui.adsabs.harvard.edu/abs/2018ApJ...853..179T}
}

@ARTICLE{Worthey1994,
       author = {{Worthey}, Guy},
        title = "{Comprehensive Stellar Population Models and the Disentanglement of Age and Metallicity Effects}",
      journal = {\apjs},
         year = 1994,
        month = nov,
       volume = {95},
        pages = {107},
          doi = {10.1086/192096},
       adsurl = {https://ui.adsabs.harvard.edu/abs/1994ApJS...95..107W}
}

@ARTICLE{Thomas1999,
       author = {{Thomas}, Daniel},
        title = "{Abundance ratios in hierarchical galaxy formation}",
      journal = {\mnras},
         year = 1999,
        month = jul,
       volume = {306},
       number = {3},
        pages = {655-661},
          doi = {10.1046/j.1365-8711.1999.02552.x},
archivePrefix = {arXiv},
       eprint = {astro-ph/9901226},
 primaryClass = {astro-ph},
       adsurl = {https://ui.adsabs.harvard.edu/abs/1999MNRAS.306..655T}
}

@ARTICLE{Trager2000b,
       author = {{Trager}, S.~C. and {Faber}, S.~M. and {Worthey}, Guy and {Gonz{\'a}lez}, J. Jes{\'u}s},
        title = "{The Stellar Population Histories of Local Early-Type Galaxies. I. Population Parameters}",
      journal = {\aj},
         year = 2000,
        month = apr,
       volume = {119},
       number = {4},
        pages = {1645-1676},
          doi = {10.1086/301299},
archivePrefix = {arXiv},
       eprint = {astro-ph/0001072},
 primaryClass = {astro-ph},
       adsurl = {https://ui.adsabs.harvard.edu/abs/2000AJ....119.1645T}
}

@ARTICLE{Walcher2009,
       author = {{Walcher}, C.~J. and {Coelho}, P. and {Gallazzi}, A. and {Charlot}, S.},
        title = "{Differential stellar population models: how to reliably measure [Fe/H] and [{\ensuremath{\alpha}}/Fe] in galaxies}",
      journal = {\mnras},
         year = 2009,
        month = sep,
       volume = {398},
       number = {1},
        pages = {L44-L48},
          doi = {10.1111/j.1745-3933.2009.00705.x},
archivePrefix = {arXiv},
       eprint = {0906.5000},
 primaryClass = {astro-ph.CO},
       adsurl = {https://ui.adsabs.harvard.edu/abs/2009MNRAS.398L..44W}
}

@ARTICLE{Abazajian2009,
       author = {{Abazajian}, Kevork N. and {Adelman-McCarthy}, Jennifer K. and {Ag{\"u}eros}, Marcel A. and {Allam}, Sahar S. and {Allende Prieto}, Carlos and {An}, Deokkeun and {Anderson}, Kurt S.~J. and {Anderson}, Scott F. and {Annis}, James and {Bahcall}, Neta A. and {Bailer-Jones}, C.~A.~L. and {Barentine}, J.~C. and {Bassett}, Bruce A. and {Becker}, Andrew C. and {Beers}, Timothy C. and {Bell}, Eric F. and {Belokurov}, Vasily and {Berlind}, Andreas A. and {Berman}, Eileen F. and {Bernardi}, Mariangela and {Bickerton}, Steven J. and {Bizyaev}, Dmitry and {Blakeslee}, John P. and {Blanton}, Michael R. and {Bochanski}, John J. and {Boroski}, William N. and {Brewington}, Howard J. and {Brinchmann}, Jarle and {Brinkmann}, J. and {Brunner}, Robert J. and {Budav{\'a}ri}, Tam{\'a}s and {Carey}, Larry N. and {Carliles}, Samuel and {Carr}, Michael A. and {Castander}, Francisco J. and {Cinabro}, David and {Connolly}, A.~J. and {Csabai}, Istv{\'a}n and {Cunha}, Carlos E. and {Czarapata}, Paul C. and {Davenport}, James R.~A. and {de Haas}, Ernst and {Dilday}, Ben and {Doi}, Mamoru and {Eisenstein}, Daniel J. and {Evans}, Michael L. and {Evans}, N.~W. and {Fan}, Xiaohui and {Friedman}, Scott D. and {Frieman}, Joshua A. and {Fukugita}, Masataka and {G{\"a}nsicke}, Boris T. and {Gates}, Evalyn and {Gillespie}, Bruce and {Gilmore}, G. and {Gonzalez}, Belinda and {Gonzalez}, Carlos F. and {Grebel}, Eva K. and {Gunn}, James E. and {Gy{\"o}ry}, Zsuzsanna and {Hall}, Patrick B. and {Harding}, Paul and {Harris}, Frederick H. and {Harvanek}, Michael and {Hawley}, Suzanne L. and {Hayes}, Jeffrey J.~E. and {Heckman}, Timothy M. and {Hendry}, John S. and {Hennessy}, Gregory S. and {Hindsley}, Robert B. and {Hoblitt}, J. and {Hogan}, Craig J. and {Hogg}, David W. and {Holtzman}, Jon A. and {Hyde}, Joseph B. and {Ichikawa}, Shin-ichi and {Ichikawa}, Takashi and {Im}, Myungshin and {Ivezi{\'c}}, {\v{Z}}eljko and {Jester}, Sebastian and {Jiang}, Linhua and {Johnson}, Jennifer A. and {Jorgensen}, Anders M. and {Juri{\'c}}, Mario and {Kent}, Stephen M. and {Kessler}, R. and {Kleinman}, S.~J. and {Knapp}, G.~R. and {Konishi}, Kohki and {Kron}, Richard G. and {Krzesinski}, Jurek and {Kuropatkin}, Nikolay and {Lampeitl}, Hubert and {Lebedeva}, Svetlana and {Lee}, Myung Gyoon and {Lee}, Young Sun and {French Leger}, R. and {L{\'e}pine}, S{\'e}bastien and {Li}, Nolan and {Lima}, Marcos and {Lin}, Huan and {Long}, Daniel C. and {Loomis}, Craig P. and {Loveday}, Jon and {Lupton}, Robert H. and {Magnier}, Eugene and {Malanushenko}, Olena and {Malanushenko}, Viktor and {Mandelbaum}, Rachel and {Margon}, Bruce and {Marriner}, John P. and {Mart{\'\i}nez-Delgado}, David and {Matsubara}, Takahiko and {McGehee}, Peregrine M. and {McKay}, Timothy A. and {Meiksin}, Avery and {Morrison}, Heather L. and {Mullally}, Fergal and {Munn}, Jeffrey A. and {Murphy}, Tara and {Nash}, Thomas and {Nebot}, Ada and {Neilsen}, Eric H., Jr. and {Newberg}, Heidi Jo and {Newman}, Peter R. and {Nichol}, Robert C. and {Nicinski}, Tom and {Nieto-Santisteban}, Maria and {Nitta}, Atsuko and {Okamura}, Sadanori and {Oravetz}, Daniel J. and {Ostriker}, Jeremiah P. and {Owen}, Russell and {Padmanabhan}, Nikhil and {Pan}, Kaike and {Park}, Changbom and {Pauls}, George and {Peoples}, John, Jr. and {Percival}, Will J. and {Pier}, Jeffrey R. and {Pope}, Adrian C. and {Pourbaix}, Dimitri and {Price}, Paul A. and {Purger}, Norbert and {Quinn}, Thomas and {Raddick}, M. Jordan and {Re Fiorentin}, Paola and {Richards}, Gordon T. and {Richmond}, Michael W. and {Riess}, Adam G. and {Rix}, Hans-Walter and {Rockosi}, Constance M. and {Sako}, Masao and {Schlegel}, David J. and {Schneider}, Donald P. and {Scholz}, Ralf-Dieter and {Schreiber}, Matthias R. and {Schwope}, Axel D. and {Seljak}, Uro{\v{s}} and {Sesar}, Branimir and {Sheldon}, Erin and {Shimasaku}, Kazu and {Sibley}, Valena C. and {Simmons}, A.~E. and {Sivarani}, Thirupathi and {Allyn Smith}, J. and {Smith}, Martin C. and {Smol{\v{c}}i{\'c}}, Vernesa and {Snedden}, Stephanie A. and {Stebbins}, Albert and {Steinmetz}, Matthias and {Stoughton}, Chris and {Strauss}, Michael A. and {SubbaRao}, Mark and {Suto}, Yasushi and {Szalay}, Alexander S. and {Szapudi}, Istv{\'a}n and {Szkody}, Paula and {Tanaka}, Masayuki and {Tegmark}, Max and {Teodoro}, Luis F.~A. and {Thakar}, Aniruddha R. and {Tremonti}, Christy A. and {Tucker}, Douglas L. and {Uomoto}, Alan and {Vanden Berk}, Daniel E. and {Vandenberg}, Jan and {Vidrih}, S. and {Vogeley}, Michael S. and {Voges}, Wolfgang and {Vogt}, Nicole P. and {Wadadekar}, Yogesh and {Watters}, Shannon and {Weinberg}, David H. and {West}, Andrew A. and {White}, Simon D.~M. and {Wilhite}, Brian C. and {Wonders}, Alainna C. and {Yanny}, Brian and {Yocum}, D.~R. and {York}, Donald G. and {Zehavi}, Idit and {Zibetti}, Stefano and {Zucker}, Daniel B.},
        title = "{The Seventh Data Release of the Sloan Digital Sky Survey}",
      journal = {\apjs},
         year = 2009,
        month = jun,
       volume = {182},
       number = {2},
        pages = {543-558},
          doi = {10.1088/0067-0049/182/2/543},
archivePrefix = {arXiv},
       eprint = {0812.0649},
 primaryClass = {astro-ph},
       adsurl = {https://ui.adsabs.harvard.edu/abs/2009ApJS..182..543A}
}

@ARTICLE{vanderWel2016,
       author = {{van der Wel}, A. and {Noeske}, K. and {Bezanson}, R. and {Pacifici}, C. and {Gallazzi}, A. and {Franx}, M. and {Mu{\~n}oz-Mateos}, J.~C. and {Bell}, E.~F. and {Brammer}, G. and {Charlot}, S. and {Chauk{\'e}}, P. and {Labb{\'e}}, I. and {Maseda}, M.~V. and {Muzzin}, A. and {Rix}, H. -W. and {Sobral}, D. and {van de Sande}, J. and {van Dokkum}, P.~G. and {Wild}, V. and {Wolf}, C.},
        title = "{The VLT LEGA-C Spectroscopic Survey: The Physics of Galaxies at a Lookback Time of 7 Gyr}",
      journal = {\apjs},
         year = 2016,
        month = apr,
       volume = {223},
       number = {2},
          eid = {29},
        pages = {29},
          doi = {10.3847/0067-0049/223/2/29},
archivePrefix = {arXiv},
       eprint = {1603.05479},
 primaryClass = {astro-ph.GA},
       adsurl = {https://ui.adsabs.harvard.edu/abs/2016ApJS..223...29V}
}

@ARTICLE{vanderWel2021,
       author = {{van der Wel}, Arjen and {Bezanson}, Rachel and {D'Eugenio}, Francesco and {Straatman}, Caroline and {Franx}, Marijn and {van Houdt}, Josha and {Maseda}, Michael V. and {Gallazzi}, Anna and {Wu}, Po-Feng and {Pacifici}, Camilla and {Barisic}, Ivana and {Brammer}, Gabriel B. and {Munoz-Mateos}, Juan Carlos and {Vervalcke}, Sarah and {Zibetti}, Stefano and {Sobral}, David and {de Graaff}, Anna and {Calhau}, Joao and {Kaushal}, Yasha and {Muzzin}, Adam and {Bell}, Eric F. and {van Dokkum}, Pieter G.},
        title = "{The Large Early Galaxy Astrophysics Census (LEGA-C) Data Release 3: 3000 High-quality Spectra of K$_{s}$-selected Galaxies at z > 0.6}",
      journal = {\apjs},
         year = 2021,
        month = oct,
       volume = {256},
       number = {2},
          eid = {44},
        pages = {44},
          doi = {10.3847/1538-4365/ac1356},
archivePrefix = {arXiv},
       eprint = {2108.00744},
 primaryClass = {astro-ph.GA},
       adsurl = {https://ui.adsabs.harvard.edu/abs/2021ApJS..256...44V}
}

@ARTICLE{Straatman2018,
       author = {{Straatman}, Caroline M.~S. and {van der Wel}, Arjen and {Bezanson}, Rachel and {Pacifici}, Camilla and {Gallazzi}, Anna and {Wu}, Po-Feng and {Noeske}, Kai and {Bari{\v{s}}i{\'c}}, Ivana and {Bell}, Eric F. and {Brammer}, Gabriel B. and {Calhau}, Jo{\~a}o and {Chauke}, Priscilla and {Franx}, Marijn and {van Houdt}, Josha and {Labb{\'e}}, Ivo and {Maseda}, Michael V. and {Mu{\~n}oz-Mateos}, Juan C. and {Muzzin}, Adam and {van de Sande}, Jesse and {Sobral}, David and {Spilker}, Justin S.},
        title = "{The Large Early Galaxy Astrophysics Census (LEGA-C) Data Release 2: Dynamical and Stellar Population Properties of z {\ensuremath{\lesssim}} 1 Galaxies in the COSMOS Field}",
      journal = {\apjs},
         year = 2018,
        month = dec,
       volume = {239},
       number = {2},
          eid = {27},
        pages = {27},
          doi = {10.3847/1538-4365/aae37a},
archivePrefix = {arXiv},
       eprint = {1809.08236},
 primaryClass = {astro-ph.GA},
       adsurl = {https://ui.adsabs.harvard.edu/abs/2018ApJS..239...27S}
}

@ARTICLE{Conroy2012,
       author = {{Conroy}, Charlie and {van Dokkum}, Pieter G.},
        title = "{The Stellar Initial Mass Function in Early-type Galaxies From Absorption Line Spectroscopy. II. Results}",
      journal = {\apj},
         year = 2012,
        month = nov,
       volume = {760},
       number = {1},
          eid = {71},
        pages = {71},
          doi = {10.1088/0004-637X/760/1/71},
archivePrefix = {arXiv},
       eprint = {1205.6473},
 primaryClass = {astro-ph.CO},
       adsurl = {https://ui.adsabs.harvard.edu/abs/2012ApJ...760...71C}
}

@INPROCEEDINGS{Chartab2023,
       author = {{Chartab}, Nima and {Newman}, Andrew and {Rudie}, Gwen and {Blanc}, Guillermo},
        title = "{The Stellar Mass-Metallicity Relation of Star-forming Galaxies at z 2.5}",
    booktitle = {American Astronomical Society Meeting Abstracts},
         year = 2023,
       series = {American Astronomical Society Meeting Abstracts},
       volume = {55},
        month = jan,
          eid = {177.53},
        pages = {177.53},
       adsurl = {https://ui.adsabs.harvard.edu/abs/2023AAS...24117753C}
}

@ARTICLE{Mason2023,
       author = {{Mason}, Andrew C. and {Crain}, Robert A. and {Schiavon}, Ricardo P. and {Weinberg}, David H. and {Pfeffer}, Joel and {Schaye}, Joop and {Schaller}, Matthieu and {Theuns}, Tom},
        title = "{Realistic simulated galaxies form [$\alpha$/Fe]-[Fe/H] knees due to a sustained decline in their star formation rates}",
      journal = {arXiv e-prints},
         year = 2023,
        month = oct,
          eid = {arXiv:2311.00041},
        pages = {arXiv:2311.00041},
          doi = {10.48550/arXiv.2311.00041},
archivePrefix = {arXiv},
       eprint = {2311.00041},
 primaryClass = {astro-ph.GA},
       adsurl = {https://ui.adsabs.harvard.edu/abs/2023arXiv231100041M}
}

@ARTICLE{Conroy2018,
       author = {{Conroy}, Charlie and {Villaume}, Alexa and {van Dokkum}, Pieter G. and {Lind}, Karin},
        title = "{Metal-rich, Metal-poor: Updated Stellar Population Models for Old Stellar Systems}",
      journal = {\apj},
         year = 2018,
        month = feb,
       volume = {854},
       number = {2},
          eid = {139},
        pages = {139},
          doi = {10.3847/1538-4357/aaab49},
archivePrefix = {arXiv},
       eprint = {1801.10185},
 primaryClass = {astro-ph.GA},
       adsurl = {https://ui.adsabs.harvard.edu/abs/2018ApJ...854..139C}
}

@ARTICLE{Foreman-Mackey2013,
       author = {{Foreman-Mackey}, Daniel and {Hogg}, David W. and {Lang}, Dustin and {Goodman}, Jonathan},
        title = "{emcee: The MCMC Hammer}",
      journal = {\pasp},
         year = 2013,
        month = mar,
       volume = {125},
       number = {925},
        pages = {306},
          doi = {10.1086/670067},
archivePrefix = {arXiv},
       eprint = {1202.3665},
 primaryClass = {astro-ph.IM},
       adsurl = {https://ui.adsabs.harvard.edu/abs/2013PASP..125..306F}
}

@ARTICLE{choi2016,
       author = {{Choi}, Jieun and {Dotter}, Aaron and {Conroy}, Charlie and {Cantiello}, Matteo and {Paxton}, Bill and {Johnson}, Benjamin D.},
        title = "{Mesa Isochrones and Stellar Tracks (MIST). I. Solar-scaled Models}",
      journal = {\apj},
         year = 2016,
        month = jun,
       volume = {823},
       number = {2},
          eid = {102},
        pages = {102},
          doi = {10.3847/0004-637X/823/2/102},
archivePrefix = {arXiv},
       eprint = {1604.08592},
 primaryClass = {astro-ph.SR},
       adsurl = {https://ui.adsabs.harvard.edu/abs/2016ApJ...823..102C}
}

@ARTICLE{Sanchez-Blazquez2006,
       author = {{S{\'a}nchez-Bl{\'a}zquez}, P. and {Peletier}, R.~F. and {Jim{\'e}nez-Vicente}, J. and {Cardiel}, N. and {Cenarro}, A.~J. and {Falc{\'o}n-Barroso}, J. and {Gorgas}, J. and {Selam}, S. and {Vazdekis}, A.},
        title = "{Medium-resolution Isaac Newton Telescope library of empirical spectra}",
      journal = {\mnras},
         year = 2006,
        month = sep,
       volume = {371},
       number = {2},
        pages = {703-718},
          doi = {10.1111/j.1365-2966.2006.10699.x},
archivePrefix = {arXiv},
       eprint = {astro-ph/0607009},
 primaryClass = {astro-ph},
       adsurl = {https://ui.adsabs.harvard.edu/abs/2006MNRAS.371..703S}
}

@ARTICLE{Villaume2017,
       author = {{Villaume}, Alexa and {Conroy}, Charlie and {Johnson}, Benjamin and {Rayner}, John and {Mann}, Andrew W. and {van Dokkum}, Pieter},
        title = "{The Extended IRTF Spectral Library: Expanded Coverage in Metallicity, Temperature, and Surface Gravity}",
      journal = {\apjs},
         year = 2017,
        month = jun,
       volume = {230},
       number = {2},
          eid = {23},
        pages = {23},
          doi = {10.3847/1538-4365/aa72ed},
archivePrefix = {arXiv},
       eprint = {1705.08906},
 primaryClass = {astro-ph.SR},
       adsurl = {https://ui.adsabs.harvard.edu/abs/2017ApJS..230...23V}
}

@ARTICLE{Beverage2023,
       author = {{Beverage}, Aliza G. and {Kriek}, Mariska and {Conroy}, Charlie and {Sandford}, Nathan R. and {Bezanson}, Rachel and {Franx}, Marijn and {van der Wel}, Arjen and {Weisz}, Daniel R.},
        title = "{From Carbon to Cobalt: Chemical Compositions and Ages of z   0.7 Quiescent Galaxies}",
      journal = {\apj},
         year = 2023,
        month = may,
       volume = {948},
       number = {2},
          eid = {140},
        pages = {140},
          doi = {10.3847/1538-4357/acc176},
archivePrefix = {arXiv},
       eprint = {2303.03412},
 primaryClass = {astro-ph.GA},
       adsurl = {https://ui.adsabs.harvard.edu/abs/2023ApJ...948..140B}
}

@ARTICLE{Beverage2021,
       author = {{Beverage}, Aliza G. and {Kriek}, Mariska and {Conroy}, Charlie and {Bezanson}, Rachel and {Franx}, Marijn and {van der Wel}, Arjen},
        title = "{Elemental Abundances and Ages of z   0.7 Quiescent Galaxies on the Mass-Size Plane: Implication for Chemical Enrichment and Star Formation Quenching}",
      journal = {\apjl},
         year = 2021,
        month = aug,
       volume = {917},
       number = {1},
          eid = {L1},
        pages = {L1},
          doi = {10.3847/2041-8213/ac12cd},
archivePrefix = {arXiv},
       eprint = {2105.12750},
 primaryClass = {astro-ph.GA},
       adsurl = {https://ui.adsabs.harvard.edu/abs/2021ApJ...917L...1B}
}

@ARTICLE{Conroy2014,
       author = {{Conroy}, Charlie and {Graves}, Genevieve J. and {van Dokkum}, Pieter G.},
        title = "{Early-type Galaxy Archeology: Ages, Abundance Ratios, and Effective Temperatures from Full-spectrum Fitting}",
      journal = {\apj},
         year = 2014,
        month = jan,
       volume = {780},
       number = {1},
          eid = {33},
        pages = {33},
          doi = {10.1088/0004-637X/780/1/33},
archivePrefix = {arXiv},
       eprint = {1303.6629},
 primaryClass = {astro-ph.CO},
       adsurl = {https://ui.adsabs.harvard.edu/abs/2014ApJ...780...33C}
}

@ARTICLE{Leroy2008,
       author = {{Leroy}, Adam K. and {Walter}, Fabian and {Brinks}, Elias and {Bigiel}, Frank and {de Blok}, W.~J.~G. and {Madore}, Barry and {Thornley}, M.~D.},
        title = "{The Star Formation Efficiency in Nearby Galaxies: Measuring Where Gas Forms Stars Effectively}",
      journal = {\aj},
         year = 2008,
        month = dec,
       volume = {136},
       number = {6},
        pages = {2782-2845},
          doi = {10.1088/0004-6256/136/6/2782},
archivePrefix = {arXiv},
       eprint = {0810.2556},
 primaryClass = {astro-ph},
       adsurl = {https://ui.adsabs.harvard.edu/abs/2008AJ....136.2782L}
}

@ARTICLE{Kroupa2001,
       author = {{Kroupa}, Pavel},
        title = "{On the variation of the initial mass function}",
      journal = {\mnras},
         year = 2001,
        month = apr,
       volume = {322},
       number = {2},
        pages = {231-246},
          doi = {10.1046/j.1365-8711.2001.04022.x},
archivePrefix = {arXiv},
       eprint = {astro-ph/0009005},
 primaryClass = {astro-ph},
       adsurl = {https://ui.adsabs.harvard.edu/abs/2001MNRAS.322..231K}
}

@ARTICLE{Kobayashi2020,
       author = {{Kobayashi}, Chiaki and {Karakas}, Amanda I. and {Lugaro}, Maria},
        title = "{The Origin of Elements from Carbon to Uranium}",
      journal = {\apj},
         year = 2020,
        month = sep,
       volume = {900},
       number = {2},
          eid = {179},
        pages = {179},
          doi = {10.3847/1538-4357/abae65},
archivePrefix = {arXiv},
       eprint = {2008.04660},
 primaryClass = {astro-ph.GA},
       adsurl = {https://ui.adsabs.harvard.edu/abs/2020ApJ...900..179K}
}

@ARTICLE{Maoz2017,
       author = {{Maoz}, Dan and {Graur}, Or},
        title = "{Star Formation, Supernovae, Iron, and {\ensuremath{\alpha}}: Consistent Cosmic and Galactic Histories}",
      journal = {\apj},
         year = 2017,
        month = oct,
       volume = {848},
       number = {1},
          eid = {25},
        pages = {25},
          doi = {10.3847/1538-4357/aa8b6e},
archivePrefix = {arXiv},
       eprint = {1703.04540},
 primaryClass = {astro-ph.HE},
       adsurl = {https://ui.adsabs.harvard.edu/abs/2017ApJ...848...25M}
}

@ARTICLE{Zahid2012,
       author = {{Zahid}, H.~J. and {Dima}, G.~I. and {Kewley}, L.~J. and {Erb}, D.~K. and {Dav{\'e}}, R.},
        title = "{A Census of Oxygen in Star-forming Galaxies: An Empirical Model Linking Metallicities, Star Formation Rates, and Outflows}",
      journal = {\apj},
         year = 2012,
        month = sep,
       volume = {757},
       number = {1},
          eid = {54},
        pages = {54},
          doi = {10.1088/0004-637X/757/1/54},
archivePrefix = {arXiv},
       eprint = {1207.5509},
 primaryClass = {astro-ph.CO},
       adsurl = {https://ui.adsabs.harvard.edu/abs/2012ApJ...757...54Z}
}

@ARTICLE{Peeples2011,
       author = {{Peeples}, Molly S. and {Shankar}, Francesco},
        title = "{Constraints on star formation driven galaxy winds from the mass-metallicity relation at z= 0}",
      journal = {\mnras},
         year = 2011,
        month = nov,
       volume = {417},
       number = {4},
        pages = {2962-2981},
          doi = {10.1111/j.1365-2966.2011.19456.x},
archivePrefix = {arXiv},
       eprint = {1007.3743},
 primaryClass = {astro-ph.CO},
       adsurl = {https://ui.adsabs.harvard.edu/abs/2011MNRAS.417.2962P}
}

@ARTICLE{Finlator2008,
       author = {{Finlator}, Kristian and {Dav{\'e}}, Romeel},
        title = "{The origin of the galaxy mass-metallicity relation and implications for galactic outflows}",
      journal = {\mnras},
         year = 2008,
        month = apr,
       volume = {385},
       number = {4},
        pages = {2181-2204},
          doi = {10.1111/j.1365-2966.2008.12991.x},
archivePrefix = {arXiv},
       eprint = {0704.3100},
 primaryClass = {astro-ph},
       adsurl = {https://ui.adsabs.harvard.edu/abs/2008MNRAS.385.2181F}
}

@ARTICLE{Sun2023,
       author = {{Sun}, Jiayi and {Leroy}, Adam K. and {Ostriker}, Eve C. and {Meidt}, Sharon and {Rosolowsky}, Erik and {Schinnerer}, Eva and {Wilson}, Christine D. and {Utomo}, Dyas and {Belfiore}, Francesco and {Blanc}, Guillermo A. and {Emsellem}, Eric and {Faesi}, Christopher and {Groves}, Brent and {Hughes}, Annie and {Koch}, Eric W. and {Kreckel}, Kathryn and {Liu}, Daizhong and {Pan}, Hsi-An and {Pety}, J{\'e}r{\^o}me and {Querejeta}, Miguel and {Razza}, Alessandro and {Saito}, Toshiki and {Sardone}, Amy and {Usero}, Antonio and {Williams}, Thomas G. and {Bigiel}, Frank and {Bolatto}, Alberto D. and {Chevance}, M{\'e}lanie and {Dale}, Daniel A. and {Gensior}, Jindra and {Glover}, Simon C.~O. and {Grasha}, Kathryn and {Henshaw}, Jonathan D. and {Jim{\'e}nez-Donaire}, Mar{\'\i}a J. and {Klessen}, Ralf S. and {Kruijssen}, J.~M. Diederik and {Murphy}, Eric J. and {Neumann}, Lukas and {Teng}, Yu-Hsuan and {Thilker}, David A.},
        title = "{Star Formation Laws and Efficiencies across 80 Nearby Galaxies}",
      journal = {\apjl},
         year = 2023,
        month = mar,
       volume = {945},
       number = {2},
          eid = {L19},
        pages = {L19},
          doi = {10.3847/2041-8213/acbd9c},
archivePrefix = {arXiv},
       eprint = {2302.12267},
 primaryClass = {astro-ph.GA},
       adsurl = {https://ui.adsabs.harvard.edu/abs/2023ApJ...945L..19S}
}

@ARTICLE{Chieffi2013,
       author = {{Chieffi}, Alessandro and {Limongi}, Marco},
        title = "{Pre-supernova Evolution of Rotating Solar Metallicity Stars in the Mass Range 13-120 M $_{{\ensuremath{\odot}}}$ and their Explosive Yields}",
      journal = {\apj},
         year = 2013,
        month = feb,
       volume = {764},
       number = {1},
          eid = {21},
        pages = {21},
          doi = {10.1088/0004-637X/764/1/21},
       adsurl = {https://ui.adsabs.harvard.edu/abs/2013ApJ...764...21C}
}

@ARTICLE{Rodriguez2023,
       author = {{Rodr{\'\i}guez}, {\'O}smar and {Maoz}, Dan and {Nakar}, Ehud},
        title = "{The Iron Yield of Core-collapse Supernovae}",
      journal = {arXiv e-prints},
         year = 2023,
        month = sep,
          eid = {arXiv:2209.05552},
        pages = {arXiv:2209.05552},
archivePrefix = {arXiv},
       eprint = {2209.05552},
 primaryClass = {astro-ph.HE},
       adsurl = {https://ui.adsabs.harvard.edu/abs/2022arXiv220905552R}
}

@ARTICLE{Andrews2017,
       author = {{Andrews}, Brett H. and {Weinberg}, David H. and {Sch{\"o}nrich}, Ralph and {Johnson}, Jennifer A.},
        title = "{Inflow, Outflow, Yields, and Stellar Population Mixing in Chemical Evolution Models}",
      journal = {\apj},
         year = 2017,
        month = feb,
       volume = {835},
       number = {2},
          eid = {224},
        pages = {224},
          doi = {10.3847/1538-4357/835/2/224},
archivePrefix = {arXiv},
       eprint = {1604.08613},
 primaryClass = {astro-ph.GA},
       adsurl = {https://ui.adsabs.harvard.edu/abs/2017ApJ...835..224A}
}

@ARTICLE{Dubay2024,
       author = {{Dubay}, Liam O. and {Johnson}, Jennifer A. and {Johnson}, James W.},
        title = "{Galactic Chemical Evolution Models Favor an Extended Type Ia Supernova Delay-Time Distribution}",
      journal = {arXiv e-prints},
         year = 2024,
        month = apr,
          eid = {arXiv:2404.08059},
        pages = {arXiv:2404.08059},
          doi = {10.48550/arXiv.2404.08059},
archivePrefix = {arXiv},
       eprint = {2404.08059},
 primaryClass = {astro-ph.GA},
       adsurl = {https://ui.adsabs.harvard.edu/abs/2024arXiv240408059D}
}

@ARTICLE{Greggio2005,
       author = {{Greggio}, L.},
        title = "{The rates of type Ia supernovae. I. Analytical formulations}",
      journal = {\aap},
         year = 2005,
        month = oct,
       volume = {441},
       number = {3},
        pages = {1055-1078},
          doi = {10.1051/0004-6361:20052926},
archivePrefix = {arXiv},
       eprint = {astro-ph/0504376},
 primaryClass = {astro-ph},
       adsurl = {https://ui.adsabs.harvard.edu/abs/2005A&A...441.1055G}
}

@ARTICLE{Schoenrich2009,
   author = {{Sch{\"o}nrich}, R. and {Binney}, J.},
    title = "{Chemical evolution with radial mixing}",
  journal = {\mnras},
archivePrefix = "arXiv",
   eprint = {0809.3006},
     year = 2009,
    month = jun,
   volume = 396,
    pages = {203-222},
      doi = {10.1111/j.1365-2966.2009.14750.x},
   adsurl = {http://adsabs.harvard.edu/abs/2009MNRAS.396..203S}
}

@ARTICLE{Weinberg2017equ,
   author = {{Weinberg}, D.~H. and {Andrews}, B.~H. and {Freudenburg}, J.
	},
    title = "{Equilibrium and Sudden Events in Chemical Evolution}",
  journal = {\apj},
archivePrefix = "arXiv",
   eprint = {1604.07435},
     year = 2017,
    month = mar,
   volume = 837,
      eid = {183},
    pages = {183},
      doi = {10.3847/1538-4357/837/2/183},
   adsurl = {http://adsabs.harvard.edu/abs/2017ApJ...837..183W}
}

@ARTICLE{Sukhbold2016,
   author = {{Sukhbold}, T. and {Ertl}, T. and {Woosley}, S.~E. and {Brown}, J.~M. and 
	{Janka}, H.-T.},
    title = "{Core-collapse Supernovae from 9 to 120 Solar Masses Based on Neutrino-powered Explosions}",
  journal = {\apj},
archivePrefix = "arXiv",
   eprint = {1510.04643},
 primaryClass = "astro-ph.HE",
     year = 2016,
    month = apr,
   volume = 821,
      eid = {38},
    pages = {38},
      doi = {10.3847/0004-637X/821/1/38},
   adsurl = {http://adsabs.harvard.edu/abs/2016ApJ...821...38S}
}

@ARTICLE{Nomoto2013,
   author = {{Nomoto}, K. and {Kobayashi}, C. and {Tominaga}, N.},
    title = "{Nucleosynthesis in Stars and the Chemical Enrichment of Galaxies}",
  journal = {\araa},
     year = 2013,
    month = aug,
   volume = 51,
    pages = {457-509},
      doi = {10.1146/annurev-astro-082812-140956},
   adsurl = {http://adsabs.harvard.edu/abs/2013ARA%26A..51..457N}
}

@ARTICLE{Griffith2021bulge,
       author = {{Griffith}, Emily and {Weinberg}, David H. and {Johnson}, Jennifer A. and {Beaton}, Rachael and {Garc{\'\i}a-Hern{\'a}ndez}, D.~A. and {Hasselquist}, Sten and {Holtzman}, Jon and {Johnson}, James W. and {J{\"o}nsson}, Henrik and {Lane}, Richard R. and {Nataf}, David M. and {Roman-Lopes}, Alexandre},
        title = "{The Similarity of Abundance Ratio Trends and Nucleosynthetic Patterns in the Milky Way Disk and Bulge}",
      journal = {\apj},
         year = 2021,
        month = mar,
       volume = {909},
       number = {1},
          eid = {77},
        pages = {77},
          doi = {10.3847/1538-4357/abd6be},
archivePrefix = {arXiv},
       eprint = {2009.05063},
 primaryClass = {astro-ph.SR},
       adsurl = {https://ui.adsabs.harvard.edu/abs/2021ApJ...909...77G}
}

@ARTICLE{Conroy2012IMF,
       author = {{Conroy}, Charlie and {van Dokkum}, Pieter G.},
        title = "{The Stellar Initial Mass Function in Early-type Galaxies From Absorption Line Spectroscopy. II. Results}",
      journal = {\apj},
         year = 2012,
        month = nov,
       volume = {760},
       number = {1},
          eid = {71},
        pages = {71},
          doi = {10.1088/0004-637X/760/1/71},
archivePrefix = {arXiv},
       eprint = {1205.6473},
 primaryClass = {astro-ph.CO},
       adsurl = {https://ui.adsabs.harvard.edu/abs/2012ApJ...760...71C}
}

@ARTICLE{BJohnson2021,
       author = {{Johnson}, Benjamin D. and {Leja}, Joel and {Conroy}, Charlie and {Speagle}, Joshua S.},
        title = "{Stellar Population Inference with Prospector}",
      journal = {\apjs},
         year = 2021,
        month = jun,
       volume = {254},
       number = {2},
          eid = {22},
        pages = {22},
          doi = {10.3847/1538-4365/abef67},
archivePrefix = {arXiv},
       eprint = {2012.01426},
 primaryClass = {astro-ph.GA},
       adsurl = {https://ui.adsabs.harvard.edu/abs/2021ApJS..254...22J}
}

@ARTICLE{Carnall2018,
       author = {{Carnall}, A.~C. and {McLure}, R.~J. and {Dunlop}, J.~S. and {Dav{\'e}}, R.},
        title = "{Inferring the star formation histories of massive quiescent galaxies with BAGPIPES: evidence for multiple quenching mechanisms}",
      journal = {\mnras},
         year = 2018,
        month = nov,
       volume = {480},
       number = {4},
        pages = {4379-4401},
          doi = {10.1093/mnras/sty2169},
archivePrefix = {arXiv},
       eprint = {1712.04452},
 primaryClass = {astro-ph.GA},
       adsurl = {https://ui.adsabs.harvard.edu/abs/2018MNRAS.480.4379C}
}

@ARTICLE{Beverage2024,
       author = {{Beverage}, Aliza G. and {Kriek}, Mariska and {Suess}, Katherine A. and {Conroy}, Charlie and {Price}, Sedona H. and {Barro}, Guillermo and {Bezanson}, Rachel and {Franx}, Marijn and {Lorenz}, Brian and {Ma}, Yilun and {Mowla}, Lamiya and {Pasha}, Imad and {van Dokkum}, Pieter and {Weisz}, Daniel},
        title = "{The Heavy Metal Survey: The Evolution of Stellar Metallicities, Abundance Ratios, and Ages of Massive Quiescent Galaxies Since z\raisebox{-0.5ex}\textasciitilde2}",
      journal = {arXiv e-prints},
         year = 2024,
        month = dec,
          eid = {arXiv:2312.05307},
        pages = {arXiv:2312.05307},
          doi = {10.48550/arXiv.2312.05307},
archivePrefix = {arXiv},
       eprint = {2312.05307},
 primaryClass = {astro-ph.GA},
       adsurl = {https://ui.adsabs.harvard.edu/abs/2023arXiv231205307B}
}

@ARTICLE{Beverage2024Suspense,
       author = {{Beverage}, Aliza G. and {Slob}, Martje and {Kriek}, Mariska and {Conroy}, Charlie and {Barro}, Guillermo and {Bezanson}, Rachel and {Brammer}, Gabriel and {Cheng}, Chloe M. and {de Graaff}, Anna and {F{\"o}rster Schreiber}, Natascha M. and {Franx}, Marijn and {Lorenz}, Brian and {Mancera Pi{\~n}a}, Pavel E. and {Marchesini}, Danilo and {Muzzin}, Adam and {Newman}, Andrew B. and {Price}, Sedona H. and {Shapley}, Alice E. and {Stefanon}, Mauro and {Suess}, Katherine A. and {van Dokkum}, Pieter and {Weinberg}, David and {Weisz}, Daniel R.},
        title = "{Carbon and Iron Deficiencies in Quiescent Galaxies at z=1-3 from JWST-SUSPENSE: Implications for the Formation Histories of Massive Galaxies}",
      journal = {arXiv e-prints},
         year = 2024,
        month = jul,
          eid = {arXiv:2407.02556},
        pages = {arXiv:2407.02556},
          doi = {10.48550/arXiv.2407.02556},
archivePrefix = {arXiv},
       eprint = {2407.02556},
 primaryClass = {astro-ph.GA},
       adsurl = {https://ui.adsabs.harvard.edu/abs/2024arXiv240702556B}
}

@ARTICLE{Kriek2024,
       author = {{Kriek}, Mariska and {Beverage}, Aliza G. and {Price}, Sedona H. and {Suess}, Katherine A. and {Barro}, Guillermo and {Bezanson}, Rachel S. and {Conroy}, Charlie and {Cutler}, Sam E. and {Franx}, Marijn and {Lin}, Jamie and {Lorenz}, Brian and {Ma}, Yilun and {Momcheva}, Ivelina G. and {Mowla}, Lamiya A. and {Pasha}, Imad and {van Dokkum}, Pieter and {Whitaker}, Katherine E.},
        title = "{The Heavy Metal Survey: Star Formation Constraints and Dynamical Masses of 21 Massive Quiescent Galaxies at z = 1.3{\textendash}2.3}",
      journal = {\apj},
         year = 2024,
        month = may,
       volume = {966},
       number = {1},
          eid = {36},
        pages = {36},
          doi = {10.3847/1538-4357/ad2df9},
archivePrefix = {arXiv},
       eprint = {2311.16232},
 primaryClass = {astro-ph.GA},
       adsurl = {https://ui.adsabs.harvard.edu/abs/2024ApJ...966...36K}
}

@ARTICLE{Slob2024,
       author = {{Slob}, Martje and {Kriek}, Mariska and {Beverage}, Aliza G. and {Suess}, Katherine A. and {Barro}, Guillermo and {Bezanson}, Rachel and {Cheng}, Chloe M. and {Conroy}, Charlie and {de Graaff}, Anna and {F{\"o}rster Schreiber}, Natascha M. and {Franx}, Marijn and {Lorenz}, Brian and {Mancera Pi{\~n}a}, Pavel E. and {Marchesini}, Danilo and {Muzzin}, Adam and {Newman}, Andrew B. and {Price}, Sedona H. and {Shapley}, Alice E. and {Stefanon}, Mauro and {van Dokkum}, Pieter and {Weisz}, Daniel R.},
        title = "{The JWST-SUSPENSE Ultradeep Spectroscopic Program: Survey Overview and Star-Formation Histories of Quiescent Galaxies at 1 < z < 3}",
      journal = {arXiv e-prints},
         year = 2024,
        month = apr,
          eid = {arXiv:2404.12432},
        pages = {arXiv:2404.12432},
          doi = {10.48550/arXiv.2404.12432},
archivePrefix = {arXiv},
       eprint = {2404.12432},
 primaryClass = {astro-ph.GA},
       adsurl = {https://ui.adsabs.harvard.edu/abs/2024arXiv240412432S}
}

@ARTICLE{Carnall2024,
       author = {{Carnall}, A.~C. and {Cullen}, F. and {McLure}, R.~J. and {McLeod}, D.~J. and {Begley}, R. and {Donnan}, C.~T. and {Dunlop}, J.~S. and {Shapley}, A.~E. and {Rowlands}, K. and {Almaini}, O. and {Arellano-C{\'o}rdova}, K.~Z. and {Barrufet}, L. and {Cimatti}, A. and {Ellis}, R.~S. and {Grogin}, N.~A. and {Hamadouche}, M.~L. and {Illingworth}, G.~D. and {Koekemoer}, A.~M. and {Leung}, H. -H. and {Lovell}, C.~C. and {P{\'e}rez-Gonz{\'a}lez}, P.~G. and {Santini}, P. and {Stanton}, T.~M. and {Wild}, V.},
        title = "{The JWST EXCELS survey: Too much, too young, too fast? Ultra-massive quiescent galaxies at 3 < z < 5}",
      journal = {arXiv e-prints},
         year = 2024,
        month = may,
          eid = {arXiv:2405.02242},
        pages = {arXiv:2405.02242},
          doi = {10.48550/arXiv.2405.02242},
archivePrefix = {arXiv},
       eprint = {2405.02242},
 primaryClass = {astro-ph.GA},
       adsurl = {https://ui.adsabs.harvard.edu/abs/2024arXiv240502242C}
}

@ARTICLE{Stanton2024,
       author = {{Stanton}, T.~M. and {Cullen}, F. and {McLure}, R.~J. and {Shapley}, A.~E. and {Arellano-C{\'o}rdova}, K.~Z. and {Begley}, R. and {Amor{\'\i}n}, R. and {Barrufet}, L. and {Calabr{\`o}}, A. and {Carnall}, A.~C. and {Cirasuolo}, M. and {Dunlop}, J.~S. and {Donnan}, C.~T. and {Hamadouche}, M.~L. and {Liu}, F. -Y. and {McLeod}, D.~J. and {Pentericci}, L. and {Pozzetti}, L. and {Sanders}, R.~L. and {Scholte}, D. and {Topping}, M.~W.},
        title = "{The NIRVANDELS Survey: the stellar and gas-phase mass-metallicity relations of star-forming galaxies at z = 3.5}",
      journal = {arXiv e-prints},
         year = 2024,
        month = may,
          eid = {arXiv:2405.00774},
        pages = {arXiv:2405.00774},
          doi = {10.48550/arXiv.2405.00774},
archivePrefix = {arXiv},
       eprint = {2405.00774},
 primaryClass = {astro-ph.GA},
       adsurl = {https://ui.adsabs.harvard.edu/abs/2024arXiv240500774S}
}

@ARTICLE{Steidel2016,
       author = {{Steidel}, Charles C. and {Strom}, Allison L. and {Pettini}, Max and {Rudie}, Gwen C. and {Reddy}, Naveen A. and {Trainor}, Ryan F.},
        title = "{Reconciling the Stellar and Nebular Spectra of High-redshift Galaxies}",
      journal = {\apj},
         year = 2016,
        month = aug,
       volume = {826},
       number = {2},
          eid = {159},
        pages = {159},
          doi = {10.3847/0004-637X/826/2/159},
archivePrefix = {arXiv},
       eprint = {1605.07186},
 primaryClass = {astro-ph.GA},
       adsurl = {https://ui.adsabs.harvard.edu/abs/2016ApJ...826..159S}
}

@ARTICLE{Topping2020,
       author = {{Topping}, Michael W. and {Shapley}, Alice E. and {Reddy}, Naveen A. and {Sanders}, Ryan L. and {Coil}, Alison L. and {Kriek}, Mariska and {Mobasher}, Bahram and {Siana}, Brian},
        title = "{The MOSDEF-LRIS Survey: the interplay between massive stars and ionized gas in high-redshift star-forming galaxies}",
      journal = {\mnras},
         year = 2020,
        month = jul,
       volume = {495},
       number = {4},
        pages = {4430-4444},
          doi = {10.1093/mnras/staa1410},
archivePrefix = {arXiv},
       eprint = {1912.10243},
 primaryClass = {astro-ph.GA},
       adsurl = {https://ui.adsabs.harvard.edu/abs/2020MNRAS.495.4430T}
}

@ARTICLE{Kashino2022,
       author = {{Kashino}, Daichi and {Lilly}, Simon J. and {Renzini}, Alvio and {Daddi}, Emanuele and {Zamorani}, Giovanni and {Silverman}, John D. and {Ilbert}, Olivier and {Peng}, Ying-jie and {Mainieri}, Vincenzo and {Bardelli}, Sandro and {Zucca}, Elena and {Kartaltepe}, Jeyhan S. and {Sanders}, David B.},
        title = "{The Stellar Mass versus Stellar Metallicity Relation of Star-forming Galaxies at 1.6 {\ensuremath{\leq}} z {\ensuremath{\leq}} 3.0 and Implications for the Evolution of the {\ensuremath{\alpha}}-enhancement}",
      journal = {\apj},
         year = 2022,
        month = jan,
       volume = {925},
       number = {1},
          eid = {82},
        pages = {82},
          doi = {10.3847/1538-4357/ac399e},
archivePrefix = {arXiv},
       eprint = {2109.06044},
 primaryClass = {astro-ph.GA},
       adsurl = {https://ui.adsabs.harvard.edu/abs/2022ApJ...925...82K}
}

@ARTICLE{vanDokkum2010,
       author = {{van Dokkum}, Pieter G. and {Conroy}, Charlie},
        title = "{A substantial population of low-mass stars in luminous elliptical galaxies}",
      journal = {\nat},
         year = 2010,
        month = dec,
       volume = {468},
       number = {7326},
        pages = {940-942},
          doi = {10.1038/nature09578},
archivePrefix = {arXiv},
       eprint = {1009.5992},
 primaryClass = {astro-ph.CO},
       adsurl = {https://ui.adsabs.harvard.edu/abs/2010Natur.468..940V}
}

@ARTICLE{vanDokkum2012,
       author = {{van Dokkum}, Pieter G. and {Conroy}, Charlie},
        title = "{The Stellar Initial Mass Function in Early-type Galaxies from Absorption Line Spectroscopy. I. Data and Empirical Trends}",
      journal = {\apj},
         year = 2012,
        month = nov,
       volume = {760},
       number = {1},
          eid = {70},
        pages = {70},
          doi = {10.1088/0004-637X/760/1/70},
archivePrefix = {arXiv},
       eprint = {1205.6471},
 primaryClass = {astro-ph.CO},
       adsurl = {https://ui.adsabs.harvard.edu/abs/2012ApJ...760...70V}
}

@ARTICLE{Conroy2017IMF,
       author = {{Conroy}, Charlie and {van Dokkum}, Pieter G. and {Villaume}, Alexa},
        title = "{The Stellar Initial Mass Function in Early-type Galaxies from Absorption Line Spectroscopy. IV. A Super-Salpeter IMF in the Center of NGC 1407 from Non-parametric Models}",
      journal = {\apj},
         year = 2017,
        month = mar,
       volume = {837},
       number = {2},
          eid = {166},
        pages = {166},
          doi = {10.3847/1538-4357/aa6190},
archivePrefix = {arXiv},
       eprint = {1612.00013},
 primaryClass = {astro-ph.GA},
       adsurl = {https://ui.adsabs.harvard.edu/abs/2017ApJ...837..166C}
}

@ARTICLE{Griffith2021snyield,
       author = {{Griffith}, Emily J. and {Sukhbold}, Tuguldur and {Weinberg}, David H. and {Johnson}, Jennifer A. and {Johnson}, James W. and {Vincenzo}, Fiorenzo},
        title = "{The Impact of Black Hole Formation on Population-averaged Supernova Yields}",
      journal = {\apj},
         year = 2021,
        month = nov,
       volume = {921},
       number = {1},
          eid = {73},
        pages = {73},
          doi = {10.3847/1538-4357/ac1bac},
archivePrefix = {arXiv},
       eprint = {2103.09837},
 primaryClass = {astro-ph.SR},
       adsurl = {https://ui.adsabs.harvard.edu/abs/2021ApJ...921...73G}
}

@INPROCEEDINGS{Wuyts2007,
       author = {{Wuyts}, S.},
        title = "{What do We Learn from IRAC Observations of Galaxies at 2<z<3.5?}",
    booktitle = {Cosmic Frontiers},
         year = 2007,
       editor = {{Metcalfe}, N. and {Shanks}, T.},
       series = {Astronomical Society of the Pacific Conference Series},
       volume = {379},
        month = dec,
        pages = {356},
       adsurl = {https://ui.adsabs.harvard.edu/abs/2007ASPC..379..356W}
}

@ARTICLE{Muzzin2013a,
       author = {{Muzzin}, Adam and {Wilson}, Gillian and {Demarco}, Ricardo and {Lidman}, Chris and {Nantais}, Julie and {Hoekstra}, Henk and {Yee}, H.~K.~C. and {Rettura}, Alessandro},
        title = "{Discovery of a Rich Cluster at z = 1.63 Using the Rest-frame 1.6 {\ensuremath{\mu}}m ``Stellar Bump Sequence'' Method}",
      journal = {\apj},
         year = 2013,
        month = apr,
       volume = {767},
       number = {1},
          eid = {39},
        pages = {39},
          doi = {10.1088/0004-637X/767/1/39},
archivePrefix = {arXiv},
       eprint = {1301.5905},
 primaryClass = {astro-ph.CO},
       adsurl = {https://ui.adsabs.harvard.edu/abs/2013ApJ...767...39M}
}

@ARTICLE{Johnson2020,
       author = {{Johnson}, James W. and {Weinberg}, David H.},
        title = "{The impact of starbursts on element abundance ratios}",
      journal = {\mnras},
         year = 2020,
        month = oct,
       volume = {498},
       number = {1},
        pages = {1364-1381},
          doi = {10.1093/mnras/staa2431},
archivePrefix = {arXiv},
       eprint = {1911.02598},
 primaryClass = {astro-ph.GA},
       adsurl = {https://ui.adsabs.harvard.edu/abs/2020MNRAS.498.1364J}
}

@ARTICLE{Conroy2022,
       author = {{Conroy}, Charlie and {Weinberg}, David H. and {Naidu}, Rohan P. and {Buck}, Tobias and {Johnson}, James W. and {Cargile}, Phillip and {Bonaca}, Ana and {Caldwell}, Nelson and {Chandra}, Vedant and {Han}, Jiwon Jesse and {Johnson}, Benjamin D. and {Speagle}, Joshua S. and {Ting}, Yuan-Sen and {Woody}, Turner and {Zaritsky}, Dennis},
        title = "{Birth of the Galactic Disk Revealed by the H3 Survey}",
      journal = {arXiv e-prints},
         year = 2022,
        month = apr,
          eid = {arXiv:2204.02989},
        pages = {arXiv:2204.02989},
archivePrefix = {arXiv},
       eprint = {2204.02989},
 primaryClass = {astro-ph.GA},
       adsurl = {https://ui.adsabs.harvard.edu/abs/2022arXiv220402989C}
}

@ARTICLE{Johnson2021,
       author = {{Johnson}, James W. and {Weinberg}, David H. and {Vincenzo}, Fiorenzo and {Bird}, Jonathan C. and {Loebman}, Sarah R. and {Brooks}, Alyson M. and {Quinn}, Thomas R. and {Christensen}, Charlotte R. and {Griffith}, Emily J.},
        title = "{Stellar migration and chemical enrichment in the milky way disc: a hybrid model}",
      journal = {\mnras},
         year = 2021,
        month = dec,
       volume = {508},
       number = {3},
        pages = {4484-4511},
          doi = {10.1093/mnras/stab2718},
archivePrefix = {arXiv},
       eprint = {2103.09838},
 primaryClass = {astro-ph.GA},
       adsurl = {https://ui.adsabs.harvard.edu/abs/2021MNRAS.508.4484J}
}

@ARTICLE{Dave2012,
       author = {{Dav{\'e}}, Romeel and {Finlator}, Kristian and {Oppenheimer}, Benjamin D.},
        title = "{An analytic model for the evolution of the stellar, gas and metal content of galaxies}",
      journal = {\mnras},
         year = 2012,
        month = mar,
       volume = {421},
       number = {1},
        pages = {98-107},
          doi = {10.1111/j.1365-2966.2011.20148.x},
archivePrefix = {arXiv},
       eprint = {1108.0426},
 primaryClass = {astro-ph.CO},
       adsurl = {https://ui.adsabs.harvard.edu/abs/2012MNRAS.421...98D}
}

@ARTICLE{Park2023,
       author = {{Park}, Minjung and {Belli}, Sirio and {Conroy}, Charlie and {Tacchella}, Sandro and {Leja}, Joel and {Cutler}, Sam E. and {Johnson}, Benjamin D. and {Nelson}, Erica J. and {Emami}, Razieh},
        title = "{Rapid Quenching of Galaxies at Cosmic Noon}",
      journal = {\apj},
         year = 2023,
        month = aug,
       volume = {953},
       number = {1},
          eid = {119},
        pages = {119},
          doi = {10.3847/1538-4357/acd54a},
archivePrefix = {arXiv},
       eprint = {2210.03747},
 primaryClass = {astro-ph.GA},
       adsurl = {https://ui.adsabs.harvard.edu/abs/2023ApJ...953..119P}
}

@ARTICLE{Leung2023,
       author = {{Leung}, Ho-Hin and {Wild}, Vivienne and {Papathomas}, Michail and {Carnall}, Adam and {Zheng}, Yirui and {Boardman}, Nicholas and {Wang}, Cara and {Johansson}, Peter H.},
        title = "{Chemical evolution of local post-starburst galaxies: Implications for the mass-metallicity relation}",
      journal = {arXiv e-prints},
         year = 2023,
        month = sep,
          eid = {arXiv:2309.16626},
        pages = {arXiv:2309.16626},
          doi = {10.48550/arXiv.2309.16626},
archivePrefix = {arXiv},
       eprint = {2309.16626},
 primaryClass = {astro-ph.GA},
       adsurl = {https://ui.adsabs.harvard.edu/abs/2023arXiv230916626L}
}

@ARTICLE{Leethochawalit2018,
       author = {{Leethochawalit}, Nicha and {Kirby}, Evan N. and {Moran}, Sean M. and {Ellis}, Richard S. and {Treu}, Tommaso},
        title = "{Evolution of the Stellar Mass-Metallicity Relation. I. Galaxies in the z {\ensuremath{\sim}} 0.4 Cluster Cl0024}",
      journal = {\apj},
         year = 2018,
        month = mar,
       volume = {856},
       number = {1},
          eid = {15},
        pages = {15},
          doi = {10.3847/1538-4357/aab26a},
archivePrefix = {arXiv},
       eprint = {1802.09560},
 primaryClass = {astro-ph.GA},
       adsurl = {https://ui.adsabs.harvard.edu/abs/2018ApJ...856...15L}
}

@ARTICLE{Johansson2012,
       author = {{Johansson}, Jonas and {Thomas}, Daniel and {Maraston}, Claudia},
        title = "{Chemical element ratios of Sloan Digital Sky Survey early-type galaxies}",
      journal = {\mnras},
         year = 2012,
        month = apr,
       volume = {421},
       number = {3},
        pages = {1908-1926},
          doi = {10.1111/j.1365-2966.2011.20316.x},
archivePrefix = {arXiv},
       eprint = {1112.0322},
 primaryClass = {astro-ph.CO},
       adsurl = {https://ui.adsabs.harvard.edu/abs/2012MNRAS.421.1908J}
}

@ARTICLE{Smith2009,
       author = {{Smith}, Russell J. and {Lucey}, John R. and {Hudson}, Michael J.},
        title = "{Ages and metallicities for quiescent galaxies in the Shapley supercluster: driving parameters of the stellar populations}",
      journal = {\mnras},
         year = 2009,
        month = dec,
       volume = {400},
       number = {4},
        pages = {1690-1705},
          doi = {10.1111/j.1365-2966.2009.15580.x},
archivePrefix = {arXiv},
       eprint = {0908.2990},
 primaryClass = {astro-ph.CO},
       adsurl = {https://ui.adsabs.harvard.edu/abs/2009MNRAS.400.1690S}
}

@ARTICLE{Graves2009,
       author = {{Graves}, Genevieve J. and {Faber}, S.~M. and {Schiavon}, Ricardo P.},
        title = "{Dissecting the Red Sequence. II. Star Formation Histories of Early-Type Galaxies Throughout the Fundamental Plane}",
      journal = {\apj},
         year = 2009,
        month = jun,
       volume = {698},
       number = {2},
        pages = {1590-1608},
          doi = {10.1088/0004-637X/698/2/1590},
archivePrefix = {arXiv},
       eprint = {0903.3603},
 primaryClass = {astro-ph.CO},
       adsurl = {https://ui.adsabs.harvard.edu/abs/2009ApJ...698.1590G}
}

@ARTICLE{Schiavon2007,
       author = {{Schiavon}, Ricardo P.},
        title = "{Population Synthesis in the Blue. IV. Accurate Model Predictions for Lick Indices and UBV Colors in Single Stellar Populations}",
      journal = {\apjs},
         year = 2007,
        month = jul,
       volume = {171},
       number = {1},
        pages = {146-205},
          doi = {10.1086/511753},
archivePrefix = {arXiv},
       eprint = {astro-ph/0611464},
 primaryClass = {astro-ph},
       adsurl = {https://ui.adsabs.harvard.edu/abs/2007ApJS..171..146S}
}

@ARTICLE{Thomas2005,
       author = {{Thomas}, Daniel and {Maraston}, Claudia and {Bender}, Ralf and {Mendes de Oliveira}, Claudia},
        title = "{The Epochs of Early-Type Galaxy Formation as a Function of Environment}",
      journal = {\apj},
         year = 2005,
        month = mar,
       volume = {621},
       number = {2},
        pages = {673-694},
          doi = {10.1086/426932},
archivePrefix = {arXiv},
       eprint = {astro-ph/0410209},
 primaryClass = {astro-ph},
       adsurl = {https://ui.adsabs.harvard.edu/abs/2005ApJ...621..673T}
}

@ARTICLE{Balogh1999,
       author = {{Balogh}, Michael L. and {Morris}, Simon L. and {Yee}, H.~K.~C. and {Carlberg}, R.~G. and {Ellingson}, Erica},
        title = "{Differential Galaxy Evolution in Cluster and Field Galaxies at z\raisebox{-0.5ex}\textasciitilde0.3}",
      journal = {\apj},
         year = 1999,
        month = dec,
       volume = {527},
       number = {1},
        pages = {54-79},
          doi = {10.1086/308056},
archivePrefix = {arXiv},
       eprint = {astro-ph/9906470},
 primaryClass = {astro-ph},
       adsurl = {https://ui.adsabs.harvard.edu/abs/1999ApJ...527...54B}
}

@BOOK{Binney1998,
       author = {{Binney}, James and {Merrifield}, Michael},
        title = "{Galactic Astronomy}",
         year = 1998,
       adsurl = {https://ui.adsabs.harvard.edu/abs/1998gaas.book.....B}
}

@ARTICLE{Zabludoff1996,
       author = {{Zabludoff}, Ann I. and {Zaritsky}, Dennis and {Lin}, Huan and {Tucker}, Douglas and {Hashimoto}, Yasuhiro and {Shectman}, Stephen A. and {Oemler}, Augustus and {Kirshner}, Robert P.},
        title = "{The Environment of ``E+A'' Galaxies}",
      journal = {\apj},
         year = 1996,
        month = jul,
       volume = {466},
        pages = {104},
          doi = {10.1086/177495},
archivePrefix = {arXiv},
       eprint = {astro-ph/9512058},
 primaryClass = {astro-ph},
       adsurl = {https://ui.adsabs.harvard.edu/abs/1996ApJ...466..104Z}
}

@ARTICLE{Dressler1983,
       author = {{Dressler}, A. and {Gunn}, J.~E.},
        title = "{Spectroscopy of galaxies in distant clusters. II. The population of the 3C 295 cluster.}",
      journal = {\apj},
         year = 1983,
        month = jul,
       volume = {270},
        pages = {7-19},
          doi = {10.1086/161093},
       adsurl = {https://ui.adsabs.harvard.edu/abs/1983ApJ...270....7D}
}

@ARTICLE{Hartwick1976,
       author = {{Hartwick}, F.~D.~A.},
        title = "{The Chemical Evolution of the Galactic Halo}",
      journal = {\apj},
         year = 1976,
        month = oct,
       volume = {209},
        pages = {418-423},
          doi = {10.1086/154735},
       adsurl = {https://ui.adsabs.harvard.edu/abs/1976ApJ...209..418H}
}

@ARTICLE{Talbot1971,
       author = {{Talbot}, Raymond J., Jr. and {Arnett}, W. David},
        title = "{The Evolution of Galaxies. I. Formulation and Mathematical Behavior of the One-Zone Model}",
      journal = {\apj},
         year = 1971,
        month = dec,
       volume = {170},
        pages = {409},
          doi = {10.1086/151228},
       adsurl = {https://ui.adsabs.harvard.edu/abs/1971ApJ...170..409T}
}

@ARTICLE{Wiseman2021,
       author = {{Wiseman}, P. and {Sullivan}, M. and {Smith}, M. and {Frohmaier}, C. and {Vincenzi}, M. and {Graur}, O. and {Popovic}, B. and {Armstrong}, P. and {Brout}, D. and {Davis}, T.~M. and {Galbany}, L. and {Hinton}, S.~R. and {Kelsey}, L. and {Kessler}, R. and {Lidman}, C. and {M{\"o}ller}, A. and {Nichol}, R.~C. and {Rose}, B. and {Scolnic}, D. and {Toy}, M. and {Zontou}, Z. and {Asorey}, J. and {Carollo}, D. and {Glazebrook}, K. and {Lewis}, G.~F. and {Tucker}, B.~E. and {Abbott}, T.~M.~C. and {Aguena}, M. and {Allam}, S. and {Andrade-Oliveira}, F. and {Annis}, J. and {Bacon}, D. and {Bertin}, E. and {Brooks}, D. and {Buckley-Geer}, E. and {Burke}, D.~L. and {Carnero Rosell}, A. and {Carrasco Kind}, M. and {Carretero}, J. and {Costanzi}, M. and {da Costa}, L.~N. and {Pereira}, M.~E.~S. and {Desai}, S. and {Diehl}, H.~T. and {Doel}, P. and {Everett}, S. and {Ferrero}, I. and {Flaugher}, B. and {Fosalba}, P. and {Frieman}, J. and {Garc{\'\i}a-Bellido}, J. and {Gaztanaga}, E. and {Giannantonio}, T. and {Gruen}, D. and {Gruendl}, R.~A. and {Gschwend}, J. and {Gutierrez}, G. and {Hollowood}, D.~L. and {Honscheid}, K. and {Hoyle}, B. and {James}, D.~J. and {Krause}, E. and {Kuehn}, K. and {Kuropatkin}, N. and {Maia}, M.~A.~G. and {Marshall}, J.~L. and {Martini}, P. and {Menanteau}, F. and {Miquel}, R. and {Morgan}, R. and {Ogando}, R.~L.~C. and {Palmese}, A. and {Paz-Chinch{\'o}n}, F. and {Petravick}, D. and {Pieres}, A. and {Plazas Malag{\'o}n}, A.~A. and {Romer}, A.~K. and {Sanchez}, E. and {Scarpine}, V. and {Schubnell}, M. and {Serrano}, S. and {Sevilla-Noarbe}, I. and {Soares-Santos}, M. and {Suchyta}, E. and {Swanson}, M.~E.~C. and {Tarle}, G. and {Thomas}, D. and {To}, C. and {Varga}, T.~N. and {Walker}, A.~R. and {DES Collaboration}},
        title = "{Rates and delay times of Type Ia supernovae in the Dark Energy Survey}",
      journal = {\mnras},
         year = 2021,
        month = sep,
       volume = {506},
       number = {3},
        pages = {3330-3348},
          doi = {10.1093/mnras/stab1943},
archivePrefix = {arXiv},
       eprint = {2105.11954},
 primaryClass = {astro-ph.GA},
       adsurl = {https://ui.adsabs.harvard.edu/abs/2021MNRAS.506.3330W}
}

@ARTICLE{Palicio2024,
       author = {{Palicio}, P.~A. and {Matteucci}, F. and {Della Valle}, M. and {Spitoni}, E.},
        title = "{Cosmic Type Ia SN rate and constraints on SN Ia progenitors}",
      journal = {arXiv e-prints},
         year = 2024,
        month = feb,
          eid = {arXiv:2402.16635},
        pages = {arXiv:2402.16635},
          doi = {10.48550/arXiv.2402.16635},
archivePrefix = {arXiv},
       eprint = {2402.16635},
 primaryClass = {astro-ph.CO},
       adsurl = {https://ui.adsabs.harvard.edu/abs/2024arXiv240216635P}
}

@ARTICLE{Strolger2020,
       author = {{Strolger}, Louis-Gregory and {Rodney}, Steven A. and {Pacifici}, Camilla and {Narayan}, Gautham and {Graur}, Or},
        title = "{Delay Time Distributions of Type Ia Supernovae from Galaxy and Cosmic Star Formation Histories}",
      journal = {\apj},
         year = 2020,
        month = feb,
       volume = {890},
       number = {2},
          eid = {140},
        pages = {140},
          doi = {10.3847/1538-4357/ab6a97},
archivePrefix = {arXiv},
       eprint = {2001.05967},
 primaryClass = {astro-ph.GA},
       adsurl = {https://ui.adsabs.harvard.edu/abs/2020ApJ...890..140S}
}

@ARTICLE{Greene2019,
       author = {{Greene}, Jenny E. and {Veale}, Melanie and {Ma}, Chung-Pei and {Thomas}, Jens and {Quenneville}, Matthew E. and {Blakeslee}, John P. and {Walsh}, Jonelle L. and {Goulding}, Andrew and {Ito}, Jennifer},
        title = "{The MASSIVE Survey. XII. Connecting Stellar Populations of Early-type Galaxies to Kinematics and Environment}",
      journal = {\apj},
         year = 2019,
        month = mar,
       volume = {874},
       number = {1},
          eid = {66},
        pages = {66},
          doi = {10.3847/1538-4357/ab01e3},
archivePrefix = {arXiv},
       eprint = {1901.01271},
 primaryClass = {astro-ph.GA},
       adsurl = {https://ui.adsabs.harvard.edu/abs/2019ApJ...874...66G}
}

@article{Gountanis2024,
  author    = {Gountanis, N. M. and Weinberg, D. H. and Sandford, N. and Beverage, A.},
  title     = {fanCE},
  journal   = {The Astrophysical Journal},
  year      = {2024},
  publisher = {Zenodo},
  note      = {fanCE v1.0.0},
  doi       = {10.5281/zenodo.13845124}
}
\bibliographystyle{aasjournal}

\end{document}